# Type-Preserving Flow Analysis and Interprocedural Unboxing

Extended Version

Neal Glew
Intel Labs
neal.glew@intel.com

Leaf Petersen
Intel Labs
leaf.petersen@intel.com

## Abstract

Interprocedural flow analysis can be used to eliminate otherwise unnecessary heap allocated objects (*unboxing*), and in previous work we have shown how to do so while maintaining correctness with respect to the garbage collector. In this paper, we extend the notion of flow analysis to incorporate types, enabling analysis and optimization of typed programs. We apply this typed analysis to specify a type preserving interprocedural unboxing optimization, and prove that the optimization preserves both type and GC safety along with program semantics. We also show that the unboxing optimization can be applied independently to separately compiled program modules, and prove via a contextual equivalence result that unboxing a module in isolation preserves program semantics.

## 1. Introduction

Many languages and compilers use a *uniform object representation* in which every source level object is represented at least initially by a heap allocated object. Such a representation allows polymorphic functions to be compiled once and enables the implementation of features that use runtime type information. In this representation machine integers and floating-point numbers are placed in a single-field object, a box, and this operation is called boxing. Operations such as addition require first projecting the number from the box (unboxing), followed by the actual addition, followed by the creation of a new box for the result (boxing). Boxing and unboxing operations add considerable overhead, and thus it is highly desirable to remove them when possible – e.g. when polymorphism or features requiring runtime type information are not being used. We refer to the general class of optimizations that attempt to remove unnecessary box and unbox operations as *unboxing* optimizations. We refer to unboxing optimizations that attempt to eliminate boxing and unboxing across function boundaries as *interprocedural unboxing*. We also include in this latter category optimizations (such as the one given in this paper) which attempt to unbox objects written to and read from other objects in the heap.

Interprocedural unboxing presents additional challenges in a typed setting, since type information must be updated to reflect any unboxing. A box might flow to an argument in an application, and the parameter of the called function might flow to an unbox operation. If the optimization decides to remove the box and unbox operations then it must also remove the box type on the parameter. In other words, typed unboxing requires not just rewriting uses and definitions in the traditional sense, but also rewriting intermediate points in the program through which the unboxed values flow. At a high level then, the optimization can be viewed as selecting a set of box operations, unbox operations, and box types to remove. Such a selection has a global consistency requirement—a box type should only be removed if all boxes that flow to it are removed, a box operation should only be removed if all unbox operations it flows to are removed, and so on. Thus choosing a set of boxed objects to eliminate and rewriting the program to reflect this choice in a consistent manner requires knowing what things flow to what points in the program, a question that flow analyses are designed to answer. In this paper we use the results of flow analysis to formulate correctness conditions for unboxing and then prove that those conditions ensure correct optimization.

In previous work [8] we considered a simpler problem, that of rewriting garbage-collector (GC) metadata rather than full types. An accurate GC requires specifying for each field of each object and each slot of each stack activation frame whether it contains a pointer into the GC heap or not (contains a machine integer, floating-point number, etc.). As with types, when interprocedurally unboxing such metadata must be rewritten in a globally consistent manner. Our previous paper showed how to do this rewriting correctly using the results of a flow analysis, in a whole program setting. In this paper we extend these ideas to develop a methodology for dealing with interprocedural optimization of statically typed languages (including universal polymorphism) in a type preserving fashion. We also show that this methodology does not depend on whole program compilation, and extends easily to support modular compilation.

In the following sections, we begin by defining a core language with a polymorphic type system that has box and unbox operations. As in our previous paper we formalize a notion of GC safety for our language—that the GC metadata is currently correct—and show that well-typed programs are GC safe throughout execution. Next we specify a set of abstract conditions that a reasonably flow analysis must satisfy, with the property that any flow analysis that satisfies these conditions can be used in our framework to optimize programs. The main section of the paper defines an unboxing optimization parameterized over a choice of objects to unbox, and gives a set of correctness conditions under which such a choice is guaranteed to preserve typing and preserve semantics. We show that this set of correctness conditions is satisfiable by constructing a simple unboxing algorithm which satisfies these conditions. Finally we extend the system slightly by defining a notion of unboxing for modules and show that it is correct in the sense that a module is contextually equivalent to its unboxing.

While our paper is specifically about the concrete optimization of unboxing, the ideas used here generalize naturally to other op-timizations that change the representation of objects in a non-local fashion. For example, dead-field elimination and dead-parameter elimination impose similar requirements for rewriting types and GC metadata in a globally consistent fashion. Flow analyses can be used to specify and implement these (and others), and we believe (based on practical experience in our compiler) that the framework presented here extends naturally to such optimizations. As far as we know, this and our previous paper are the first to use a flow analysis to rewrite types and GC metadata in a globally consistent fashion, and to use a flow analysis to formulate correctness conditions for this rewriting process and prove these conditions sound.

## 2. A type and GC safe core language

Consider the following untyped program (using informal notation), where `box` denotes a boxing operation that wraps its argument in a heap-allocated structure, and `unbox` denotes its elimination form that projects out the boxed item from the box:

$$\texttt{let}\ f\ =\ \lambda x.(\texttt{box}\,x)\ \ \texttt{in}\,\texttt{unbox}(\texttt{unbox}(f\ (\texttt{box}\,3)))$$

The only definition reaching the variable $x$ is the boxed machine integer $3$. Information from an interprocedural analysis can be used to rewrite this program to eliminate the boxing as follows:

$$\texttt{let}\ f\ =\ \lambda x.x\ \ \texttt{in}\,f\ 3$$

This second version is much better in that it does less allocation, and executes fewer instructions. In this optimized version of the program however, an important property has changed that is not reflected in this untyped synatax. Specifically, the GC status of values reaching $x$ has changed: whereas in the original program all values reaching $x$ are represented as heap allocated pointers, in the second program all values reaching $x$ are represented as machine integers. From the standpoint of a garbage collector, a garbage collection occuring while $x$ is live must treat $x$ as a root in the first program, and must ignore $x$ in the second program.

The question of which variables should be treated as roots by the garbage collector is a subtle but crucial one for the purposes of optimization and compiler correctness. Consider a modification of the previous example in which the function $f$ is used polymorphically:

$$\texttt{let}\ f\ =\ \lambda x.(\texttt{box}\,x)\ \ \texttt{in}\,\texttt{unbox}(\texttt{unbox}((\texttt{unbox}(f\ f))\ (\texttt{box}\,3)))$$

In this variant, $f$ is applied to itself and the boxed result (itself) is unboxed and applied to a boxed integer. The resulting doubly boxed integer is then unboxed. Assuming that functions are represented as heap-allocated objects, each variable in this program has a concrete and statically known status as either a GC root or GC non-root, since all objects passed to $f$ are heap references. However, an attempt to unbox this program as with the previous example results in $f$ being applied to both heap references ($f$) and non-heap references (3).

$$\texttt{let}\ f\ =\ \lambda x.x\ \ \texttt{in}(f\ f)\ 3$$

Consequently, a correct optimizer must decline to unbox this program (at least in entirety) to avoid incorrect GC behavior [1].

In our previous work[8] we developed a core language capturing the essential issues of GC safety, along with an analysis and optimization framework for reasoning about and optimizing GC safe programs in an untyped setting. This framework allows us to show that given a GC safe program, our unboxing optimization will always produce a semantically equivalent GC safe program. However, the framework is essentially limited to untyped programs and consequently it does not scale to typed core languages in which one must be able to *check* the well-typedness (and hence the GC safety) of programs before and after optimization[6]. In this paper, we intend to develop a methodology for addressing this style of optimization in a strongly typed setting.

### 2.1 Type safety

How does the problem of unboxing change in a typed setting? Consider again the first example from this section using a still informal but now typed notation:

$$\begin{array}{l}\texttt{let}\\ \quad f\colon \texttt{box}(\texttt{int}) \to \texttt{box}(\texttt{box}(\texttt{int})) = \lambda x\colon \texttt{box}(\texttt{int}).(\texttt{box}\,x)\\ \texttt{in}\,\texttt{unbox}(\texttt{unbox}(f(\texttt{box}\,3)))\end{array}$$

As before, it is apparent that the only definition reaching the variable $x$ is the boxed machine integer 3, and as before we can consider rewriting this program to eliminate (interprocedurally) the boxing. However, simply rewriting the terms of the program is inadequate from the standpoint of type preserving compilation, since the result is not well-typed:

$$\begin{array}{l}\texttt{let}\\ \quad f\colon \texttt{box}(\texttt{int}) \to \texttt{box}(\texttt{box}(\texttt{int})) = \lambda x\colon \texttt{box}(\texttt{int}).x\\ \texttt{in}\,f\ 3\end{array}$$

The types of both the actual argument and the return value of $f$ have changed, and are no longer consistent with the type annotation for $f$ and $x$. In order to correctly unbox this program then, it is necessary to rewrite not just the terms, but also the types:

$$\texttt{let}\ f{:}\texttt{int} \to \texttt{int} = \lambda x{:}\texttt{int}.x\ \ \texttt{in}\,f\ 3$$

This requirement is a more substantial change than might at first be apparent. In the original (untyped) setting, it was sufficient to have information only about the direct definitions (boxes) and uses (unboxes) of objects. Rewriting the types in this fashion requires information not just about the uses and definitions, but also about intermediate program points (and other objects) through which the boxed objects flow.

In addition to incurring these additional rewriting requirements, the typed setting must still account for GC safety. Consider again the polymorphic variant of the previous untyped example (naming the first application of $f$ for clarity).

$$\begin{array}{l}\texttt{let}\\ \quad f{:}\forall\alpha.\alpha \to \texttt{box}(\alpha) = \Lambda\alpha.\lambda x{:}\alpha.(\texttt{box}\,x)\\ \quad g\colon \texttt{box}(\forall\alpha.\alpha \to \texttt{box}(\alpha)) = f[\forall\alpha.\alpha \to \texttt{box}(\alpha)](f)\\ \texttt{in}\,\texttt{unbox}(\texttt{unbox}((\texttt{unbox}\,g)[\texttt{box}(\texttt{int})](\texttt{box}\,3)))\end{array}$$

Here we have $f$ applied to itself at a universal type to produce a boxed version of itself ($g$), which is then unboxed and applied to a boxed integer, at the boxed integer type. Attempting to unbox this example (rewriting types as necessary) immediately illuminates the problem.

$$\begin{array}{l}\texttt{let}\\ \quad f{:}\forall\alpha.\alpha \to \alpha = \Lambda\alpha.\lambda x{:}\alpha.x\\ \quad g{:}\forall\alpha.\alpha \to \alpha = f[\forall\alpha.\alpha \to \alpha](f)\\ \texttt{in}\,g[\texttt{int}](3)\end{array}$$

The function $f$ is instantiated directly at a universal function type, and via its alias ($g$) at a machine integer type. As with the untyped example in the previous section, the compiler cannot assign a concrete GC status to the variable $x$. For correctness then, the compiler must not (fully) unbox this example, and must leave at least the boxing operation on the integer parameter to $f$ [2].

[1] A conservative GC, or a GC implementation which tags pointers to distinguish them from non-pointers would not impose this restriction. See Section 2.2 for more discussion of the GC model.

[2] It is worth noting that an optimizing compiler might choose to duplicate the body of $f$ to make it monomorphic, and hence allow it to be unboxed. It is also possible to use a runtime type passing interpretation to relax the

| | | | |
|---|---|---|---|
| Traceabilities | $t$ | ::= | $\mathtt{b} \mid \mathtt{r}$ |
| Labels | $i$ | ::= | $0, 1, \ldots$ |
| Type variables | | | $\alpha, \beta$ |
| Labeled Types | $\tau$ | ::= | $\sigma^i$ |
| Types | $\sigma$ | ::= | $\alpha \mid \mathtt{B} \mid \forall\alpha.\tau_1 \to \tau_2 \mid \mathtt{box}(\tau)$ |
| Term variables | | | $f, x, y, z$ |
| Constants | | | $c$ |
| Labeled Terms | $e$ | ::= | $m^i \mid v^i$ |
| Terms | $m$ | ::= | $x \mid \mathtt{fix}\ f[\alpha](x{:}\tau_1){:}\tau_2.e \mid e_1[\tau]\ e_2 \mid$ $\mathtt{box}_\tau\, e \mid \mathtt{unbox}\, e \mid \rho(e)$ |
| Values | $v$ | ::= | $c \mid \langle \rho, \mathtt{fix}\ f[\alpha](x{:}\tau_1){:}\tau_2.e\rangle \mid \langle v^i{:}\tau\rangle$ |
| Environments | $\rho$ | ::= | $x_1{:}\tau_1 = {v_1}^{j_1}, \ldots, x_n{:}\tau_n = {v_n}^{j_n}$ |
| States | $M$ | ::= | $(\rho, e)$ |

**Figure 1.** Syntax

In the rest of this paper we make these issues concrete and formal, and we show how to deal with them issues by extending the notion of flow analysis to incorporate types, thereby generating the necessary flow information to correctly rewrite types and terms in a consistent fashion. While we focus on a concrete optimization (unboxing), we believe that these ideas are generally applicable to flow analysis based representation optimizations in typed intermediate languages.

### 2.2 A core language for GC safety

In order to give a precise account of typed flow analysis and interprocedural unboxing, we begin by defining a type safe core language incorporating the essential features of GC safety. The motivation for the (small) idiosyncracies of this language lies in the requirements of the underlying model of garbage collection. We assume that pointers cannot be intrinsically distinguished from non-pointers, and hence the compiler is required to statically annotate the program with garbage collection meta-data such that at any garbage collection point the garbage collector can reconstruct exactly which live variables are roots. Typically, this takes the form of annotations on variables and temporaries indicating which contain heap-pointers (the roots) and which do not (the non-roots), along with information at every allocation site indicating which fields of the allocated object contain traceable data. This approach is common in modern systems, and it is this approach that we target in this paper.

Figure 1 defines the syntax of our core language. The essence of the language is a standard polymorphic lambda calculus extended with a fix point operator, implemented via an explicit environment semantics. For the purposes of the semantics, we also include a form of degenerate type information we call *traceabilities*. Traceabilities describe the GC status of variables: the traceability $\mathtt{b}$ (for bits) indicates something that should be ignored by the garbage collector, while the traceability $\mathtt{r}$ (for reference) indicates a GC-managed pointer. The traceability $\mathtt{b}$ is inhabited by an unspecified set of constants $c$ while the traceability $\mathtt{r}$ is inhabited by functions (anticipating their implementation by heap-allocated closures) and by boxed objects. Anticipating the needs of the flow analysis, we label each type, term, value, and variable binding site with an integer label. We do not assume that labels or variables are unique within a program.

Types $\sigma$ consist of type variables, the base type of constants, $\mathtt{B}$, function types, $\forall\alpha.\tau_1 \to \tau_2$, and boxed types $\mathtt{box}(\tau)$. In order to provide a concrete implementation strategy for the garbage collector, we insist that every type correspond to a traceability so that we can extract the necessary garbage collection meta-data. Types are mapped to traceabilities using the function $tr(\tau)$, defined in Figure 2. Polymorphic functions are restricted by well-formedness rules to only be instantiated with types with the traceability $\mathtt{r}$, and consequently $tr(\alpha) = \mathtt{r}$. We define substitution of types in the standard way and define $\tau[\sigma^i/\alpha] = \tau[\sigma/\alpha]$.

Expressions $e$ consist of labeled terms $m^i$ and labeled values $v^i$. The terms $m$ consist of variables, functions, applications, box introductions, box eliminations, and frames. Functions $\mathtt{fix}\ f[\alpha](x{:}\tau_1){:}\tau_2.e$ are polymorphic and recursive and variable binding sites are decorated with types. We represent heap allocation in the language via the $\mathtt{box}_\tau\, e$ term, which corresponds to allocating a heap cell containing the value for $e$. The type $\tau$ is used by the dynamic semantics to provide the meta-data with which the heap-cell will be tagged, allowing the garbage collector to trace the cell. However, only the top-level traceability of the type (given by the $tr()$ function in Figure 2) is actually required by the dynamic semantics, and so the language can be erased into an untyped language in the obvious way. Objects can be projected out of an allocated object by the $\mathtt{unbox}\, e$ operation. Frames $\rho(e)$ are discussed further below.

Values consist of either constants, closures, or heap-allocated boxes. We distinguish between the introduction form ($\mathtt{box}_\tau\ e$) and the value form ($\langle v^i{:}\tau\rangle$) for allocated objects. The introduction form corresponds to the allocation instruction, whereas the value form corresponds to the allocated heap value. This distinction is key for the formulation of GC safety and the dynamic semantics. For the purposes of the dynamic semantics we also distinguish between functions ($\mathtt{fix}\ f[\alpha](x{:}\tau_1){:}\tau_2.e$) and the heap allocated closures that represent them at runtime ($\langle \rho, \mathtt{fix}\ f[\alpha](x{:}\tau_1){:}\tau_2.e\rangle$).

For notational convenience, we will sometimes use the notation $v_\mathtt{b}$ to indicate that a value $v$ is a non-heap-allocated value (i.e. a constant $c$), and $v_\mathtt{r}$ to indicate that a value $v$ is a heap-allocated value (i.e. either a function value or a boxed value). If $t$ is a traceability meta-variable, then we use $v_t$ to indicate that $v$ is a value of the same traceability as $t$.

In examples, we use a derived `let` expression, taking it to be syntactic sugar for application in the usual manner. Environments $\rho$ map variables to values. The term $\rho(e)$ executes $e$ in the environment $\rho$ rather than the outer environment – all of the free variables of $e$ are provided by $\rho$. The nested set of these environments at any point can be thought of as the activation stack frames of the executing program. The traceability of the typing annotations on variables in the environments play the role of stack-frame GC meta-data, indicating which slots of the frame are roots (traceability $\mathtt{r}$). The environments buried in closures ($\langle \rho, \mathtt{fix}\ f[\alpha](x{:}\tau_1){:}\tau_2.e\rangle$) similarly provide the traceabilities of values reachable from the closure via the type annotations on the variables in the environment, and hence provide the GC meta-data for tracing through closures. While we do not make the process of garbage collection explicit, it should be clear how to extract the appropriate set of GC roots from the environment and any active frames.

This core language contains the appropriate information to formalize a notion of GC safety consisting of two complementing pieces. First we define a dynamic semantics in which reductions that might lead to undefined garbage-collector behavior are explicitly undefined. Programs that take steps in this semantics do not introduce ill-formed heap objects. Secondly, we define a notion of a traceable program: one in which all heap values have valid GC meta-data. Reduction steps in the semantics can then be shown to maintain the traceability property in addition to the usual well-typedness property. The GC correctness criteria for a compiler optimization then becomes simply the usual one: that the optimization map well-typed programs to semantically equivalent well-typed constraints on the garbage collector sufficiently to permit this example[1]. These optimizations are orthogonal (but complementary) to the issues addressed by this paper.

programs. By showing that typable programs are both traceable programs and have well-defined semantics, we thereby show that GC correctness for a compiler optimization can be achieved simply by preserving well-typedness.

It is worth noting that in our implementation, the compiler intermediate language under consideration is substantially more low-level: a control-flow graph based, static single assignment intermediate representation. We believe however that all of the key issues are captured faithfully in this higher-level representation, and with greater clarity of presentation.

### 2.3 Operational semantics

We choose to use an explicit environment semantics rather than a standard substitution semantics since this makes the GC meta-data (implicit in the types) for stack frames and closures explicit in the semantics. Thus a machine state $(\rho, e)$ supplies an environment $\rho$ for $e$ that provides the values of the free variables of $e$ during execution. Environments contain typing annotations on each of the variables mapped by the environment, which provide the traceabilities of the variables.

Reduction in this language is for the most part fairly standard. We deviate somewhat in that we explicitly model the allocation of heap objects as a reduction step—hence there is an explicit reduction mapping a function term $\texttt{fix}\ \ f[\alpha](x{:}\tau_1){:}\tau_2.e$ to an allocated closure $\langle\rho, \texttt{fix}\ \ f[\alpha](x{:}\tau_1){:}\tau_2.e\rangle$, and similarly for boxed objects and values. More notably, beta-reduction is restricted to only permit construction of a stack frame when the type for the parameter variable has an appropriate traceability for the actual argument value. This captures the requirement that stack frames have correct meta-data for the garbage collector. In actual practice, incorrect meta-data for stack frames leads to undefined behavior (since incorrect meta-data may cause arbitrary memory corruption by the garbage collector)—similarly here in the meta-theory we leave the behavior of such programs undefined. In a similar fashion, we only define the reduction of the allocation operation to an allocated value ($\texttt{box}_\tau\ v_t \longmapsto \langle v_t{:}\tau\rangle$) when the operation meta-data is appropriate for the value (i.e. $tr(\tau) = t$).

It is important to note that this semantics does not model a dynamically checked language, in which there is an explicit check of the meta-data associated with these reductions. The point is simply that the semantics only specifies how programs behave when these conditions are met—in all other cases the behavior of the program is undefined.

### 2.4 Traceability

The operational semantics ensures that no reduction step introduces mis-tagged values. In order to make use of this, we define a judgment for checking that a program does not have a mis-tagged value in the first place. Implicitly this judgement defines what a well-formed heap and activation stack looks like; however, since our heap and stack are implicit in our machine states, it takes the form of a judgement on terms, values, environments, and machine states.

The value judgement $\vdash_{\text{v}} v{:}t$ asserts that a value $v$ is well-formed, and has traceability $t$. In this simple language, this corresponds to having the types on the variables in the environment of each function value have traceabilities which are consistent with the values to which they are bound, and the type on each boxed value be consistent with the traceability of the object nested in the box. An environment is consistent, $\vdash \rho\ \mathbf{tr}$, when the annotation on each variable agrees with the traceability of the value it is bound to. The term judgement $\vdash e\ \mathbf{tr}$ and machine state judgement $\vdash M\ \mathbf{tr}$ simply check that all values and environments (and hence stack frames) contained in the term or machine state are well-formed.

$$\begin{array}{lcl} tr(\sigma^i) & = & tr(\sigma) \\ tr(\alpha) & = & \texttt{r} \\ tr(\texttt{B}) & = & \texttt{b} \\ tr(\forall\alpha.\tau_1 \to \tau_2) & = & \texttt{r} \\ tr(\texttt{box}(\tau)) & = & \texttt{r} \end{array}$$

$$\frac{x{:}\tau = v^j \in \rho}{(\rho, x^k) \longmapsto (\rho, v^j)}$$

$$\frac{}{(\rho, (\texttt{fix}\ \ f[\alpha](x{:}\tau_1){:}\tau_2.e)^j) \longmapsto (\rho, \langle\rho, \texttt{fix}\ \ f[\alpha](x{:}\tau_1){:}\tau_2.e\rangle^j)}$$

$$\frac{tr(\tau) = t}{(\rho, (\texttt{box}_\tau\ {v_t}^i)^j) \longmapsto (\rho, \langle {v_t}^i{:}\tau\rangle^j)}$$

$$\frac{(\rho, e_1) \longmapsto (\rho, e_1')}{(\rho, (e_1[\tau]\ e_2)^i) \longmapsto (\rho, (e_1'[\tau]\ e_2)^i)}$$

$$\frac{(\rho, e_2) \longmapsto (\rho, e_2')}{(\rho, (v^i[\tau]\ e_2)^j) \longmapsto (\rho, (v^i[\tau]\ e_2')^j)}$$

$$\frac{v_f = \langle\rho', \texttt{fix}\ \ f[\alpha](x{:}\tau_1){:}\tau_2.e\rangle \quad \tau' = (\forall\alpha.\tau_1 \to \tau_2)^i \quad \tau_1' = \tau_1[\tau/\alpha] \quad tr(\tau_1') = t}{(\rho, ({v_f}^i[\tau]\ {v_t}^j)^k) \longmapsto (\rho, (\rho', f{:}\tau' = {v_f}^i, x{:}\tau_1' = {v_t}^j)(e[\tau/\alpha])^k)}$$

$$\frac{(\rho, e) \longmapsto (\rho, e')}{(\rho, (\texttt{box}_\tau\ e)^i) \longmapsto (\rho, (\texttt{box}_\tau\ e')^i)}$$

$$\frac{(\rho, e) \longmapsto (\rho, e')}{(\rho, (\texttt{unbox}\, e)^i) \longmapsto (\rho, (\texttt{unbox}\, e')^i)}$$

$$\frac{}{(\rho, (\texttt{unbox}\ \langle v^i{:}\tau\rangle^j)^k) \longmapsto (\rho, v^i)}$$

$$\frac{(\rho', e) \longmapsto (\rho', e')}{(\rho, \rho'(e)^i) \longmapsto (\rho, \rho'(e')^i)} \qquad \frac{}{(\rho, \rho'(v^i)^j) \longmapsto (\rho, v^i)}$$

**Figure 2.** Operational Semantics

The key result for traceability is that it is preserved under reduction. That is, if a traceable term takes a well-defined reduction step, then the resulting term will be traceable.

**Lemma 1 (Preservation of traceability)**
*If* $\vdash M\ \mathbf{tr}$ *and* $M \longmapsto M'$ *then* $\vdash M'\ \mathbf{tr}$.

**Proof:** If $\vdash (\rho, e)\ \mathbf{tr}$ then $\vdash \rho\ \mathbf{tr}$ and $\vdash e\ \mathbf{tr}$. If $(\rho, e) \longmapsto (\rho, e')$ then the result follows if we can show $\vdash e'\ \mathbf{tr}$. The proof of that is by induction on the derivation of $(\rho, e) \longmapsto (\rho, e')$. Consider the cases for the last rule used to derive it (the cases are in the same order as in the figure):

- In this case, $e = x^k$ for some $x$ and $k$, and $e' = v^j$ where $x{:}\tau = v^j \in \rho$ for some $\tau$, $v$, and $j$. Since $\vdash \rho\ \mathbf{tr}$ then $\vdash_{\text{v}} v{:}tr(\tau)$, so by the traceability rules $\vdash v^j\ \mathbf{tr}$ as required.

**Labeled Terms** $\boxed{\vdash e\ \mathbf{tr}}$

$$\frac{\vdash m\ \mathbf{tr}}{\vdash m^i\ \mathbf{tr}} \qquad \frac{\vdash_{\mathrm{v}} v{:}t}{\vdash v^i\ \mathbf{tr}}$$

**Terms** $\boxed{\vdash m\ \mathbf{tr}}$

$$\frac{}{\vdash x\ \mathbf{tr}} \qquad \frac{\vdash e\ \mathbf{tr}}{\vdash \mathtt{fix}\ f[\alpha](x{:}\tau_1){:}\tau_2.e\ \mathbf{tr}}$$

$$\frac{\vdash e_1\ \mathbf{tr} \quad \vdash e_2\ \mathbf{tr}}{\vdash e_1[\tau]\ e_2\ \mathbf{tr}} \qquad \frac{\vdash e\ \mathbf{tr}}{\vdash \mathtt{box}_\tau\ e\ \mathbf{tr}}$$

$$\frac{\vdash e\ \mathbf{tr}}{\vdash \mathtt{unbox}\, e\ \mathbf{tr}} \qquad \frac{\vdash \rho\ \mathbf{tr} \quad \vdash e\ \mathbf{tr}}{\vdash \rho(e)\ \mathbf{tr}}$$

**Values** $\boxed{\vdash_{\mathrm{v}} v{:}t}$

$$\frac{}{\vdash_{\mathrm{v}} c{:}\mathtt{b}} \qquad \frac{\vdash_{\mathrm{v}} v{:}tr(\tau)}{\vdash_{\mathrm{v}} \langle v^i{:}\tau\rangle{:}\mathtt{r}}$$

$$\frac{\vdash \rho\ \mathbf{tr} \quad \vdash e\ \mathbf{tr}}{\vdash_{\mathrm{v}} \langle \rho, \mathtt{fix}\ f[\alpha](x{:}\tau_1){:}\tau_2.e\rangle{:}\mathtt{r}}$$

**Environments** $\boxed{\vdash \rho\ \mathbf{tr}}$

$$\frac{\vdash_{\mathrm{v}} v_1{:}tr(\tau_1) \quad \cdots \quad \vdash_{\mathrm{v}} v_n{:}tr(\tau_n)}{\vdash x_1{:}\tau_1 = {v_1}^{j_1}, \ldots, x_n{:}\tau_n = {v_n}^{j_n}\ \mathbf{tr}}$$

**Machine States** $\boxed{\vdash M\ \mathbf{tr}}$

$$\frac{\vdash \rho\ \mathbf{tr} \quad \vdash e\ \mathbf{tr}}{\vdash (\rho, e)\ \mathbf{tr}}$$

**Figure 3.** Traceability

- In this case, $e = (\mathtt{fix}\ f[\alpha](x{:}\tau_1){:}\tau_2.e'')^j$ for some $f$, $\alpha$, $x$, $\tau_1$, $\tau_2$, $e''$, and $j$, and $e' = \langle \rho, \mathtt{fix}\ f[\alpha](x{:}\tau_1){:}\tau_2.e''\rangle^j$. The first hypothesis is that $\vdash (\mathtt{fix}\ f[\alpha](x{:}\tau_1){:}\tau_2.e'')^j\ \mathbf{tr}$. There is only one rule to derive this judgement and that rule requires that $\vdash \mathtt{fix}\ f[\alpha](x{:}\tau_1){:}\tau_2.e''\ \mathbf{tr}$, which in turn can only be derived by one rule that requires that $\vdash e''\ \mathbf{tr}$. Then, and since $\vdash \rho\ \mathbf{tr}$, by the rules for traceability, $\vdash_{\mathrm{v}} \langle \rho, \mathtt{fix}\ f[\alpha](x{:}\tau_1){:}\tau_2.e''\rangle{:}\mathtt{r}$, and by the traceability rules again $\vdash \langle \rho, \mathtt{fix}\ f[\alpha](x{:}\tau_1){:}\tau_2.e''\rangle^j\ \mathbf{tr}$, as we are required to prove.
- In this case, $e = (\mathtt{box}_\tau\, {v_t}^i)^j$ for some $\tau$, $v_t$, $i$, and $j$, $e' = \langle {v_t}^i{:}\tau\rangle^j$, and $tr(\tau) = t$. The first hypothesis is that $\vdash (\mathtt{box}_\tau\, {v_t}^i)^j\ \mathbf{tr}$. There is only one rule to derive this judgement and that rule requires that $\vdash \mathtt{box}_\tau\, {v_t}^i\ \mathbf{tr}$, which in turn can only be derived by one rule that requires that $\vdash {v_t}^i\ \mathbf{tr}$. There is only one rule to derive the latter judgement and it requires that $\vdash_{\mathrm{v}} v_t{:}t'$ for some $t'$. By inspection of the rules for value traceability, we see that $t = t'$. Since $tr(\tau) = t = t'$, by the rules for traceability, $\vdash_{\mathrm{v}} \langle {v_t}^i{:}\tau\rangle{:}\mathtt{r}$, and by the traceability rules again $\vdash \langle {v_t}^i{:}\tau\rangle^j\ \mathbf{tr}$, as we are required to prove.
- In this case, $e = (e_1[\tau]\ e_2)^i$ for some $e_1$, $\tau$, $e_2$, and $i$, $e' = (e_1'[\tau]\ e_2)^i$ for some $e_1'$, and $(\rho, e_1) \longmapsto (\rho, e_1')$ is a subderivation. The first hypothesis is that $\vdash (e_1[\tau]\ e_2)^i\ \mathbf{tr}$. There is only one rule to derive this judgement and that rule requires that $\vdash e_1[\tau]\ e_2\ \mathbf{tr}$, which in turn can only be derived by one rule that requires both $\vdash e_1\ \mathbf{tr}$ and $\vdash e_2\ \mathbf{tr}$. Thus, by the induction hypothesis, $\vdash e_1'\ \mathbf{tr}$. Then, by the rules for traceability, $\vdash e_1'[\tau]\ e_2\ \mathbf{tr}$, and by the traceability rules again, $\vdash (e_1'[\tau]\ e_2)^i\ \mathbf{tr}$, as we are required to prove.
- In this case, $e = (v^i[\tau]\ e_2)^j$ for some $v$, $i$, $\tau$, $e_2$, and $j$, $e' = (v^i[\tau]\ e_2')^j$ for some $e_2'$, and $(\rho, e_2) \longmapsto (\rho, e_2')$ is a subderivation. The first hypothesis is that $\vdash (v^i[\tau]\ e_2)^j\ \mathbf{tr}$. There is only one rule to derive this judgement and that rule requires that $\vdash v^i[\tau]\ e_2\ \mathbf{tr}$, which in turn can only be derived by one rule that requires both $\vdash v^i\ \mathbf{tr}$ and $\vdash e_2\ \mathbf{tr}$. Thus, by the induction hypothesis, $\vdash e_2'\ \mathbf{tr}$. Then, by the rules for traceability, $\vdash v^i[\tau]\ e_2'\ \mathbf{tr}$, and by the traceability rules again, $\vdash (v^i[\tau]\ e_2')^j\ \mathbf{tr}$, as we are required to prove.
- In this case:

$$\begin{array}{l} e = ({v_f}^j[\tau]\ {v_t}^k)^l \\ v_f = \langle \rho', \mathtt{fix}\ f[\alpha](x{:}\tau_1){:}\tau_2.e''\rangle \\ e' = \rho''(e''[\tau/\alpha])^l \\ \rho'' = \rho', f{:}\tau' = {v_f}^j, x{:}\tau_1' = {v_t}^k \\ \tau' = (\forall\alpha.\tau_1 \to \tau_2)^j \\ \tau_1' = \tau_1[\tau/\alpha] \\ tr(\tau_1') = t \qquad (6) \end{array}$$

for some $\rho'$, $f$, $\alpha$, $x$, $\tau_1$, $\tau_2$, $e''$, $j$, $\tau$, $v_t$, $k$, and $l$. The first hypothesis is that $\vdash ({v_f}^j[\tau]\ {v_t}^k)^l\ \mathbf{tr}$. There is only one rule to derive that judgement and that rule requires that $\vdash {v_f}^j[\tau]\ {v_t}^k\ \mathbf{tr}$, which in turn can only be derived by one rule that requires both $\vdash {v_f}^j\ \mathbf{tr}$ and $\vdash {v_t}^k\ \mathbf{tr}$. Both of these latter derivations can only be derived by one rule and those rules require that $\vdash_{\mathrm{v}} v_f{:}\mathtt{r}$ (1) and $\vdash_{\mathrm{v}} v_t{:}t$ (2) (a simple inspection reveals the traceabilities to be $\mathtt{r}$ and $t$). Judgement 1 can only be derived by one rule and that rule requires that $\vdash \rho'\ \mathbf{tr}$ (3) and $\vdash e''\ \mathbf{tr}$ (4). By (3), (1), $tr(\tau') = \mathtt{r}$, (2), and (6) we can derive $\vdash \rho''\ \mathbf{tr}$ (5). By (5) and (4) we can derive $\vdash \rho''(e'')\ \mathbf{tr}$, and then $\vdash e'\ \mathbf{tr}$, as required.
- In this case, $e = (\mathtt{box}_\tau\, e'')^i$ for some $\tau$, $e''$, and $i$, $e' = (\mathtt{box}_\tau\, e''')^i$, and $(\rho, e'') \longmapsto (\rho, e''')$ is a subderivation. The first hypothesis is that $\vdash (\mathtt{box}_\tau\, e'')^i\ \mathbf{tr}$. There is only one rule to derive this judgement and that rule requires that $\vdash \mathtt{box}_\tau\, e''\ \mathbf{tr}$, which in turn can only be derived by one rule that requires that $\vdash e''\ \mathbf{tr}$. Thus, by the induction hypothesis, $\vdash e'''\ \mathbf{tr}$. Then, by the rules for traceability, $\vdash \mathtt{box}_\tau\, e'''\ \mathbf{tr}$, and by the traceability rules again, $\vdash (\mathtt{box}_\tau\, e''')^i\ \mathbf{tr}$, as we are required to prove.
- In this case, $e = (\mathtt{unbox}\, e'')^i$ for some $e''$ and $i$, $e' = (\mathtt{unbox}\, e''')^i$, and $(\rho, e'') \longmapsto (\rho, e''')$ is a subderivation. The first hypothesis is that $\vdash (\mathtt{unbox}\, e'')^i\ \mathbf{tr}$. There is only one rule to derive this judgement and that rule requires that $\vdash \mathtt{unbox}\, e''\ \mathbf{tr}$, which in turn can only be derived by one rule that requires that $\vdash e''\ \mathbf{tr}$. Thus, by the induction hypothesis, $\vdash e'''\ \mathbf{tr}$. Then, by the rules for traceability, $\vdash \mathtt{unbox}\, e'''\ \mathbf{tr}$, and by the traceability rules again, $\vdash (\mathtt{unbox}\, e''')^i\ \mathbf{tr}$, as we are required to prove.
- In this case, $e = (\mathtt{unbox}\, \langle v^i{:}\tau\rangle^j)^k$ for some $\tau$, $v$, $i$, $j$, and $k$, and $e' = v^i$. The first hypothesis is that $\vdash (\mathtt{unbox}\, \langle v^i{:}\tau\rangle^j)^k\ \mathbf{tr}$. There is only one rule to derive this judgement and that rule requires that $\vdash \mathtt{unbox}\, \langle v^i{:}\tau\rangle^j\ \mathbf{tr}$, which in turn can only be

$$
\begin{array}{rcl}
\Delta & ::= & \alpha_1, \ldots, \alpha_n \\
\Gamma & ::= & x_1{:}\tau_1, \ldots, x_n{:}\tau_n
\end{array}
$$

$\boxed{\Delta \vdash \tau\ \mathit{wf}}$

$$\frac{\mathit{ftv}(\tau) \subseteq \Delta}{\Delta \vdash \tau\ \mathit{wf}}$$

$\boxed{\vdash \tau_1 = \tau_2}$

$$\frac{}{\vdash \alpha^i = \alpha^j} \qquad \frac{}{\vdash \mathtt{B}^i = \mathtt{B}^j}$$

$$\frac{\vdash \tau_{11} = \tau_{21} \quad \vdash \tau_{12} = \tau_{22}}{\vdash (\forall\alpha.\tau_{11} \to \tau_{12})^i = (\forall\alpha.\tau_{21} \to \tau_{22})^j}$$

$$\frac{\vdash \tau_1 = \tau_2}{\vdash \mathtt{box}(\tau_1)^i = \mathtt{box}(\tau_2)^j}$$

$\boxed{\vdash \rho : \Gamma}$

$$\frac{\begin{array}{c}\emptyset \vdash \tau_1\ \mathit{wf} \quad \cdots \quad \emptyset \vdash \tau_n\ \mathit{wf} \\ \emptyset;\emptyset \vdash {v_1}^{i_1} : \tau_1' \quad \cdots \quad \emptyset;\emptyset \vdash {v_n}^{i_n} : \tau_n' \\ \vdash \tau_1 = \tau_1' \quad \cdots \quad \vdash \tau_n = \tau_n'\end{array}}{\vdash x_1{:}\tau_1 = {v_1}^{i_1}, \ldots, x_n{:}\tau_n = {v_n}^{i_n} : x_1{:}\tau_1, \ldots, x_n{:}\tau_n}$$

$\boxed{\vdash M : \tau}$

$$\frac{\vdash \rho : \Gamma \quad \emptyset;\Gamma \vdash e : \tau}{\vdash (\rho, e) : \tau}$$

**Figure 4.** Type rules, other constructs

$\boxed{\Delta;\Gamma \vdash e : \tau}$

$$\frac{x{:}\tau \in \Gamma}{\Delta;\Gamma \vdash x^i : \tau}$$

$$\frac{\begin{array}{c}\Delta \vdash (\forall\alpha.\tau_1 \to \tau_2)^i\ \mathit{wf} \\ \Delta,\alpha;\Gamma, f{:}(\forall\alpha.\tau_1 \to \tau_2)^i, x{:}\tau_1 \vdash e : \tau_2\end{array}}{\Delta;\Gamma \vdash (\mathtt{fix}\ \ f[\alpha](x{:}\tau_1){:}\tau_2.e)^i : (\forall\alpha.\tau_1 \to \tau_2)^i}$$

$$\frac{\begin{array}{c}\Delta;\Gamma \vdash e_1 : (\forall\alpha.\tau_1 \to \tau')^j \quad \Delta;\Gamma \vdash e_2 : \tau_2 \\ \Delta \vdash \tau\ \mathit{wf} \quad \mathit{tr}(\tau) = \mathtt{r} \quad \vdash \tau_1[\tau/\alpha] = \tau_2\end{array}}{\Delta;\Gamma \vdash (e_1[\tau]\ e_2)^i : \tau'[\tau/\alpha]}$$

$$\frac{\Delta \vdash \tau\ \mathit{wf} \quad \Delta;\Gamma \vdash e : \tau' \quad \vdash \tau = \tau'}{\Delta;\Gamma \vdash (\mathtt{box}_\tau\ e)^i : \mathtt{box}(\tau)^i}$$

$$\frac{\Delta;\Gamma \vdash e : \mathtt{box}(\tau)^j}{\Delta;\Gamma \vdash (\mathtt{unbox}\,e)^i : \tau}$$

$$\frac{\vdash \rho : \Gamma' \quad \emptyset;\Gamma' \vdash e : \tau}{\Delta;\Gamma \vdash \rho(e)^i : \tau}$$

$$\frac{}{\Delta;\Gamma \vdash c^i : \mathtt{B}^i}$$

$$\frac{\begin{array}{c}\vdash \rho : \Gamma' \quad \emptyset \vdash (\forall\alpha.\tau_1 \to \tau_2)^i\ \mathit{wf} \\ \alpha;\Gamma', f{:}(\forall\alpha.\tau_1 \to \tau_2)^i, x{:}\tau_1 \vdash e : \tau_2\end{array}}{\Delta;\Gamma \vdash \langle \rho, \mathtt{fix}\ \ f[\alpha](x{:}\tau_1){:}\tau_2.e\rangle^i : (\forall\alpha.\tau_1 \to \tau_2)^i}$$

$$\frac{\emptyset \vdash \tau\ \mathit{wf} \quad \Delta;\Gamma \vdash v^j : \tau' \quad \vdash \tau = \tau'}{\Delta;\Gamma \vdash \langle v^j{:}\tau\rangle^i : \mathtt{box}(\tau)^i}$$

**Figure 5.** Type rules, expressions

derived by one rule that requires that $\vdash \langle v^i{:}\tau\rangle^j\ \mathbf{tr}$. There is only one rule to derive this latter judgement and that rule requires that $\vdash_{\mathrm{v}} \langle v^i{:}\tau\rangle{:}t$ for some $t$, which in turn can only be derived by one rule that requires that $\vdash_{\mathrm{v}} v{:}\mathit{tr}(\tau)$. Then, by the rules for traceability, $\vdash v^i\ \mathbf{tr}$, as we are required to prove.

- In this case, $e = \rho'(e'')^i$ for some $\rho'$, $e''$ and $i$, $e' = \rho'(e''')^i$ for some $e'''$, and $(\rho', e'') \longmapsto (\rho', e''')$. The hypothesis $\vdash e\ \mathbf{tr}$ can only be derived in a certain way, unpacking that we see that $\vdash \rho'\ \mathbf{tr}$ and $\vdash e''\ \mathbf{tr}$. Then by the induction hypothesis, $\vdash e'''\ \mathbf{tr}$. So applyling the rules, we derive that $\vdash \rho'(e''')\ \mathbf{tr}$ and then $\vdash e'\ \mathbf{tr}$, as required.
- In this case, $e = \rho'(v^i)^j$ for some $\rho'$, $v$, $i$, and $j$, and $e' = v^i$. The hypothesis, $\vdash e\ \mathbf{tr}$ can only be derived in one way and unpacking that we see that $\vdash v^i\ \mathbf{tr}$, which is what we are required to prove.

■

There is no corresponding progress property for our notion of traceability, since in the absence of further guarantees, programs can go wrong. However, typable programs are both traceable and do not go wrong, as we will see in the next section, and so preserving typability ensures GC correctness.

### 2.5 Typing

The typing rules appear in Figures 4 and 5. They are for the most part standard except for three modifications. First, as types are labelled, we must sometimes ignore the labels in typing. Judgement $\vdash \tau_1 = \tau_2$ states that types $\tau_1$ and $\tau_2$ are syntactically equivalent except that the labels on their sub-terms might differ. This is important in (for example) the rule for application, where we require only that the parameter type $\tau_1$ and the actual argument type $\tau_2$ satisfy $\vdash \tau_1 = \tau_2$ rather than $\tau_1 = \tau_2$; similarly in the rule for environments. Second, in the rules for boxes we require that the traceability of the box equal the traceability of the type of the thing being boxed. This is essential for showing that well-typedness implies traceability. Finally, the instantiation rule for polymorphic functions enforces the property that the type argument have traceability $\mathtt{r}$.

One particularly important aspect of our language is that we assume a type erasure semantics. For this interpretation to be correct, we must show that we can compute the correct GC metadata when erasing types. The operational semantics have the application of a polymorphic function step to a frame where the annotation on the function's parameter is a substituted type. We need that the GC metadata for this substituted type equal the GC metadata for the unsubstituted parameter type of the function. The requirement $\mathit{tr}(\tau) = \mathtt{r}$ in the typing rule for application is crucial to that equality, and the following lemma proves it.

**Lemma 2**
*If $tr(\tau) = \mathtt{r}$ then $tr(\tau') = tr(\tau'[\tau/\alpha])$.*

**Proof:** The proof is by inspection of the definitions. ∎

We can prove type safety for this language in the standard way, via progress and preservations lemmas. First we need several lemmas: that type equality is an equivalence relation, that equal types have the same traceabilities, that a well-typed value has the same traceability as its type, that type equality respects type substitution, that value typing is independent of the typing context, and a type substitution lemma.

**Lemma 3**
*Type equality is an equivalence relation, that is, $\vdash \tau = \tau$, $\vdash \tau_1 = \tau_2$ implies $\vdash \tau_2 = \tau_1$, and $\vdash \tau_1 = \tau_2$ and $\vdash \tau_2 = \tau_3$ implies $\vdash \tau_1 = \tau_3$.*

**Proof:** The proof is by a simple induction on the structure of $\tau$ for reflexivity or the structure of the derivation(s) for symmetry and transitivity and inspection of the rules. ∎

**Lemma 4**
*If $\vdash \tau_1 = \tau_2$ then $tr(\tau_1) = tr(\tau_2)$.*

**Proof:** The proof is by inspection of the last rule used. ∎

**Lemma 5**
*If $\vdash \tau_1 = \tau_2$ then $\vdash \tau_1[\tau/\alpha] = \tau_2[\tau/\alpha]$.*

**Proof:** The proof is by an easy induction on the derivation of $\vdash \tau_1 = \tau_2$. ∎

**Lemma 6**
*If $\Delta; \Gamma \vdash {v_t}^i : \tau$ then $tr(\tau) = t$.*

**Proof:** The proof is by inspection of the three rules for value typing. ∎

**Lemma 7**
*If $\Delta; \Gamma \vdash v^i : \tau$ then $\Delta'; \Gamma' \vdash v^i : \tau$ for any $\Delta'$ and $\Gamma'$.*

**Proof:** The proof is by any easy induction on the typing derivation and inspection of the three rules for value typing. ∎

**Lemma 8**
*If $\Delta, \alpha, \Delta'; \Gamma \vdash e : \tau$, $\Delta \vdash \tau'$ wf, and $tr(\tau') = \mathtt{r}$ then $\Delta, \Delta'; \Gamma[\tau'/\alpha] \vdash e[\tau'/\alpha] : \tau[\tau'/\alpha]$.*

**Proof:** The proof is a straight forward induction over the derivation of $\Delta, \alpha, \Delta'; \Gamma \vdash e : \tau$. It uses Lemma 2 in the case of the rule for application. ∎

With all these lemmas we can prove Type Preservation and Progress.

**Lemma 9 (Type Preservation)**
*If $\vdash M_1 : \tau_1$ and $M_1 \longmapsto M_2$ then $\vdash M_2 : \tau_2$ and $\vdash \tau_1 = \tau_2$ for some $\tau_2$.*

**Proof:** Assume that $\vdash (\rho, e_1) : \tau_1$ and $(\rho, e_1) \longmapsto (\rho, e_2)$. We will show by induction on the derivation of the latter that $\vdash (\rho, e_2) : \tau_2$ and $\vdash \tau_1 = \tau_2$ for some $\tau_2$. By the typing rules, $\vdash \rho : \Gamma$ and $\emptyset; \Gamma \vdash e_1 : \tau_1$ for some $\Gamma$. By the typing rules, we just need to show that $\emptyset; \Gamma \vdash e_2 : \tau_2$ and $\vdash \tau_1 = \tau_2$ for some $\tau_2$. Consider the cases, in the same rule as the figure, for the last rule used to derive the reduction:

- (Variable) In this case, $e_1 = x^i$, $e_2 = v^j$, and $x{:}\tau' = v^j \in \rho$ for some $x$, $i$, $v$, $j$, and $\tau'$. The typing judgement can only be derived with one rule and it requires that $x{:}\tau \in \Gamma$. The typing judgement (for $\rho$) can only be derived in one way and it requires that $\tau = \tau'$, $\emptyset; \emptyset \vdash v^j : \tau''$, and $\vdash \tau = \tau''$. Thus the desired $\tau_2$ is $\tau''$. We just need to show that $\emptyset; \Gamma \vdash v^j : \tau_2$, which follows by Lemma 7.
- (Fix expression) In this case, $e_1 = (\mathtt{fix}\ f[\alpha](x{:}\tau_1'){:}\tau_2'.e')^i$ and $e_2 = \langle \rho, \mathtt{fix}\ f[\alpha](x{:}\tau_1'){:}\tau_2'.e' \rangle^i$. The typing judgement can only be derived with one rule and it requires that $\emptyset \vdash \tau_1$ *wf*, $\alpha; \Gamma, f{:}\tau_1, x{:}\tau_1' \vdash e' : \tau_2'$ and $\tau_1 = (\forall \alpha.\tau_1' \to \tau_2')^i$. Thus by the typing rules, $\emptyset; \Gamma \vdash e_2 : \tau_1$. By Lemma 3, $\vdash \tau_1 = \tau_1$, so the result follows by setting $\tau_2 = \tau_1$.
- (Box expression) In this case, $e_1 = (\mathtt{box}_\tau\, v^i)^j$ and $e_2 = \langle v^i{:}\tau \rangle^j$ for some $\tau$, $v$, $i$, and $j$. The typing judgement can only be derived with one rule and it requires that $\emptyset \vdash \tau$ *wf*, $\emptyset; \Gamma \vdash v^i : \tau'$, $\vdash \tau = \tau'$, and $\tau_1 = \mathtt{box}(\tau)^j$. By the typing rules, $\emptyset; \Gamma \vdash e_2 : \tau_1$. By Lemma 3, $\vdash \tau_1 = \tau_1$, so the result follows by setting $\tau_2 = \tau_1$.
- (Application function) In this case, $e_1 = (e_3[\tau]\ e_4)^i$, $e_2 = (e_5[\tau]\ e_4)^i$, and $(\rho, e_3) \longmapsto (\rho, e_5)$ for some $e_3$, $\tau$, $e_4$, and $e_5$. The typing judgement can only be derived with one rule and it requires that $\emptyset; \Gamma \vdash e_3 : (\forall \alpha.\tau_3 \to \tau')^j$, $\emptyset; \Gamma \vdash e_4 : \tau_4$, $\emptyset \vdash \tau$ *wf*, $tr(\tau) = \mathtt{r}$, $\vdash \tau_3[\tau/\alpha] = \tau_4$, and $\tau = \tau'[\tau/\alpha]$ for some $\tau_3$, $\tau'$, $j$, and $\tau_4$. By the induction hypothesis, $\emptyset; \Gamma \vdash e_5 : \tau_5$ and $\vdash (\forall \alpha.\tau_3 \to \tau')^j = \tau_5$ for some $\tau_5$. There is only one rule to derive the latter and it requires that $\tau_5 = (\forall \alpha.\tau_{51} \to \tau_{52})^k$, $\vdash \tau_3 = \tau_{51}$, and $\vdash \tau' = \tau_{52}$ for some $\tau_{51}$, $\tau_{52}$, and $k$. By Lemma 5, $\vdash \tau_3[\tau/\alpha] = \tau_{51}[\tau/\alpha]$ and $\vdash \tau'[\tau/\alpha] = \tau_{52}[\tau/\alpha]$. By Lemma 3, $\vdash \tau_{51}[\tau/\alpha] = \tau_4$. So by the typing rules, $\emptyset; \Gamma \vdash e_2 : \tau_{52}[\tau/\alpha]$. The result follows by setting $\tau_2 = \tau_{52}[\tau/\alpha]$.
- (Application argument) In this case, $e_1 = (e_3[\tau]\ e_4)^i$, $e_2 = (e_3[\tau]\ e_5)^i$, and $(\rho, e_3) \longmapsto (\rho, e_5)$ for some $e_3$, $e_4$, and $e_5$. The typing judgement can only be derived with one rule and it requires that $\emptyset; \Gamma \vdash e_3 : (\forall \alpha.\tau_3 \to \tau')^j$, $\emptyset; \Gamma \vdash e_4 : \tau_4$, $\emptyset \vdash \tau$ *wf*, $tr(\tau) = \mathtt{r}$, $\vdash \tau_3[\tau/\alpha] = \tau_4$, and $\tau_1 = \tau'[\tau/\alpha]$ for some $\tau_3$, $\tau'$, $j$, and $\tau_4$. By the induction hypothesis, $\emptyset; \Gamma \vdash e_5 : \tau_5$ and $\vdash \tau_4 = \tau_5$. By Lemma 3, $\vdash \tau_3[\tau/\alpha] = \tau_5$. So by the typing rules, $\emptyset; \Gamma \vdash e_2 : \tau_1$. By Lemma 3, $\vdash \tau_1 = \tau_1$, so the result follows by setting $\tau_2 = \tau_1$.
- (Application beta) In this case:

$$\begin{aligned}
e_1 &= ({v_1}^i[\tau]\ {v_2}^j)^k \\
v_1 &= \langle \rho', \mathtt{fix}\ f[\alpha](x{:}\tau_1'){:}\tau_2'.e' \rangle \\
e_2 &= \rho''(e'[\tau/\alpha])^k \\
\rho'' &= \rho', f{:}\tau' = {v_1}^i, x{:}\tau_1'[\tau/\alpha] = {v_2}^j \\
\tau' &= (\forall \alpha.\tau_1' \to \tau_2')^i
\end{aligned}$$

for some $\rho'$, $f$, $\alpha$, $x$, $\tau_1'$, $\tau_2'$, $e'$, $i$, $v_2$, $j$, and $k$. Unpacking the typing judgement, which can only be derived in one way, $\emptyset; \Gamma \vdash {v_1}^i : \tau'$ (1), $\vdash \rho' : \Gamma'$ (2), $\emptyset \vdash \tau'$ *wf* (12), $\alpha; \Gamma', f{:}\tau', x{:}\tau_1' \vdash e' : \tau_2'$ (3), $\tau_1 = \tau_2'[\tau/\alpha]$ (4), $\emptyset; \Gamma \vdash {v_2}^j : \tau_1''$ (5), $\vdash \tau_1'[\tau/\alpha] = \tau_1''$ (6), $\emptyset \vdash \tau$ *wf*, and $tr(\tau) = \mathtt{r}$ for some $\Gamma'$ and $\tau_2''$. By (1) and Lemma 7, $\emptyset; \emptyset \vdash {v_1}^i : \tau'$ (7). By Lemma 3, $\vdash \tau' = \tau'$ (8). By (5) and Lemma 7, $\emptyset; \emptyset \vdash {v_2}^j : \tau_2''$ (9). By (2), (7), (8), (9), and (6), the typing rules give $\vdash \rho'' : \Gamma', f{:}\tau', x{:}\tau_1'[\tau/\alpha]$ (10). By (3) and Lemma 8, $\emptyset; (\Gamma', f{:}\tau', x{:}\tau_1')[\tau/\alpha] \vdash e'[\tau/\alpha] : \tau_2'[\tau/\alpha]$ (11). By (2) and (12), by inspection of the typing rules, $\Gamma'[\tau/\alpha] = \Gamma'$ and $\tau'[\tau/\alpha] = \tau'$. Thus, $\emptyset; \Gamma', f{:}\tau', x{:}\tau_1'[\tau/\alpha] \vdash e'[\tau/\alpha] : \tau_2'[\tau/\alpha]$ (13). By (10) and (13), the typing rules give $\emptyset; \Gamma \vdash$ $\rho''(e'[\tau/\alpha])^k : \tau_2'[\tau/\alpha]$ (14). By (4), the result follows by setting $\tau_2 = \tau_2'[\tau/\alpha]$.

- (Box argument) In this case, $e_1 = (\mathtt{box}_\tau\ e)^i$, $e_2 = (\mathtt{box}_\tau\ e')^i$, and $(\rho, e) \longmapsto (\rho, e')$ for some $\tau$, $e$, $i$, and $e'$. The typing judgement can only be derived with one rule and it requires that $\emptyset \vdash \tau\ wf$, $\emptyset; \Gamma \vdash e : \tau'$, $\vdash \tau = \tau'$ and $\tau_1 = \mathtt{box}(\tau)^i$ for some $\tau'$. By the induction hypothesis, $\emptyset; \Gamma \vdash e' : \tau''$ and $\vdash \tau' = \tau''$ for some $\tau''$. By Lemma 3, $\vdash \tau = \tau''$. By the typing rules, $\emptyset; \Gamma \vdash e_2 : \mathtt{box}(\tau)^i$. By Lemma 3, $\vdash \tau_1 = \tau_1$, so the result follows by setting $\tau_2 = \tau_1$.
- (Unbox argument) In this case, $e_1 = (\mathtt{unbox}\, e)^i$, $e_2 = (\mathtt{unbox}\, e')^i$, and $(\rho, e) \longmapsto (\rho, e')$ for some $e$, $i$, and $e'$. The typing judgement can only be derived with one rule and it requires that $\emptyset; \Gamma \vdash e : \mathtt{box}(\tau_1)^j$ for some $j$. By the induction hypothesis, $\emptyset; \Gamma \vdash e' : \tau'$ and $\vdash \mathtt{box}(\tau_1)^j = \tau'$ for some $\tau'$. The latter can only be derived with one rule and it requires that $\tau' = \mathtt{box}(\tau'')^k$ and $\vdash \tau_1 = \tau''$ for some $\tau''$ and $k$. By the typing rules, $\emptyset; \Gamma \vdash e_2 : \tau''$, so the result follows by setting $\tau_2 = \tau''$.
- (Unbox beta) In this case, $e_1 = (\mathtt{unbox}\, \langle v^i{:}\tau\rangle^j)^k$ and $e_2 = v^i$ for some $\tau$, $v$, $i$, $j$, and $k$. The typing judgement can only be derived with one rule and it requires that $\emptyset; \Gamma \vdash \langle v^i{:}\tau\rangle^j : \mathtt{box}(\tau_1)^l$ for some $l$. The latter can only be derived with one rule and it requires $\tau = \tau_1$, $\emptyset; \Gamma \vdash v^i : \tau'$, and $\vdash \tau = \tau'$. So the result follows by setting $\tau_2 = \tau'$.
- (Frame step) In this case, $e_1 = \rho'(e)^i$, $e_2 = \rho'(e')^i$, and $(\rho', e) \longmapsto (\rho', e')$ for some $\rho'$, $e$, $i$, and $e'$. The typing judgement can only be derived with one rule and it requires that $\rho' \vdash \Gamma'$ : and $\emptyset; \Gamma' \vdash e : \tau_1$ for some $\Gamma'$. By the induction hypothesis, $\emptyset; \Gamma' \vdash e' : \tau'$ and $\vdash \tau_1 = \tau'$ for some $\tau'$. By the typing rules, $\emptyset; \Gamma \vdash e_2 : \tau'$, so the result follows by setting $\tau_2 = \tau'$.
- (Frame return) In this case, $e_1 = \rho'(v^i)^j$ and $e_2 = v^i$ for some $\rho'$, $v$, $i$, and $j$. The typing judgement can only be derived with one rule and it requires that $\rho' \vdash \Gamma'$ : and $\emptyset; \Gamma' \vdash v^i : \tau_1$. By Lemma 7, $\emptyset; \Gamma \vdash v^i : \tau_1$. By Lemma 3, $\vdash \tau_1 = \tau_1$, so the result follows by setting $\tau_2 = \tau_1$.

■

**Lemma 10 (Progress)**
*If* $\vdash M : \tau$ *then either* $M$ *has the form* $(\rho, v^i)$ *or* $M \longmapsto M'$ *for some* $M'$.

**Proof:** The result follows from: If $\vdash \rho : \Gamma$ and $\emptyset; \Gamma \vdash e : \tau$ then either $e$ has the form $v^i$ or $(\rho, e) \longmapsto (\rho, e')$ for some $e'$. We will prove this by induction on the typing derivation for $e$. Consider the last rule, in the same order as the figure, used to derive the judgement:

- (Variable) In this case $e = x^i$ and $x{:}\tau \in \Gamma$. There is only one rule to derive $\vdash \rho : \Gamma$ and it requires that $x{:}\tau = v^j \in \rho$ and other conditions for some $v$ and $j$. Then by the variable rule, $(\rho, e) \longmapsto v^j$, as required.
- (Fix expression) In this case $e = (\mathtt{fix}\ f[\alpha](x{:}\tau_1){:}\tau_2.e')^i$. Clearly by the fix expression rule:

  $$(\rho, e) \longmapsto (\rho, \langle \rho, \mathtt{fix}\ f[\alpha](x{:}\tau_1){:}\tau_2.e'\rangle^i)$$
- (Application) In this case, $e = (e_1[\tau']\ e_2)^i$. The typing rule requires that $\emptyset; \Gamma \vdash e_1 : (\forall\alpha.\tau_1 \to \tau_3)^j$ (1), $\emptyset; \Gamma \vdash e_2 : \tau_2$ (2), and $\vdash \tau_1[\tau'/\alpha] = \tau_2$ (3) for some $\tau_1$, $j$, and $\tau_2$. By the induction hypothesis, either $e_1$ is a value or reduces, and $e_2$ is a value or reduces. There are three subcases:
  - Case 1, $e_1$ reduces: In this case there is $e_1'$ such that $(\rho, e_1) \longmapsto (\rho, e_1')$. Then by the application function rule, $(\rho, e) \longmapsto (\rho, (e_1'[\tau']\ e_2)^i)$, as required.
  - Case 2, $e_1$ is a value and $e_2$ reduces: In this case there is $e_2'$ such that $(\rho, e_2) \longmapsto (\rho, e_2')$. Then by the application function rule, $(\rho, e) \longmapsto (\rho, (e_1[\tau']\ e_2')^i)$, as required.
  - Case 3, $e_1 = {v_1}^k$ and $e_2 = {v_2}^l$ for some $v_1$, $k$, $v_2$, and $l$: There is only one typing rule to derive (1) and it requires that $v_1$ have the form $\langle \rho', \mathtt{fix}\ f[\alpha](x{:}\tau_1){:}\tau_3.e'\rangle$ for some $\rho'$, $f$, $x$, and $e'$. Let $t$ be the traceability of $v_2$. By Lemma 6 and (2), $tr(\tau_2) = t$. By Lemma 4 and (3), $tr(\tau_1[\tau'/\alpha]) = t$. Then by the application beta rule:

    $$(\rho, e) \longmapsto$$
    $$(\rho, (\rho', f{:}\tau' = {v_1}^k, x{:}\tau_1[\tau'/\alpha] = {v_2}^l)(e'[\tau'/\alpha])^i)$$

    where $\tau' = (\forall\alpha.\tau_1 \to \tau_3)^k$, as required.
- (Box expression) In this case, $e = (\mathtt{box}_{\tau'}\ e')^i$ for some $\tau'$, $e'$, and $i$. The typing rule requires that $\tau = \mathtt{box}(\tau')^i$, $\emptyset; \Gamma \vdash e' : \tau''$ (1), and $\vdash \tau' = \tau''$ (2) for some $\tau''$. By the induction hypothesis, either $e'$ is a value or reduces:
  - If $e' = {v_t}^j$ then by Lemma 6 and (1), $tr(\tau'') = t$. By (2) and Lemma 4, $tr(\tau') = t$. So by the box reduction rule, $(\rho, e) \longmapsto (\rho, \langle {v_{t'}}^j{:}\tau'\rangle^i)$, as required.
  - If $(\rho, e') \longmapsto (\rho, e'')$ then $(\rho, e) \longmapsto (\rho, (\mathtt{box}_{\tau'}\ e'')^i)$, as required.
- (Unbox) In this case, $e = (\mathtt{unbox}\, e')^i$ for some $e'$ and $i$. The typing rule requires that $\emptyset; \Gamma \vdash e' : \mathtt{box}(\tau)^j$ (1) for some $j$. By the induction hypothesis, $e'$ is a value or reduces:
  - If $e' = v^k$ then (1) can be derived by only one rule and it requires that $v = \langle {v'}^l{:}\tau'\rangle$ for some $v'$, $l$, and $\tau'$. By the unbox beta rule, $(\rho, e) \longmapsto (\rho, {v'}^l)$, as required.
  - If $(\rho, e') \longmapsto (\rho, e'')$ then by the unbox argument rule, $(\rho, e) \longmapsto (\rho, (\mathtt{unbox}\, e'')^i)$, as required.
- (Frame) In this case, $e = \rho'(e')^i$ for some $\rho'$, $e'$, and $i$. The typing rule requires that $\vdash \rho' : \Gamma'$ and $\emptyset; \Gamma' \vdash e' : \tau$ for some $\Gamma'$. By the induction hypothesis, $e'$ is a value or reduces:
  - If $e' = v^j$ then by the frame return rule, $(\rho, e) \longmapsto (\rho, v^j)$, as required.
  - If $(\rho, e') \longmapsto (\rho, e'')$ then by the frame step rule, $(\rho, e) \longmapsto (\rho, \rho'(e'')^i)$, as required.
- (Constant) In this case $e = c^i$ for some $c$ and $i$ and is clearly a value.
- (Fix value) In this case $e = \langle \rho', \mathtt{fix}\ f[\alpha](x{:}\tau_1){:}\tau_2.e'\rangle^i$ for some $\rho'$, $f$, $x$, $\tau_1$, $\tau_2$, $e'$ and $i$ and is clearly a value.
- (Box value) In this case $e = \langle v^i{:}\tau'\rangle^j$ for some $\tau'$, $v$, $i$, and $j$ and is clearly a value.

■

We can also prove that typability implies traceability and thus typable programs are GC safe and remain so throughout execution.

**Lemma 11**

- *If* $\vdash M : \tau$ *then* $\vdash M\ \mathbf{tr}$.
- *If* $\vdash \rho : \Gamma$ *then* $\vdash \rho\ \mathbf{tr}$.
- *If* $\Delta; \Gamma \vdash e : \tau$ *then* $\vdash e\ \mathbf{tr}$.

- *If* $\Delta;\Gamma \vdash v^i : \tau$ *then* $\vdash_{\text{v}} v{:}tr(\tau)$.

**Proof:** The results are proven simultaneously by induction on the structure of the typing derivation. The cases for the last rule used, in the same order as the figure, are:

- (Variable) In this case clearly $\vdash e\ \mathbf{tr}$.
- (Fix expression) In this case $e = (\mathtt{fix}\ f[\alpha](x{:}\tau_1){:}\tau_2.e')^i$ for some $x$, $\alpha$, $\tau_1$, $\tau_2$, $e'$, and $i$. Then by the typing rule, $\Delta, \alpha; \Gamma, f{:}\tau, x{:}\tau_1 \vdash e' : \tau_2$ is a subderivation. By the induction hypothesis, $\vdash e'\ \mathbf{tr}$. So by the rules for traceability, $\vdash e\ \mathbf{tr}$, as required.
- (Application) In this case $e = (e_1[\tau']\ e_2)^i$ for some $e_1$, $\tau'$, $e_2$, and $i$. By the typing rule, $\Delta;\Gamma \vdash e_1 : \tau_1$ and $\Delta;\Gamma \vdash e_2 : \tau_2$ for some $\tau_1$ and $\tau_2$. By the induction hypothesis, $\vdash e_1\ \mathbf{tr}$ and $\vdash e_2\ \mathbf{tr}$. By the rules for traceability $\vdash e\ \mathbf{tr}$, as required.
- (Box expression) In this case $e = (\mathtt{box}_{\tau'}\ e')^i$ for some $\tau'$, $e'$, and $i$. By the typing rule, $\Delta;\Gamma \vdash e' : \tau''$ for some $\tau''$. By the induction hypothesis, $\vdash e'\ \mathbf{tr}$. By the rules for traceability, $\vdash e'\ \mathbf{tr}$, as required.
- (Unbox) In this case $e = (\mathtt{unbox}\, e')^i$ for some $e'$ and $i$. By the typing rule, $\Delta;\Gamma \vdash e' : \tau'$ for some $\tau'$. By the induction hypothesis, $\vdash e'\ \mathbf{tr}$. By the rules for traceability, $\vdash e\ \mathbf{tr}$, as required.
- (Frame) In this case, $e = \rho(e')^i$ for some $\rho$, $e'$, and $i$. By the typing rule, $\vdash \rho : \Gamma'$ and $\emptyset;\Gamma' \vdash e' : \tau$ for some $\Gamma'$. By the induction hypothesis, $\vdash \rho\ \mathbf{tr}$ and $\vdash e'\ \mathbf{tr}$. By the rules for traceability, $\vdash e\ \mathbf{tr}$, as required.
- (Constant) In this case $e = c^i$. By the typing rule, $\tau = \mathtt{B}^i$ and so clearly $tr(\tau) = \mathtt{b}$. By the rules for traceability, $\vdash_{\text{v}} c{:}\mathtt{b}$, proving the fourth result. By the rules for traceability again, $\vdash e\ \mathbf{tr}$, proving the third result.
- (Fix value) In this case $e = \langle \rho, \mathtt{fix}\ f[\alpha](x{:}\tau_1){:}\tau_2.e' \rangle^i$ for some $\rho$, $f$, $\alpha$, $x$, $\tau_1$, $\tau_2$, $e'$, and $i$. By the typing rule, $\vdash \rho : \Gamma'$ and $\alpha;\Gamma', f{:}\tau, x{:}\tau_1 \vdash e' : \tau_2$ for some $\Gamma'$. Also by the typing rules, $\tau$ is a function type, so $tr(\tau) = \mathtt{r}$. By the induction hypothesis, $\vdash \rho\ \mathbf{tr}$ and $\vdash e'\ \mathbf{tr}$. By the rules for traceability, $\vdash_{\text{v}} \langle \rho, \mathtt{fix}\ \ f[\alpha](x{:}\tau_1){:}\tau_2.e' \rangle{:}\mathtt{r}$, proving the fourth result. By the rules for traceability again, $\vdash e\ \mathbf{tr}$, proving the third result.
- (Box value) In this case $e = \langle v^i{:}\tau' \rangle^j$. By the typing rule, $\Delta;\Gamma \vdash v^i : \tau''$ and $\vdash \tau' = \tau''$ for some $\tau''$. Also by the typing rule, $\tau$ is a box type, so $tr(\tau) = \mathtt{r}$. By the induction hypothesis, $\vdash_{\text{v}} v{:}tr(\tau'')$. By Lemma 4, $\vdash_{\text{v}} v{:}tr(\tau')$. By the rules for traceability, $\vdash_{\text{v}} \langle v^i{:}\tau' \rangle{:}\mathtt{r}$, proving the third result. By the rules for traceability again, $\vdash e\ \mathbf{tr}$, as required.
- (Environment) In this case $\rho = x_1{:}\tau_1 = {v_1}^{i_1}, \ldots, x_n{:}\tau_n = {v_n}^{i_n}$. By the typing rule, $\emptyset;\emptyset \vdash {v_j}^{i_j} : \tau'_j$ and $\vdash \tau_j = \tau'_j$ for $1 \leq j \leq n$ and some $\tau'_j$s. By the induction hypothesis, $\vdash_{\text{v}} v_j{:}tr(\tau'_j)$ for $1 \leq j \leq n$. By Lemma 4, $tr(\tau_j) = tr(\tau'_j)$ for $1 \leq j \leq n$. Thus $\vdash_{\text{v}} v_j{:}tr(\tau_j)$ for $1 \leq j \leq n$. By the traceability rules, $\vdash \rho\ \mathbf{tr}$, as required.
- (Machine state) In this case $M = (\rho, e)$. By the typing rule, $\vdash \rho : \Gamma$ and $\emptyset;\Gamma \vdash e : \tau$ for some $\Gamma$. By the induction hypothesis, $\vdash \rho\ \mathbf{tr}$ and $\vdash e\ \mathbf{tr}$. By the rules for traceability, $\vdash M\ \mathbf{tr}$, as required.

■

## 3. Flow analysis

Our original motivation for this work was to apply interprocedural analysis to the problem of eliminating unnecessary boxing in programs. There is a vast body of literature on interprocedural analysis and optimization, and it is generally fairly straightforward to use these approaches to obtain information about what terms flow to what use sites. This paper is not intended to provide any contribution to the algorithmic side of this body of work, which we will broadly refer to as *flow analysis*. Our contribution in this paper lies in showing how to extend flow analysis to the type level, and showing that any generic flow analysis so extended can be used to implement an unboxing optimization that preserves type safety.

In order to do this, we must provide some framework for describing what information a flow analysis must provide. For the purposes of our unboxing optimization, we are interested in finding (interprocedurally) for every $(\mathtt{unbox}\, v^j)^i$ operation the set of $(\mathtt{box}_\tau\ e)^k$ terms that could possibly reach $v$. Under appropriate conditions, we can then eliminate both the box introductions and the box elimination, thereby improving the program. The core language defined in Section 2 provides labels serving as proxies for the terms, types, and variables on which they occur – the question above can therefore be re-stated as finding the set of labels $k$ that reach the position labeled with $j$.

More generally, following previous work we begin by defining an abstract notion of analysis. We say that an analysis is a pair $(\mathrm{C}, \varrho)$. Binding environments $\varrho$ simply serve to map variables to the label of their binding sites. The mappings are, as usual, global for the program. Consequently, a given environment may not apply to alpha-variants of a term. We do not require that labels be unique within a program—as usual however, analyses will be more precise if this is the case. Variables are also not required to be unique (since reduction may duplicate terms and hence binding sites). However, duplicate variable bindings in a program must be labeled consistently according to $\varrho$ or else no analysis of the program can be acceptable according to our definition. This can always be avoided by alpha-varying or relabeling appropriately.

A cache C is a mapping from labels to sets of shapes. Shapes are given by the grammar:

$$\text{Shapes:}\quad s \ ::=\ c^i \mid (\forall i.j \to k)^l_{\mathtt{v}} \mid (\mathtt{box}_t\ i)^j_{\mathtt{v}} \mid \mathtt{B}^i \mid (\forall i.j \to k)^l_{\mathtt{t}} \mid (\mathtt{box}\ i)^j_{\mathtt{t}}$$

There are two classes of shapes—term shapes and type shapes. The idea behind term shapes is that each shape provides a proxy for a set of terms that might flow to a given location, describing both the shape of the values that might flow there and the labels of the sub-components of those values. For example, for an analysis $(\mathrm{C}, \varrho)$, $c^i \in \mathrm{C}(k)$ indicates that (according to the analysis) the constant $c$, labeled with $i$, might flow to a location labeled with $k$. Similarly, if $(\forall i'.i \to j)^k_{\mathtt{v}} \in \mathrm{C}(l)$, then the analysis specifies that among the values flowing to locations labeled with $l$ might be functions labeled with $k$, whose type parameter is labeled with $i'$, parameter type is labeled with $i$, and whose bodies are labeled with $j$. If $(\mathtt{box}_t\ k)^i_{\mathtt{v}} \in \mathrm{C}(l)$ then among the values that might flow to $l$ (according to the analysis) are boxed values labeled with $i$, with meta-data $t$ and whose bodies are labeled by some $j$ such that $\mathrm{C}(j) \subseteq \mathrm{C}(k)$.

Where term shapes provide a proxy for the set of values that might flow to a given location, type shapes provide a proxy for the types of the locations that values might flow through to get to a given location. For example, for an analysis $(\mathrm{C}, \varrho)$, $\mathtt{B}^i \in \mathrm{C}(k)$ indicates that (according to the analysis) objects that reach location $k$ might flow through a variable or term of type $\mathtt{B}$, labeled with $i$. The function type and box type shapes similarly correspond to the flow of values through locations labeled with function or box types.

It is important to note that the shapes in the cache may not correspond exactly to the terms in the program, since reduction may change program terms (e.g. by instantiating variables with values).

However, reduction does not change the outer shape and labeling of values—it is this reduction invariant information that is captured by shapes.

Clearly, not every choice of analysis pairs is meaningful for program optimization. While in general it is reasonable (indeed, unavoidable) for an analysis to overestimate the set of terms associated with a label, it is unacceptable for an analysis to underestimate the set of terms that flow to a label—most optimizations will produce incorrect results, since they are designed around the idea that the analysis is telling them everything that could possibly flow to them. In order to capture the notion of when an analysis pair gives a suitable approximation of the flow of values in a program we follow the general spirit of Nielson et al. [7], and define a notion of an *acceptable analysis*. That is, we give a declarative specification that gives sufficient conditions for specifying when a given analysis does not underestimate the set of terms flowing to a label, without committing to a particular analysis. We arrange the subsequent meta-theory such that our results apply to any analysis that is *acceptable*. In this way, we completely decouple our optimization from the particulars of how the analysis is computed.

Our acceptable-analysis relation is given in Figures 6 and 7—the judgement $\mathrm{C}; \varrho \vdash (\rho, e)$ determines that an analysis pair $(\mathrm{C}, \varrho)$ is *acceptable* for a machine state $(\rho, e)$, and similarly for the environment and expression forms of the judgement. We use the notation $\mathrm{lbl}(e)$ to denote the outermost label of $e$: that is, $i$ where $e$ is of the form $m^i$ or $v^i$. The acceptability judgement generally indicates for each syntactic form what the flow of values is. For example, in the application rule, the judgment insists that for every function value that flows to the applicand position, the set of shapes associated with the parameter of that function is a super-set of the set of shapes associated with the argument of the application; and that the set of shapes associated with the result of the function is a sub-set of the set of shapes associated with the application itself.

The judgement $\mathrm{C}; \varrho \vdash \tau$ determines that an analysis pair $(\mathrm{C}, \varrho)$ is acceptable for a labeled type $\tau$. In particular, if a function flows to a function type $\forall \tau_1 . \tau_2 \rightarrow$ then the set of values that flow to the function's parameter can flow to the argument type $\tau_1$, and the set of values that can flow from the result of the function can flow to the result type $\tau_2$. And similarly for box types.

Given this definition, we can show that the acceptability relation is preserved under reduction. First we show that the cache is only refined by reduction.

**Lemma 12 (Cache refinement under reduction)**
*If* $\mathrm{C}; \varrho \vdash \rho$, $\mathrm{C}; \varrho \vdash e_1$, *and* $(\rho, e_1) \longmapsto (\rho, e_2)$ *then* $\mathrm{C}(\mathrm{lbl}(e_1)) \supseteq \mathrm{C}(\mathrm{lbl}(e_2))$.

**Proof:** The proof is by induction on the derivation of $(\rho, e_1) \longmapsto (\rho, e_2)$. Consider the cases for the last rule used to it (the cases are in the same order as in the figure):

- (Variable instantiation.) In this case, $e_1 = x^k$, $e_2 = v^j$, and $x{:}\tau = v^j \in \rho$. The assumption $\mathrm{C}; \varrho \vdash \rho$ requires that $\mathrm{C}(j) \subseteq \mathrm{C}(\mathrm{lbl}(\tau))$ and $\varrho(x) = \mathrm{lbl}(\tau)$. The assumption $\mathrm{C}; \varrho \vdash e_1$ requires that $\mathrm{C}(\varrho(x)) \subseteq \mathrm{C}(k)$. Thus $\mathrm{C}(j) \subseteq \mathrm{C}(k)$. Clearly, $\mathrm{lbl}(e_1) = k$ and $\mathrm{lbl}(e_2) = j$ and the result follows.
- (Fix introduction.) In this case, clearly $\mathrm{lbl}(e_1) = \mathrm{lbl}(e_2)$ and the result immediately follows.
- (Box introduction.) In this case, clearly $\mathrm{lbl}(e_1) = \mathrm{lbl}(e_2)$ and the result immediately follows.
- (Application left.) In this case, clearly $\mathrm{lbl}(e_1) = \mathrm{lbl}(e_2)$ and the result immediately follows.
- (Application right.) In this case, clearly $\mathrm{lbl}(e_1) = \mathrm{lbl}(e_2)$ and the result immediately follows.

$\boxed{funC(i,j,k,l) \quad boxC(i,j)}$

$$\begin{aligned}
&funC(i,j,k,l) = \\
&\quad \wedge \forall (\forall j'.k' \rightarrow l')^{i'}_{\mathsf{v}} \in \mathrm{C}(i) : \\
&\qquad \mathrm{C}(j) = \mathrm{C}(j') \wedge \mathrm{C}(k) \subseteq \mathrm{C}(k') \wedge \mathrm{C}(l') \subseteq \mathrm{C}(l) \\
&\quad \wedge \forall (\forall j'.k' \rightarrow l')^{i'}_{\mathsf{t}} \in \mathrm{C}(i) : \\
&\qquad \mathrm{C}(j) = \mathrm{C}(j') \wedge \mathrm{C}(k) \subseteq \mathrm{C}(k') \wedge \mathrm{C}(l') \subseteq \mathrm{C}(l) \\
&boxC(i,j) = \\
&\quad \wedge \forall (\mathtt{box}_t\, j')^{i'}_{\mathsf{v}} \in \mathrm{C}(i) : \mathrm{C}(j') \subseteq \mathrm{C}(j) \\
&\quad \wedge \forall (\mathtt{box}\, j')^{i'}_{\mathsf{t}} \in \mathrm{C}(i) : \mathrm{C}(j') \subseteq \mathrm{C}(j)
\end{aligned}$$

$\boxed{\mathrm{C}; \varrho \vdash e}$

$$\frac{\mathrm{C}(\varrho(x)) \subseteq \mathrm{C}(i)}{\mathrm{C}; \varrho \vdash x^i}$$

$$\frac{\begin{array}{c}\varrho(f) = i \quad \varrho(x) = \mathrm{lbl}(\tau_1) \quad \mathrm{C}; \varrho \vdash (\forall \alpha.\tau_1 \rightarrow \tau_2)^i \\ \mathrm{C}; \varrho \vdash e \quad (\forall \varrho(\alpha).\mathrm{lbl}(\tau_1) \rightarrow \mathrm{lbl}(e))^i_{\mathsf{v}} \in \mathrm{C}(i)\end{array}}{\mathrm{C}; \varrho \vdash (\mathtt{fix}\ f[\alpha](x{:}\tau_1){:}\tau_2.e)^i}$$

$$\frac{\begin{array}{c}\mathrm{C}; \varrho \vdash e_1 \quad \mathrm{C}; \varrho \vdash \tau \quad \mathrm{C}; \varrho \vdash e_2 \\ funC(\mathrm{lbl}(e_1), \mathrm{lbl}(\tau), \mathrm{lbl}(e_2), i)\end{array}}{\mathrm{C}; \varrho \vdash (e_1[\tau]\ e_2)^i}$$

$$\frac{\begin{array}{c}\mathrm{C}; \varrho \vdash \mathtt{box}(\tau)^i \quad \mathrm{C}; \varrho \vdash e \\ (\mathtt{box}_{tr(\tau)}\, j)^i_{\mathsf{v}} \in \mathrm{C}(i) \quad \mathrm{C}(\mathrm{lbl}(e)) \subseteq \mathrm{C}(j)\end{array}}{\mathrm{C}; \varrho \vdash (\mathtt{box}_\tau\ e)^i}$$

$$\frac{\mathrm{C}; \varrho \vdash e \quad boxC(\mathrm{lbl}(e), i)}{\mathrm{C}; \varrho \vdash (\mathtt{unbox}\ e)^i}$$

$$\frac{\mathrm{C}; \varrho \vdash \rho \quad \mathrm{C}; \varrho \vdash e \quad \mathrm{C}(\mathrm{lbl}(e)) \subseteq \mathrm{C}(i)}{\mathrm{C}; \varrho \vdash \rho(e)^i}$$

$$\frac{\mathrm{C}; \varrho \vdash \mathtt{B}^i \quad c^i \in \mathrm{C}(i)}{\mathrm{C}; \varrho \vdash c^i}$$

$$\frac{\begin{array}{c}\varrho(f) = i \quad \varrho(x) = \mathrm{lbl}(\tau_1) \quad \mathrm{C}; \varrho \vdash (\forall \alpha.\tau_1 \rightarrow \tau_2)^i \\ \mathrm{C}; \varrho \vdash \rho \quad \mathrm{C}; \varrho \vdash e \\ (\forall \varrho(\alpha).\mathrm{lbl}(\tau_1) \rightarrow \mathrm{lbl}(e))^i_{\mathsf{v}} \in \mathrm{C}(i)\end{array}}{\mathrm{C}; \varrho \vdash \langle \rho, \mathtt{fix}\ f[\alpha](x{:}\tau_1){:}\tau_2.e \rangle^i}$$

$$\frac{\begin{array}{c}\mathrm{C}; \varrho \vdash \mathtt{box}(\tau)^i \quad \mathrm{C}; \varrho \vdash v^j \\ (\mathtt{box}_{tr(\tau)}\, k)^i_{\mathsf{v}} \in \mathrm{C}(i) \quad \mathrm{C}(j) \subseteq \mathrm{C}(k)\end{array}}{\mathrm{C}; \varrho \vdash \langle v^j{:}\tau \rangle^i}$$

**Figure 6.** Acceptable Analysis, Expressions

$\boxed{C; \varrho \vdash \tau}$

$$\frac{C(\varrho(\alpha)) = C(i)}{C; \varrho \vdash \alpha^i}$$

$$\frac{\mathtt{B}^j \in C(i)}{C; \varrho \vdash \mathtt{B}^i}$$

$$\frac{\begin{array}{c}C; \varrho \vdash \tau_1 \quad C; \varrho \vdash \tau_2 \\ (\forall \varrho(\alpha).\mathrm{lbl}(\tau_1) \to \mathrm{lbl}(\tau_2))^j_{\mathtt{t}} \in C(i) \\ \mathit{funC}(i, \varrho(\alpha), \mathrm{lbl}(\tau_1), \mathrm{lbl}(\tau_2))\end{array}}{C; \varrho \vdash (\forall \alpha.\tau_1 \to \tau_2)^i}$$

$$\frac{C; \varrho \vdash \tau \quad (\mathtt{box}\,\mathrm{lbl}(\tau))^j_{\mathtt{t}} \in C(i) \quad \mathit{boxC}(i, \mathrm{lbl}(\tau))}{C; \varrho \vdash \mathtt{box}(\tau)^i}$$

$\boxed{C; \varrho \vdash \Gamma}$

$$\frac{\forall 1 \le j \le n : \varrho(x_j) = \mathrm{lbl}(\tau_j) \wedge C; \varrho \vdash \tau_j}{C; \varrho \vdash x_1{:}\tau_1, \ldots, x_n{:}\tau_n}$$

$\boxed{C; \varrho \vdash \rho}$

$$\frac{\begin{array}{c}C; \varrho \vdash x_1{:}\tau_1, \ldots, x_n{:}\tau_n \\ \forall 1 \le j \le n : C(i_j) \subseteq C(\mathrm{lbl}(\tau_j)) \wedge C; \varrho \vdash {v_k}^{i_k}\end{array}}{C; \varrho \vdash x_1{:}\tau_1 = {v_1}^{i_1}, \ldots, x_n{:}\tau_n = {v_n}^{i_n}}$$

$\boxed{C; \varrho \vdash M}$

$$\frac{C; \varrho \vdash \rho \quad C; \varrho \vdash e}{C; \varrho \vdash (\rho, e)}$$

**Figure 7.** Acceptable Analysis, Other Constructs

- (Application beta.) In this case, clearly $\mathrm{lbl}(e_1) = \mathrm{lbl}(e_2)$ and the result immediately follows.
- (Under box.) In this case, clearly $\mathrm{lbl}(e_1) = \mathrm{lbl}(e_2)$ and the result immediately follows.
- (Under unbox.) In this case, clearly $\mathrm{lbl}(e_1) = \mathrm{lbl}(e_2)$ and the result immediately follows.
- (Unbox beta.) In this case, $e_1 = (\mathtt{unbox}\, \langle v^i{:}\tau \rangle^j)^k$ and $e_2 = v^i$. The first hypothesis can be derived only by one rule and it requires that $C; \varrho \vdash \langle v^i{:}\tau \rangle^j$ (1), and $\mathit{boxC}(j, k)$ (2). Judgement 1 can only be derived by one rule and it requires that $C; \varrho \vdash v^i$ (4), $(\mathtt{box}_{tr(\tau)}\, i'')^j_{\mathtt{v}} \in C(j)$ (5) for some $i''$, and $C(i) \subseteq C(i'')$ (6). Instantiating Fact 2 with Fact 5 we get that $C(i'') \subseteq C(k)$ (7). Combining Facts 6 and 7, $C(i) \subseteq C(k)$, as we are required to prove.
- (Under frame.) In this case, clearly $\mathrm{lbl}(e_1) = \mathrm{lbl}(e_2)$ and the result immediately follows.
- (Frame return.) In this case, $e_1 = \rho'(v^i)^j$ and $e_2 = v^i$. The assumption $C; \varrho \vdash e_1$ requires that $C(i) \subseteq C(j)$. Since $\mathrm{lbl}(e_1) = j$ and $\mathrm{lbl}(e_2) = i$, the result is immediate.

■

Next we show a type substitution lemma for acceptability.

**Lemma 13**
*If* $C; \varrho \vdash \tau$ *and* $C(\mathrm{lbl}(\tau)) = C(\varrho(\alpha))$ *then:*

- *If* $C; \varrho \vdash \tau'$ *then* $C; \varrho \vdash \tau'[\tau/\alpha]$.
- *If* $C; \varrho \vdash \Gamma$ *then* $C; \varrho \vdash \Gamma[\tau/\alpha]$.
- *If* $C; \varrho \vdash e$ *then* $C; \varrho \vdash e[\tau/\alpha]$.
- *If* $C; \varrho \vdash \rho$ *then* $C; \varrho \vdash \rho[\tau/\alpha]$.

**Proof:** The proof is by induction on the derivation of the $C; \varrho \vdash \tau'$ and $C; \varrho \vdash e$. Consider the cases for the rules used to derive it (in the same order as in the figures):

- The cases for expressions are straight forward.
- (Type variable) In this case $\tau' = \beta^i$. If $\beta \neq \alpha$ then $\tau'[\tau/\alpha] = \tau'$ and the result is immediate. Otherwise, by the rules for acceptability, $C(\varrho(\alpha)) = C(i)$. If $\tau = \sigma^j$ then $\tau'[\tau/\alpha] = \sigma^i$. Consider the cases for $\sigma$:
  - Subcase 1, $\sigma = \alpha'$: Then by the rules for acceptability, $C(\varrho(\alpha')) = C(j)$. Since $C(j) = C(\varrho(\alpha)) = C(i)$, $C(\varrho(\alpha')) = C(i)$, and thus $C; \varrho \vdash {\alpha'}^i$, as required.
  - Subcase 2, $\sigma = \forall \alpha'.\tau_1 \to \tau_2$: Since $C; \varrho \vdash \tau$, the rules require:

    $C; \varrho \vdash \tau_1$ (1)
    $C; \varrho \vdash \tau_2$ (2)
    $(\forall \varrho(\alpha').\mathrm{lbl}(\tau_1) \to \mathrm{lbl}(\tau_2))^k_{\mathtt{t}} \in C(j)$ (3)
    $\mathit{funC}(j, \varrho(\alpha'), \mathrm{lbl}(\tau_1), \mathrm{lbl}(\tau_2))$ (4)

    By (3) and $C(j) = C(i)$:

    $(\forall \varrho(\alpha').\mathrm{lbl}(\tau_1) \to \mathrm{lbl}(\tau_2))^k_{\mathtt{t}} \in C(i)$ (5)

    By (4) and $C(j) = C(i)$:

    $\mathit{funC}(i, \varrho(\alpha'), \mathrm{lbl}(\tau_1), \mathrm{lbl}(\tau_2))$ (6)

    By (1), (2), (5), and (6), by the rules for acceptability, $C; \varrho \vdash \sigma^i$.
  - Subcase 3, $\sigma = \mathtt{box}(\tau'')$: Since $C; \varrho \vdash \tau$, the rules require:

    $C; \varrho \vdash \tau''$ (1)
    $(\mathtt{box}\,\mathrm{lbl}(\tau''))^k_{\mathtt{t}} \in C(j)$ (2)
    $\mathit{boxC}(j, \mathrm{lbl}(\tau''))$ (3)

    By (2) and $C(j) = C(i)$, $(\mathtt{box}\,\mathrm{lbl}(\tau''))^k_{\mathtt{t}} \in C(i)$ (4). By (3) and $C(j) = C(i)$, $\mathit{boxC}(i, \mathrm{lbl}(\tau''))$ (5). By (1), (4), and (5), by the rules for acceptability, $C; \varrho \vdash \sigma^i$, as required.
- (Base type) In this case $\tau'[\tau/\alpha] = \tau$ and the result is immediate.
- (Function type) In this case $\tau' = (\forall \alpha'.\tau_1 \to \tau_2)^i$. The rules for acceptability require:

  $C; \varrho \vdash \tau_1$ (1)
  $C; \varrho \vdash \tau_2$ (2)
  $(\forall \varrho(\alpha').\mathrm{lbl}(\tau_1) \to \mathrm{lbl}(\tau_2))^k_{\mathtt{t}} \in C(i)$ (3)
  $\mathit{funC}(i, \varrho(\alpha'), \mathrm{lbl}(\tau_1), \mathrm{lbl}(\tau_2))$ (4)

  By (1), (2), and the induction hypothesis:

  $C; \varrho \vdash \tau_1[\tau/\alpha]$ (5)
  $C; \varrho \vdash \tau_2[\tau/\alpha]$ (6)

  Since $\mathrm{lbl}(\tau_1[\tau/\alpha]) = \mathrm{lbl}(\tau_1)$ and $\mathrm{lbl}(\tau_1[\tau/\alpha]) = \mathrm{lbl}(\tau_1)$:

  $(\forall \varrho(\alpha').\mathrm{lbl}(\tau_1[\tau/\alpha]) \to \mathrm{lbl}(\tau_2[\tau/\alpha]))^k_{\mathtt{t}} \in C(i)$ (7)
  $\mathit{funC}(i, \varrho(\alpha'), \mathrm{lbl}(\tau_1[\tau/\alpha]), \mathrm{lbl}(\tau_2[\tau/\alpha]))$ (8)

  Since $\mathrm{lbl}(\tau'[\tau/\alpha]) = i$, by (5), (6), (7), and (8), $C; \varrho \vdash \tau'[\tau/\alpha]$, as required.

- (Box type) In this case, $\tau' = \texttt{box}(\tau'')^i$. The rules for acceptability require:

$$\begin{array}{ll} \mathrm{C}; \varrho \vdash \tau'' & (1) \\ (\texttt{box}\,\mathrm{lbl}(\tau''))^k_{\mathsf{t}} \in \mathrm{C}(i) & (2) \\ \mathit{boxC}(i, \mathrm{lbl}(\tau'')) & (3) \end{array}$$

By (1) and the induction hypothesis:

$$\mathrm{C}; \varrho \vdash \tau''[\tau/\alpha] \quad (4)$$

Since $\mathrm{lbl}(\tau''[\tau/\alpha]) = \mathrm{lbl}(\tau'')$:

$$\begin{array}{ll} (\texttt{box}\,\mathrm{lbl}(\tau''[\tau/\alpha]))^k_{\mathsf{t}} \in \mathrm{C}(i) & (5) \\ \mathit{boxC}(i, \mathrm{lbl}(\tau''[\tau/\alpha])) & (6) \end{array}$$

Since $\mathrm{lbl}(\tau'[\tau/\alpha]) = i$, by (4), (5), and (6), $\mathrm{C}; \varrho \vdash \tau'[\tau/\alpha]$, as required.

- The cases for type and value environments are straight forward.

■

With these lemmas we can prove that reduction preserves acceptability of the flow analysis.

**Lemma 14 (Preservation of acceptability under reduction)** *If* $\mathrm{C}; \varrho \vdash M$ *and* $M \longmapsto M'$ *then* $\mathrm{C}; \varrho \vdash M'$.

**Proof:** If $\mathrm{C}; \varrho \vdash (\rho, e)$ then $\mathrm{C}; \varrho \vdash \rho$ and $\mathrm{C}; \varrho \vdash e$. If $(\rho, e) \longmapsto (\rho, e')$ then the result follows if we show that $\mathrm{C}; \varrho \vdash e'$. The proof of the latter is by induction on the derivation of $(\rho, e) \longmapsto (\rho, e')$. Consider the cases for the last rule used to derive it (the cases are in the same order as in the figure):

- In this case $e = x^k$, $e' = v^j$, and $x{:}\tau = v^j \in \rho$. The assumption $\mathrm{C}; \varrho \vdash \rho$ requires that $\mathrm{C}; \varrho \vdash v^j$, which is what we need to prove.

- In this case $e = (\texttt{fix}\ f[\alpha](x{:}\tau_1){:}\tau_2.e'')^j$ for some $f$, $\alpha$, $x$, $\tau_1$, $\tau_2$, $e''$, and $j$, and $e' = \langle\rho, \texttt{fix}\ f[\alpha](x{:}\tau_1){:}\tau_2.e''\rangle^j$. Let $i = \mathrm{lbl}(\tau_1)$. The first hypothesis can only be derived by one rule and it requires that $\varrho(f) = j$, $\varrho(x) = i$, $\mathrm{C}; \varrho \vdash \tau$ where $\tau = (\forall\alpha.\tau_1 \to \tau_2)^j$, $\mathrm{C}; \varrho \vdash e''$, and $(\forall\varrho(\alpha).i \to \mathrm{lbl}(e''))^j_{\mathsf{v}} \in \mathrm{C}(j)$. Then, and noting $\mathrm{C}; \varrho \vdash \rho$ by assumption, by the rules for acceptable analysis, $\mathrm{C}; \varrho \vdash \langle\rho, \texttt{fix}\ f[\alpha](x{:}\tau_1){:}\tau_2.e''\rangle^j$, as we are required to prove.

- In this case $e = (\texttt{box}_\tau\ v^i)^j$ for some $\tau$, $v$, $i$, and $j$, and $e' = \langle v^i{:}\tau\rangle^j$. The first hypothesis can only be derived by one rule and it requires that $\mathrm{C}; \varrho \vdash \texttt{box}(\tau)^j$, $\mathrm{C}; \varrho \vdash v^i$, $(\texttt{box}_{tr(\tau)}\ k)^j_{\mathsf{v}} \in \mathrm{C}(j)$ for some $k$, and $\mathrm{C}(i) \subseteq \mathrm{C}(k)$. Then by the rules for acceptable analysis, $\mathrm{C}; \varrho \vdash \langle v^i{:}\tau\rangle^j$, as we are required to prove.

- In this case $e = (e_1[\tau]\ e_2)^i$ for some $e_1$, $\tau$, $e_2$, and $i$, $e' = (e'_1[\tau]\ e_2)^i$, and $(\rho, e_1) \longmapsto (\rho, e'_1)$ is a subderivation. The first hypothesis can only be derived by one rule and it requires that $\mathrm{C}; \varrho \vdash e_1$ (1), $\mathrm{C}; \varrho \vdash \tau$ (7), $\mathrm{C}; \varrho \vdash e_2$ (2), and $\mathit{funC}(\mathrm{lbl}(e_1), \mathrm{lbl}(\tau), \mathrm{lbl}(e_2), i)$ (3). By the induction hypothesis and Judgement 1, $\mathrm{C}; \varrho \vdash e'_1$ (4). By Lemma 12, $\mathrm{C}(\mathrm{lbl}(e'_1)) \subseteq \mathrm{C}(\mathrm{lbl}(e_1))$ (5). Combining Facts 3 and 5, $\mathit{funC}(\mathrm{lbl}(e'_1), \mathrm{lbl}(\tau), \mathrm{lbl}(e_2), i)$ (6). Combining Facts 4, 7, 2, and 6, and using the rules for acceptable analysis, we see that $\mathrm{C}; \varrho \vdash (e'_1[\tau]\ e_2)^i$, as we are required to prove.

- In this case $e = (v^j[\tau]\ e_2)^i$ for some $v$, $j$, $\tau$, $e_2$, and $i$, $e' = (v^j[\tau]\ e'_2)^i$, and $(\rho, e_2) \longmapsto (\rho, e'_2)$ is a subderivation. The first hypothesis can only be derived by one rule and it requires that $\mathrm{C}; \varrho \vdash v^j$ (1), $\mathrm{C}; \varrho \vdash \tau$ (7), $\mathrm{C}; \varrho \vdash e_2$ (2), and $\mathit{funC}(j, \mathrm{lbl}(\tau), \mathrm{lbl}(e_2), i)$ (3). By the induction hypothesis and Judgement 1, $\mathrm{C}; \varrho \vdash e'_2$ (4). By Lemma 12, $\mathrm{C}(\mathrm{lbl}(e'_2)) \subseteq \mathrm{C}(\mathrm{lbl}(e_2))$ (5). Combining Facts 3 and 5, $\mathit{funC}(j, \mathrm{lbl}(\tau), \mathrm{lbl}(e'_2), i)$ (6). Combining Facts 1, 7, 4, and 6, and using the rules for acceptable analysis, we see that $\mathrm{C}; \varrho \vdash (v^j[\tau]\ e'_2)^i$, as we are required to prove.

- In this case:

$$\begin{array}{l} e = ({v_1}^j[\tau]\ {v_2}^k)^l \\ v_1 = \langle\rho', \texttt{fix}\ f[\alpha](x{:}\tau_1){:}\tau_2.e''\rangle \\ e' = \rho''(e''[\tau/\alpha])^l \\ \rho'' = \rho', f{:}\tau' = {v_1}^j, x{:}\tau_1[\tau/\alpha] = {v_2}^k \\ \tau' = (\forall\alpha.\tau_1 \to \tau_2)^j \end{array}$$

for some $\rho'$, $f$, $\alpha$, $x$, $\tau_1$, $\tau_2$, $e''$, $j$, $\tau$, $v_2$, $k$, and $l$. The first hypothesis can only be derived by one rule and it requires that $\mathrm{C}; \varrho \vdash {v_1}^j$ (1), $\mathrm{C}; \varrho \vdash \tau$ (2), $\mathrm{C}; \varrho \vdash {v_2}^k$ (3), and $\mathit{funC}(j, \mathrm{lbl}(\tau), k, l)$ (4). Let $i = \mathrm{lbl}(\tau_1)$. Judgement 1 can only be derived by one rule and it requires that $\varrho(f) = j$ (5), $\varrho(x) = i$ (6), $\mathrm{C}; \varrho \vdash \tau'$ (7), $\mathrm{C}; \varrho \vdash e''$ (8), and $(\forall i.\varrho(\alpha) \to \mathrm{lbl}(e''))^j_{\mathsf{v}} \in \mathrm{C}(j)$ (9). Instantiating Fact 4 with Fact 9, $\mathrm{C}(\mathrm{lbl}(\tau)) = \mathrm{C}(\varrho(\alpha))$ (10), $\mathrm{C}(k) \subseteq \mathrm{C}(i)$ (11), and $\mathrm{C}(\mathrm{lbl}(e'')) \subseteq \mathrm{C}(l)$ (12). Judgement 7 requires that $\mathrm{C}; \varrho \vdash \tau_1$ (13). By (13), (2), and (10), by Lemma 13, $\mathrm{C}; \varrho \vdash \tau_1[\tau/\alpha]$ (14). Since $\mathrm{C}; \varrho \vdash \rho$, (5), (7), $\mathrm{C}(j) \subseteq \mathrm{C}(j)$, (1), $\mathrm{lbl}(\tau_1[\tau/\alpha]) = \mathrm{lbl}(\tau_1) = i$ and (6), (14), (11), and (3), we can derive $\mathrm{C}; \varrho \vdash \rho''$ (15). By (13), (2), and (10), by Lemma 13, $\mathrm{C}; \varrho \vdash e''[\tau/\alpha]$ (16). By (15), (16), and $\mathrm{lbl}(e''[\tau/\alpha]) = \mathrm{lbl}(e'')$ and (12), we can derive $\mathrm{C}; \varrho \vdash e'$, as required.

- In this case $e = (\texttt{box}_\tau\ e_1)^i$ for some $t$, $e_1$, and $i$, $e' = (\texttt{box}_\tau\ e_2)^i$ for some $e_2$, and $(\rho, e_1) \longmapsto (\rho, e_2)$ is a subderivation. The first hypothesis can only be derived by one rule and it requires that $\mathrm{C}; \varrho \vdash \texttt{box}(\tau)^i$ (7), $\mathrm{C}; \varrho \vdash e_1$ (1), $(\texttt{box}_{tr(\tau)}\ j)^i_{\mathsf{v}} \in \mathrm{C}(i)$ (2) for some $j$, and $\mathrm{C}(\mathrm{lbl}(e_1)) \subseteq \mathrm{C}(j)$ (3). By the induction hypothesis and Judgement 1, $\mathrm{C}; \varrho \vdash e_2$ (4). By Lemma 12, $\mathrm{C}(\mathrm{lbl}(e_2)) \subseteq \mathrm{C}(\mathrm{lbl}(e_1))$ (5). Combining Facts 3 and 5 gives $\mathrm{C}(\mathrm{lbl}(e_2)) \subseteq \mathrm{C}(j)$ (6). Then by Facts 7, 4, 2, and 6, and using the rules for acceptable analysis, $\mathrm{C}; \varrho \vdash (\texttt{box}_\tau\ e_2)^i$, as we are required to prove.

- In this case $e = (\texttt{unbox}\,e_1)^i$ for some $e_1$ and $i$, $e' = (\texttt{unbox}\,e_2)^i$ for some $e_2$, and $(\rho, e_1) \longmapsto (\rho, e_2)$ is a subderivation. The first hypothesis can only be derived by one rule and it requires that $\mathrm{C}; \varrho \vdash e_1$ (1) and $\mathit{boxC}(\mathrm{lbl}(e_1), i)$ (2). By Judgement (1) and the induction hypothesis, $\mathrm{C}; \varrho \vdash e_2$ (3). By Lemma 12, $\mathrm{C}(\mathrm{lbl}(e_2)) \subseteq \mathrm{C}(\mathrm{lbl}(e_1))$ (4). Combining Facts 4 and 2, $\mathit{boxC}(\mathrm{lbl}(e_2), i)$ (5). Combining Facts 3 and 5, by the rules for acceptable analysis, $\mathrm{C}; \varrho \vdash (\texttt{unbox}\,e_2)^i$, as we are required to prove.

- In this case $e = (\texttt{unbox}\,\langle v^i{:}\tau\rangle^j)^k$ for some $\tau$, $v$, $i$, $j$, and $k$, and $e' = v^i$. The first hypothesis can only be derived by one rule and it requires that $\mathrm{C}; \varrho \vdash \langle v^i{:}\tau\rangle^j$, which in turn can only be derived by one rule that requires that $\mathrm{C}; \varrho \vdash v^i$, as we are required to prove.

- In this case $e = \rho'(e'')^i$, $e' = \rho'(e''')^i$, and $(\rho', e'') \longmapsto (\rho', e''')$ is a subderivation. Assumption $\mathrm{C}; \varrho \vdash e$ requires that $\mathrm{C}; \varrho \vdash \rho'$ (1), $\mathrm{C}; \varrho \vdash e''$ (2), and $\mathrm{C}(\mathrm{lbl}(e'')) \subseteq \mathrm{C}(i)$ (3). By (1), (2), and the induction hypothesis, $\mathrm{C}; \varrho \vdash e'''$ (4). By Lemma 12, $\mathrm{C}(\mathrm{lbl}(e''')) \subseteq \mathrm{C}(\mathrm{lbl}(e''))$ (5). Combining (3) and (5), $\mathrm{C}(\mathrm{lbl}(e''')) \subseteq \mathrm{C}(i)$ (6). Using (1), (4), and (6) we derive $\mathrm{C}; \varrho \vdash \rho'(e''')^i$, as required.

- In this case $e = \rho'(v^i)^j$ and $e' = v^i$. The assumption $\mathrm{C}; \varrho \vdash e$ unpacks to requiring that $\mathrm{C}; \varrho \vdash v^i$, as required.

■

**Lemma 15 (Many-step reduction preserves acceptability)**
*If* $\mathrm{C}; \varrho \vdash M$ *and* $M \longmapsto^* M'$ *then* $\mathrm{C}; \varrho \vdash M'$.

**Proof:** The proof is by a straightforward induction on the length of the reduction sequence and Lemma 14. ■

We can also show an important connection between typing and acceptable flow analysis—namely that the cache of an expression's type is a contained in the cache of that expression.

**Lemma 16**
*If* $\Delta; \Gamma \vdash e : \tau$, $\mathrm{C}; \varrho \vdash \Gamma$, *and* $\mathrm{C}; \varrho \vdash e$ *then* $\mathrm{C}(\mathrm{lbl}(\tau)) \subseteq \mathrm{C}(\mathrm{lbl}(e))$ *and* $\mathrm{C}; \varrho \vdash \tau$.

**Proof:** The proof is by induction on the derivation of $\Gamma \vdash e : \tau$. Consider the cases for the last rule used (in same order as figure):

- (Variable) In this case $e = x^i$ and $x{:}\tau \in \Gamma$. By the rules for acceptable analysis, $\varrho(x) = \mathrm{lbl}(\tau)$ (1), $\mathrm{C}; \varrho \vdash \tau$ (2), and $\mathrm{C}(\varrho(x)) \subseteq \mathrm{C}(i)$ (3). By (1) and (3), $\mathrm{C}(\mathrm{lbl}(\tau)) \subseteq \mathrm{C}(i)$ (4). The result is (4) and (2).
- (Fix expression) In this case, $e = (\mathtt{fix}\ f[\alpha](x{:}\tau_1){:}\tau_2.e')^i$ and $\tau = (\forall\alpha.\tau_1 \to \tau_2)^i$. The first part is immediate since $\mathrm{lbl}(\tau) = \mathrm{lbl}(e)$. The second part is required by $\mathrm{C}; \varrho \vdash e$.
- (Application) In this case:

  $$e = (e_1[\tau']\ e_2)^i \quad \Delta; \Gamma \vdash e_1 : (\forall\alpha.\tau_1 \to \tau_3)^j \quad (1) \quad \tau = \tau_3[\tau'/\alpha]$$

  By the rules for acceptability, $\mathrm{C}; \varrho \vdash e_1$ (2), $\mathrm{C}; \varrho \vdash \tau'$ (3), $\mathrm{C}; \varrho \vdash e_2$, and $\mathit{funC}(\mathrm{lbl}(e_1), \mathrm{lbl}(\tau'), \mathrm{lbl}(e_2), i)$ (4). By (1), (2), and the induction hypothesis, $\mathrm{C}(j) \subseteq \mathrm{C}(\mathrm{lbl}(e_1))$ (5) and $\mathrm{C}; \varrho \vdash (\forall\alpha.\tau_1 \to \tau_3)^j$ (6). By (6) and the rules for acceptability, $\mathrm{C}; \varrho \vdash \tau_3$ (7) and $(\forall\varrho(\alpha).\mathrm{lbl}(\tau_1) \to \mathrm{lbl}(\tau_3))^k_{\mathtt{t}} \in \mathrm{C}(j)$ (8). By (5), instantiating (4) with (8), $\mathrm{C}(\mathrm{lbl}(\tau')) = \mathrm{C}(\varrho(\alpha))$ (9) and $\mathrm{C}(\mathrm{lbl}(\tau_3)) \subseteq \mathrm{C}(i)$, so since $\mathrm{lbl}(\tau) = \mathrm{lbl}(\tau_3)$, $\mathrm{C}(\mathrm{lbl}(\tau)) \subseteq \mathrm{C}(i)$ (10). By (3), (7), and (9), $\mathrm{C}; \varrho \vdash \tau_3[\tau'/\alpha]$ (11). The result is (10) and (11).
- (Box expression) In this case $e = (\mathtt{box}_{\tau'}\ e')^i$ and $\tau = \mathtt{box}(\tau')^i$. The first part holds as $\mathrm{lbl}(e) = \mathrm{lbl}(\tau)$. The second part is required by $\mathrm{C}; \varrho \vdash e$.
- (Unbox) In this case $e = (\mathtt{unbox}\, e')^i$ and $\Delta; \Gamma \vdash e' : \mathtt{box}(\tau)^j$ (1) is a subderivation. By the rules for acceptability, $\mathrm{C}; \varrho \vdash e'$ (2) and $\mathit{boxC}(\mathrm{lbl}(e'), i)$ (3). By (1), (2), and the induction hypothesis, $\mathrm{C}(j) \subseteq \mathrm{C}(\mathrm{lbl}(e'))$ (4) and $\mathrm{C}; \varrho \vdash \mathtt{box}(\tau)^j$ (5). By (5) and the rules for acceptability, $\mathrm{C}; \varrho \vdash \tau$ (6) and $(\mathtt{box}\,\mathrm{lbl}(\tau))^k_{\mathtt{t}} \in \mathrm{C}(j)$ (7). By (4), instantiating (3) with (7), $\mathrm{C}(\mathrm{lbl}(\tau)) \subseteq \mathrm{C}(i)$ (8). The result is (8) and (6).
- (Frame) In this case $e = \rho(e')^i$, $\vdash \rho : \Gamma'$ (1), and $\emptyset; \Gamma' \vdash e' : \tau$ (2). By the rules for acceptability, $\mathrm{C}; \varrho \vdash \rho$ (3), $\mathrm{C}; \varrho \vdash e'$ (4), and $\mathrm{C}(\mathrm{lbl}(e')) \subseteq \mathrm{C}(i)$ (5). By (1), (3), the rules for typing, and the rules for acceptability, $\mathrm{C}; \varrho \vdash \Gamma'$ (6). By (6), (2), (4), and the induction hypothesis, $\mathrm{C}(\mathrm{lbl}(\tau)) \subseteq \mathrm{C}(\mathrm{lbl}(e'))$ (7) and $\mathrm{C}; \varrho \vdash \tau$ (8). By (7) and (5), $\mathrm{C}(\mathrm{lbl}(\tau)) \subseteq \mathrm{C}(i)$ (9). The result is (9) and (8).
- (Constant) In this case $e = c^i$ and $\tau = \mathtt{B}^i$. The first part clearly holds as $\mathrm{lbl}(e) = \mathrm{lbl}(\tau)$. The second part is required by $\mathrm{C}; \varrho \vdash e$.
- (Fix value) In this case, $e = \langle \rho, \mathtt{fix}\ f[\alpha](x{:}\tau_1){:}\tau_2.e' \rangle^i$, $\tau = (\forall\alpha.\tau_1 \to \tau_2)^i$. The first part holds as $\mathrm{lbl}(e) = \mathrm{lbl}(\tau)$. The second part is required by $\mathrm{C}; \varrho \vdash e$.
- (Box value) In this case $e = \langle v^j{:}\tau' \rangle^i$ and $\tau = \mathtt{box}(\tau')^i$. The first part holds as $\mathrm{lbl}(e) = \mathrm{lbl}(\tau)$. The second part is required by $\mathrm{C}; \varrho \vdash e$.

■

## 4. Unboxing

The goal of the unboxing optimization is to use the information provided by a flow analysis to replace a boxed object with the contents of the box. Doing so may change the traceability, since the object in the box may not be a GC-managed reference. Moreover, the object in the box may itself be a candidate for unboxing; consequently, determining the traceability of boxed objects depends on exactly which objects are unboxed. Function parameters may be instantiated with objects from multiple different definition sites, some of which may be unboxed and some of which may not.

Consider again the first example from Section 2, written out with explicit type information and labels:

$$\begin{array}{l}\mathtt{let}\\ \quad \mathtt{fix}\ f[](x{:}\mathtt{box}(\mathtt{B}^0)^1){:}\mathtt{box}(\mathtt{box}(\mathtt{B}^2)^3)^4.(\mathtt{box}_{\mathtt{box}(\mathtt{B}^5)^6}\ x^7)^8\\ \quad z_1 = (\mathtt{box}_{\mathtt{B}^9}\ 3^{10})^{11}\\ \quad z_2 = f[](z_1{}^{12})^{13}\\ \mathtt{in}\,(\mathtt{unbox}\,(\mathtt{unbox}\, z_2{}^{15})^{16})^{17}\end{array}$$

It is fairly easy to see that this program is unboxable. The binding site for $x$ is only reached by the term labeled with 11 (the outer box introduction), and hence there should be no problems with changing its type annotation. Each box elimination is reached only by a single box introduction, and hence the box/unbox pairs in this program should be eliminable, yielding an optimized program:

$$\begin{array}{l}\mathtt{let}\\ \quad \mathtt{fix}\ f[](x{:}\mathtt{B}^0){:}\mathtt{B}^2.x^7\\ \quad z_1 = 3^{10}\\ \quad z_2 = f[](z_1{}^{12})^{13}\\ \mathtt{in}\, z_2{}^{15}\end{array}$$

Notice that in order to rewrite the program, we have had to change the types of both $f$ and $x$, since we have eliminated the box introductions on the argument and in the body of $f$. The change in type of $x$ has changed its traceability from $\mathtt{r}$ to $\mathtt{b}$. If we choose (perhaps because of limitations on the precision of the analysis, or perhaps because of other constraints) to only eliminate the outer box/unbox pair, then we must similarly adjust types on the the remaining box introduction (labeled with 8).

$$\begin{array}{l}\mathtt{let}\\ \quad \mathtt{fix}\ f[](x{:}\mathtt{B}^0){:}\mathtt{box}(\mathtt{B}^2)^3.(\mathtt{box}_{\mathtt{B}^5}\ x^7)^8\\ \quad z_1 = 3^{10}\\ \quad z_2 = f[](z_1{}^{12})^{13}\\ \mathtt{in}\,(\mathtt{unbox}\, z_2{}^{15})^{16}\end{array}$$

Clearly then, to optimize these programs in a type preserving fashion, we must rewrite types along with the terms in the program. In the rest of this section, we first develop a framework for specifying an unboxing assignment regardless of any correctness concerns, and then separately define a judgement specifying when such an assignment is a reasonable one.

$\boxed{\rfloor \tau \lfloor_{\Upsilon}}$

$$
\begin{array}{lcll}
\rfloor \alpha^i \lfloor_{\Upsilon} & = & \alpha^i & \\
\rfloor \mathtt{B}^i \lfloor_{\Upsilon} & = & \mathtt{B}^i & \\
\rfloor (\forall \alpha.\tau_1 \to \tau_2)^i \lfloor_{\Upsilon} & = & (\forall \alpha.\rfloor \tau_1 \lfloor_{\Upsilon} \to \rfloor \tau_2 \lfloor_{\Upsilon})^i & \\
\rfloor \mathtt{box}(\tau)^i \lfloor_{\Upsilon} & = & \rfloor \tau \lfloor_{\Upsilon} & i \in \Upsilon \\
\rfloor \mathtt{box}(\tau)^i \lfloor_{\Upsilon} & = & \mathtt{box}(\rfloor \tau \lfloor_{\Upsilon})^i & i \notin \Upsilon
\end{array}
$$

$\boxed{\rfloor \Gamma \lfloor_{\Upsilon}}$

$$
\rfloor x_1{:}\tau_1, \ldots, x_n{:}\tau_n \lfloor_{\Upsilon} = x_1{:}\rfloor \tau_1 \lfloor_{\Upsilon}, \ldots, x_n{:}\rfloor \tau_n \lfloor_{\Upsilon}
$$

$\boxed{\rfloor e \lfloor_{\Upsilon}}$

$$
\begin{array}{lcll}
\rfloor x^i \lfloor_{\Upsilon} & = & x^i & \\
\rfloor m^i \lfloor_{\Upsilon} & = & (\mathtt{fix}\ f[\alpha](x{:}\rfloor \tau_1 \lfloor_{\Upsilon}){:}\rfloor \tau_2 \lfloor_{\Upsilon}.\rfloor e \lfloor_{\Upsilon})^i & \\
 & & \text{where } m = \mathtt{fix}\ f[\alpha](x{:}\tau_1){:}\tau_2.e & \\
\rfloor (e_1[\tau]\ e_2)^i \lfloor_{\Upsilon} & = & (\rfloor e_1 \lfloor_{\Upsilon}[\rfloor \tau \lfloor_{\Upsilon}]\ \rfloor e_2 \lfloor_{\Upsilon})^i & \\
\rfloor (\mathtt{box}_{\tau}\ e)^i \lfloor_{\Upsilon} & = & \rfloor e \lfloor_{\Upsilon} & i \in \Upsilon \\
 & = & (\mathtt{box}_{\rfloor \tau \lfloor_{\Upsilon}}\ \rfloor e \lfloor_{\Upsilon})^i & i \notin \Upsilon \\
\rfloor (\mathtt{unbox}\, e)^i \lfloor_{\Upsilon} & = & \rfloor e \lfloor_{\Upsilon} & \mathrm{lbl}(e) \in \Upsilon \\
 & = & (\mathtt{unbox}\, \rfloor e \lfloor_{\Upsilon})^i & \mathrm{lbl}(e) \notin \Upsilon \\
\rfloor \rho(e)^i \lfloor_{\Upsilon} & = & \rfloor \rho \lfloor_{\Upsilon}(\rfloor e \lfloor_{\Upsilon})^i & \\
\rfloor c^i \lfloor_{\Upsilon} & = & c^i & \\
\rfloor v^i \lfloor_{\Upsilon} & = & \langle \rfloor \rho \lfloor_{\Upsilon}, \mathtt{fix}\ f[\alpha](x{:}\rfloor \tau_1 \lfloor_{\Upsilon}){:}\rfloor \tau_2 \lfloor_{\Upsilon}.\rfloor e \lfloor_{\Upsilon} \rangle^i & \\
 & & \text{where } v = \langle \rho, \mathtt{fix}\ f[\alpha](x{:}\tau_1){:}\tau_2.e \rangle & \\
\rfloor \langle v^j{:}\tau \rangle^i \lfloor_{\Upsilon} & = & \rfloor v^j \lfloor_{\Upsilon} & i \in \Upsilon \\
 & = & \langle \rfloor v^j \lfloor_{\Upsilon}{:}\rfloor \tau \lfloor_{\Upsilon} \rangle^i & i \notin \Upsilon
\end{array}
$$

$\boxed{\rfloor \rho \lfloor_{\Upsilon}}$

$$
\begin{array}{l}
\rfloor x_1{:}\tau_1 = {v_1}^{j_1}, \ldots, x_n{:}\tau_n = {v_n}^{j_n} \lfloor_{\Upsilon} = \\
\quad x_1{:}\rfloor \tau_1 \lfloor_{\Upsilon} = \rfloor {v_1}^{j_1} \lfloor_{\Upsilon}, \ldots, x_n{:}\rfloor \tau_n \lfloor_{\Upsilon} = \rfloor {v_n}^{j_n} \lfloor_{\Upsilon}
\end{array}
$$

$\boxed{\rfloor M \lfloor_{\Upsilon}}$

$$
\rfloor (\rho, e) \lfloor_{\Upsilon} = (\rfloor \rho \lfloor_{\Upsilon}, \rfloor e \lfloor_{\Upsilon})
$$

**Figure 8.** Unboxing

### 4.1 The unboxing optimization

We specify a particular choice of unboxing via an unboxing set $\Upsilon$ which contains the set of labels of terms and types to be unboxed. A choice of a particular $\Upsilon$ then induces an unboxing function as defined in Figure 8. The unboxing function is defined in a straightforward compositional manner. Box introductions are dropped when their labels are in the unboxing set, box type constructors are dropped when their labels are in the unboxing set, box eliminations are dropped when the labels of their arguments are in the unboxing set, and all other terms and types are left unchanged.

An important observation about the unboxing optimization as we have defined it is unlike many previous interprocedural approaches (Section 7), it only improves programs and never introduces instructions or allocation. This is easy to see, since the unboxing function only removes boxes (which allocate and have an instruction cost), and unboxes (which have an instruction cost) and never introduces any new operations at all.

### 4.2 Acceptable unboxings

While any choice of $\Upsilon$ defines an unboxing, not every unboxing set is reasonable in the sense that it defines a type and semantics preserving optimization. Just as we defined a notion of acceptable analysis in Section 3, we will define a judgement that captures sufficient conditions for ensuring correctness of an unboxing, without specifying a particular method of choosing such an unboxing. By using analyses of different precisions or choosing different optimization strategies we may end up with quite different choices of unboxings; however, so long as they satisfy our notion of acceptability we can be sure that they will preserve correctness.

Informally, a choice of an unboxing set is reasonable if it meets two criteria. Firstly, it must make uniform choices in the sense that if a box introduction is eliminated, then all of the types and elimination forms to which it flows must also be unboxed, and vice versa. Secondly, we must ensure that types remain consistent with their uses in polymorphic instantiations, since we do not allow polymorphism over base types.

We use the notation $i \overset{\Upsilon}{\simeq} j$ to indicate when an unboxing *agrees* at two labels $i$ and $j$.

$$
i \overset{\Upsilon}{\simeq} j \quad \text{iff} \quad \text{either } i, j \in \Upsilon \text{ or } i, j \notin \Upsilon
$$

The first requirement is then specified via the *cache consistency* judgement, which enforces that for any label $i$, the unboxing set must agree on $i$ and the labels of any shapes in the cache of $i$.

$$
\frac{\forall i, s : s \in \mathrm{C}(i) \implies i \overset{\Upsilon}{\simeq} \mathrm{lbl}(s)}{\mathrm{C} \vdash \Upsilon}
$$

The second requirement is specified via the *consistent unboxing* judgement of Figure 9. The type rules determine the traceability of the unboxed type: that is, the judgement $\Upsilon \vdash \tau : t$ indicates that unboxing $\tau$ with $\Upsilon$ will result in a type of traceability $t$. The key use of the type judgement is in the term level polymorphic instantiation rule, which requires that the traceability of the unboxed type be $\mathtt{r}$.

### 4.3 Type Preservation

Our goal is to show that the unboxing function induced by any acceptable unboxing is in some sense correct as an optimization. The first part of this is to show that unboxing preserves typing. One key property is that types have non-empty caches.

**Lemma 17 (Type Inhabitance)**
*If $\tau$ is not a type variable and $\mathrm{C}; \varrho \vdash \tau$ then $\mathrm{C}(\mathrm{lbl}(\tau)) \neq \emptyset$.*

**Proof:** The proof is by inspection of the rules for acceptability. ■

We also need several technical properties: labels agree when their caches intersect, unboxing preserves type well formedness, type traceability, and type equality, and unboxing commutes with type subsitution.

**Lemma 18 (Agreement)**
*If $\mathrm{C} \vdash \Upsilon$ and $\mathrm{C}(i) \cap \mathrm{C}(j) \neq \emptyset$ then $i \overset{\Upsilon}{\simeq} j$.*

**Proof:** The proof is by inspection of the rules for cache consistency. ■

**Lemma 19**
*If $\Delta \vdash \tau\ wf$ then $\Delta \vdash \rfloor \tau \lfloor_{\Upsilon}\ wf$.*

**Proof:** The proof is a straight forward induction on the structure of $\tau$. ■

**Lemma 20**
*If $\mathrm{C}; \varrho \vdash \tau_2$ and $\mathrm{C}(\mathrm{lbl}(\tau_2)) = \mathrm{C}(\varrho(\alpha))$ then:*

- *If $\mathrm{C}; \varrho \vdash \tau_1$ then $\rfloor \tau_1[\tau_2/\alpha] \lfloor_{\Upsilon} = \rfloor \tau_1 \lfloor_{\Upsilon}[\rfloor \tau_2 \lfloor_{\Upsilon}/\alpha]$.*
- *If $\mathrm{C}; \varrho \vdash e$ then $\rfloor e[\tau_2/\alpha] \lfloor_{\Upsilon} = \rfloor e \lfloor_{\Upsilon}[\rfloor \tau_2 \lfloor_{\Upsilon}/\alpha]$.*

**Proof:**

$\boxed{\Upsilon \vdash \tau : t}$

$$\frac{}{\Upsilon \vdash \alpha^i : \mathtt{r}} \qquad \frac{}{\Upsilon \vdash \mathtt{B}^i : \mathtt{b}}$$

$$\frac{}{\Upsilon \vdash (\forall \alpha.\tau_1 \to \tau_2)^i : \mathtt{r}}$$

$$\frac{i \in \Upsilon \quad \Upsilon \vdash \tau : t}{\Upsilon \vdash \mathtt{box}(\tau)^i : t} \qquad \frac{i \notin \Upsilon}{\Upsilon \vdash \mathtt{box}(\tau)^i : \mathtt{r}}$$

$\boxed{\Upsilon \vdash e}$

$$\frac{}{\Upsilon \vdash x^i} \qquad \frac{\Upsilon \vdash e}{\Upsilon \vdash (\mathtt{fix}\ f[\alpha](x{:}\tau_1){:}\tau_2.e)^i}$$

$$\frac{\Upsilon \vdash e_1 \quad \Upsilon \vdash e_2 \quad \Upsilon \vdash \tau : \mathtt{r}}{\Upsilon \vdash (e_1[\tau]\ e_2)^i}$$

$$\frac{\Upsilon \vdash e}{\Upsilon \vdash (\mathtt{box}_\tau\ e)^i} \qquad \frac{\Upsilon \vdash e}{\Upsilon \vdash (\mathtt{unbox}\ e)^i}$$

$$\frac{\Upsilon \vdash \rho \quad \Upsilon \vdash e}{\Upsilon \vdash \rho(e)^i} \qquad \frac{}{\Upsilon \vdash c^i}$$

$$\frac{\Upsilon \vdash \rho \quad \Upsilon \vdash e}{\Upsilon \vdash \langle \rho, \mathtt{fix}\ f[\alpha](x{:}\tau_1){:}\tau_2.e\rangle^i} \qquad \frac{\Upsilon \vdash v^i}{\Upsilon \vdash \langle v^i{:}\tau\rangle^j}$$

$\boxed{\Upsilon \vdash \rho}$

$$\frac{\forall 1 \le j \le n : \Upsilon \vdash {v_j}^{i_j}}{\Upsilon \vdash x_1{:}\tau_1 = {v_1}^{i_1}, \ldots, x_n{:}\tau_n = {v_n}^{i_n}}$$

$\boxed{\Upsilon \vdash M}$

$$\frac{\Upsilon \vdash \rho \quad \Upsilon \vdash e}{\Upsilon \vdash (\rho, e)}$$

**Figure 9.** Consistent unboxing

- The proof is by induction on the structure of $\tau_1$. Consider the cases for $\tau_1$:
  - Case 1, $\tau_1 = \alpha^i$: If $\tau_2 = \sigma^j$ and $\downharpoonleft\sigma^j\downharpoonright_\Upsilon = \sigma'^k$ then $\tau_1[\tau_2/\alpha] = \sigma^i$, thus $\downharpoonleft\tau_1[\tau_2/\alpha]\downharpoonright_\Upsilon = \downharpoonleft\sigma^i\downharpoonright_\Upsilon$, and also $\downharpoonleft\tau_1\downharpoonright_\Upsilon[\downharpoonleft\tau_2\downharpoonright_\Upsilon/\alpha] = \sigma'^i$. Thus I need to show that $\downharpoonleft\sigma^i\downharpoonright_\Upsilon = \sigma'^i$. When $\sigma$ is not a box type, this condition follows easily from the definitions. When $\sigma$ is a box type, this condition follows if $i \overset{\Upsilon}{\simeq} j$. By $\mathrm{C}; \varrho \vdash \tau_2$ and Lemma 17, $\mathrm{C}(j) \neq \emptyset$. By $\mathrm{C}; \varrho \vdash \tau_1$, $\mathrm{C}(\mathrm{lbl}(\tau_2)) = \mathrm{C}(\varrho(\alpha))$, and the rules for acceptability, $\mathrm{C}(i) = \mathrm{C}(j)$. By Lemma 18, $i \overset{\Upsilon}{\simeq} j$, as required.
  - Case 2, $\tau_1 = \beta^i$ and $\alpha \neq \beta$: In this case $\tau_1[\tau_2/\alpha] = \tau_1$, $\downharpoonleft\tau_1\downharpoonright_\Upsilon = \tau_1$, and the result is immediate.
  - Case 3, $\tau_1 = (\forall\alpha'.\tau_3 \to \tau_4)^i$: Then $\mathrm{C}; \varrho \vdash \tau_1$ requires $\mathrm{C}; \varrho \vdash \tau_3$ and $\mathrm{C}; \varrho \vdash \tau_4$. By the induction hypothesis, $\downharpoonleft\tau_3[\tau_2/\alpha]\downharpoonright_\Upsilon = \downharpoonleft\tau_3\downharpoonright_\Upsilon[\downharpoonleft\tau_2\downharpoonright_\Upsilon/\alpha]$ and $\downharpoonleft\tau_4[\tau_2/\alpha]\downharpoonright_\Upsilon = \downharpoonleft\tau_4\downharpoonright_\Upsilon[\downharpoonleft\tau_2\downharpoonright_\Upsilon/\alpha]$. Thus:

$$\begin{aligned}
& \downharpoonleft\tau_1[\tau_2/\alpha]\downharpoonright_\Upsilon \\
= {} & \downharpoonleft(\forall\alpha'.\tau_3[\tau_2/\alpha] \to \tau_3[\tau_2/\alpha])^i\downharpoonright_\Upsilon \\
= {} & (\forall\alpha'.\downharpoonleft\tau_3[\tau_2/\alpha]\downharpoonright_\Upsilon \to \downharpoonleft\tau_3[\tau_2/\alpha]\downharpoonright_\Upsilon)^i \\
= {} & (\forall\alpha'.\downharpoonleft\tau_3\downharpoonright_\Upsilon[\downharpoonleft\tau_2\downharpoonright_\Upsilon/\alpha] \to \downharpoonleft\tau_4\downharpoonright_\Upsilon[\downharpoonleft\tau_2\downharpoonright_\Upsilon/\alpha])^i \\
= {} & (\forall\alpha'.\downharpoonleft\tau_3\downharpoonright_\Upsilon \to \downharpoonleft\tau_4\downharpoonright_\Upsilon)^i[\downharpoonleft\tau_2\downharpoonright_\Upsilon/\alpha] \\
= {} & \downharpoonleft\tau_1\downharpoonright_\Upsilon[\downharpoonleft\tau_2\downharpoonright_\Upsilon/\alpha]
\end{aligned}$$

  - Case 4, $\tau_1 = \mathtt{box}(\tau)^i$: Then $\mathrm{C}; \varrho \vdash \tau_1$ requires $\mathrm{C}; \varrho \vdash \tau$. The induction hypothesis is $\downharpoonleft\tau[\tau_2/\alpha]\downharpoonright_\Upsilon = \downharpoonleft\tau\downharpoonright_\Upsilon[\downharpoonleft\tau_2\downharpoonright_\Upsilon/\alpha]$. If $i \in \Upsilon$ then $\downharpoonleft\tau_1\downharpoonright_\Upsilon = \downharpoonleft\tau\downharpoonright_\Upsilon$ and $\downharpoonleft\tau_1[\tau_2/\alpha]\downharpoonright_\Upsilon = \downharpoonleft\tau[\tau_2/\alpha]\downharpoonright_\Upsilon$, as required. If $i \notin \Upsilon$ then:

$$\begin{aligned}
& \downharpoonleft\tau_1[\tau_2/\alpha]\downharpoonright_\Upsilon \\
= {} & \downharpoonleft\mathtt{box}(\tau[\tau_2/\alpha])^i\downharpoonright_\Upsilon \\
= {} & \mathtt{box}(\downharpoonleft\tau[\tau_2/\alpha]\downharpoonright_\Upsilon)^i \\
= {} & \mathtt{box}(\downharpoonleft\tau\downharpoonright_\Upsilon[\downharpoonleft\tau_2\downharpoonright_\Upsilon/\alpha])^i \\
= {} & \mathtt{box}(\downharpoonleft\tau\downharpoonright_\Upsilon)^i[\downharpoonleft\tau_2\downharpoonright_\Upsilon/\alpha] \\
= {} & \downharpoonleft\tau_1\downharpoonright_\Upsilon[\downharpoonleft\tau_2\downharpoonright_\Upsilon/\alpha]
\end{aligned}$$

- The proof is a straight forward induction on the structure of $e$.

■

**Lemma 21**
*If $\Upsilon \vdash \tau : t$ then $tr(\downharpoonleft\tau\downharpoonright_\Upsilon) = t$.*

**Proof:** The proof is a straight forward induction on the derivation of $\Upsilon \vdash \tau : t$. ■

**Lemma 22**
*If $\vdash \tau_1 = \tau_2$, $\mathrm{C} \vdash \Upsilon$, $\mathrm{C}; \varrho \vdash \tau_1$, $\mathrm{C}; \varrho \vdash \tau_2$, and either $\mathrm{C}(\mathrm{lbl}(\tau_1)) \subseteq \mathrm{C}(\mathrm{lbl}(\tau_2))$ or $\mathrm{C}(\mathrm{lbl}(\tau_2)) \subseteq \mathrm{C}(\mathrm{lbl}(\tau_1))$ then $\vdash \downharpoonleft\tau_1\downharpoonright_\Upsilon = \downharpoonleft\tau_2\downharpoonright_\Upsilon$.*

**Proof:** The proof is by induction on the derivation of $\vdash \tau_1 = \tau_2$. Consider the last rule used (in the same order as the figure):

- (Type variable) In this case $\tau_1 = \alpha^i$ and $\tau_2 = \alpha^j$. By definition, $\downharpoonleft\tau_1\downharpoonright_\Upsilon = \tau_1$ and $\downharpoonleft\tau_2\downharpoonright_\Upsilon = \tau_2$, and the result is immediate.
- (Base) In this case $\tau_1 = \mathtt{B}^i$ and $\tau_2 = \mathtt{B}^j$. By definition, $\downharpoonleft\tau_1\downharpoonright_\Upsilon = \tau_1$ and $\downharpoonleft\tau_2\downharpoonright_\Upsilon = \tau_2$, and the result is immediate.
- (Function) In this case:

$$\begin{aligned}
& \tau_1 = (\forall\alpha.\tau_{11} \to \tau_{12})^{i_1} \\
& \tau_2 = (\forall\alpha.\tau_{21} \to \tau_{22})^{i_2} \\
& \vdash \tau_{11} = \tau_{21} && (1) \\
& \vdash \tau_{12} = \tau_{22} && (2)
\end{aligned}$$

  WLOG, assume $\mathrm{C}(i_1) \subseteq \mathrm{C}(i_2)$ (3). By the rules for acceptability:

$$\begin{aligned}
& \mathrm{C}; \varrho \vdash \tau_{11} && (4) \\
& \mathrm{C}; \varrho \vdash \tau_{12} && (5) \\
& (\forall\varrho(\alpha).\mathrm{lbl}(\tau_{11}) \to \mathrm{lbl}(\tau_{12}))^j_t \in \mathrm{C}(i_1) && (6) \\
& \mathrm{C}; \varrho \vdash \tau_{21} && (7) \\
& \mathrm{C}; \varrho \vdash \tau_{22} && (8) \\
& \mathit{funC}(i_2, \varrho(\alpha), \mathrm{lbl}(\tau_{21}), \mathrm{lbl}(\tau_{22})) && (9)
\end{aligned}$$

  By (6), (3), and (9), $\mathrm{C}(\mathrm{lbl}(\tau_{21})) \subseteq \mathrm{C}(\mathrm{lbl}(\tau_{11}))$ (10) and $\mathrm{C}(\mathrm{lbl}(\tau_{12})) \subseteq \mathrm{C}(\mathrm{lbl}(\tau_{22}))$ (11). By (1), (4), (7), (10), and the induction hypothesis, $\vdash \downharpoonleft\tau_{11}\downharpoonright_\Upsilon = \downharpoonleft\tau_{21}\downharpoonright_\Upsilon$ (12). By (2), (5), (8), (11), and the induction hypothesis, $\vdash \downharpoonleft\tau_{12}\downharpoonright_\Upsilon = \downharpoonleft\tau_{22}\downharpoonright_\Upsilon$ (13).

By (12), (13), and the typing rules:

$$\vdash (\forall\alpha.\lfloor\tau_{11}\rfloor_\Upsilon \to \lfloor\tau_{12}\rfloor_\Upsilon)^{i_1} = (\forall\alpha.\lfloor\tau_{21}\rfloor_\Upsilon \to \lfloor\tau_{22}\rfloor_\Upsilon)^{i_2}$$

By definition:

$$\vdash \lfloor(\forall\alpha.\tau_{11} \to \tau_{12})^{i_1}\rfloor_\Upsilon = \lfloor(\forall\alpha.\tau_{21} \to \tau_{22})^{i_2}\rfloor_\Upsilon$$

as required.

- (Box) In this case $\tau_1 = \texttt{box}(\tau_1')^{i_1}$, $\tau_2 = \texttt{box}(\tau_2')^{i_2}$, and $\vdash \tau_1' = \tau_2'$ (1). WLOG, assume $\mathrm{C}(i_1) \subseteq \mathrm{C}(i_2)$ (2). By the rules for acceptability, $\mathrm{C}; \varrho \vdash \tau_1'$ (3), $(\texttt{box}\,\mathrm{lbl}(\tau_1'))^j_{\mathtt{t}} \in \mathrm{C}(i_1)$ (4), $\mathrm{C}; \varrho \vdash \tau_2'$ (5), and $boxC(i_2, \mathrm{lbl}(\tau_2'))$ (6). By (4), (2), and (6), $\mathrm{C}(\mathrm{lbl}(\tau_1')) \subseteq \mathrm{C}(\mathrm{lbl}(\tau_2'))$ (7). By (1), (3), (5), (7), and the induction hypothesis, $\vdash \lfloor\tau_1'\rfloor_\Upsilon = \lfloor\tau_2'\rfloor_\Upsilon$ (8). By Lemmas 17 and 18, $i_1 \overset{\Upsilon}{\simeq} i_2$. There are two cases:
  - Case 1, $i_1 \in \Upsilon$: In this case, $\lfloor\tau_1\rfloor_\Upsilon = \lfloor\tau_1'\rfloor_\Upsilon$, $\lfloor\tau_2\rfloor_\Upsilon = \lfloor\tau_2'\rfloor_\Upsilon$, and the result is (8).
  - Case 2, $i_2 \notin \Upsilon$: In this case, $\lfloor\tau_1\rfloor_\Upsilon = \texttt{box}(\lfloor\tau_1'\rfloor_\Upsilon)^{i_1}$, $\lfloor\tau_2\rfloor_\Upsilon = \texttt{box}(\lfloor\tau_2'\rfloor_\Upsilon)^{i_2}$, and the result follows from (8) and the typing rules.

■

Now we can prove that unboxing preserves typing.

**Theorem 1 (Consistent unboxings preserve typing)**
*If* $\mathrm{C} \vdash \Upsilon$ *then:*

- *If* $\Delta; \Gamma \vdash e : \tau$, $\mathrm{C}; \varrho \vdash \Gamma$, $\mathrm{C}; \varrho \vdash e$, *and* $\Upsilon \vdash e$ *then* $\Delta; \lfloor\Gamma\rfloor_\Upsilon \vdash \lfloor e\rfloor_\Upsilon : \lfloor\tau\rfloor_\Upsilon$.
- *If* $\vdash \rho : \Gamma$, $\mathrm{C}; \varrho \vdash \rho$, *and* $\Upsilon \vdash \rho$ *then* $\vdash \lfloor\rho\rfloor_\Upsilon : \lfloor\Gamma\rfloor_\Upsilon$.
- *If* $\vdash M : \tau$, $\mathrm{C}; \varrho \vdash M$, *and* $\Upsilon \vdash M$ *then* $\vdash \lfloor M\rfloor_\Upsilon : \lfloor\tau\rfloor_\Upsilon$.

**Proof:** The proof is by induction on the structure of the typing judgement. Consider the cases, in the same order as the figure, for the last rule used in the derivation:

- (Variable) In this case $e = x^i$ and $x{:}\tau \in \Gamma$. Then $\lfloor e\rfloor_\Upsilon = x^i$ and clearly $x{:}\lfloor\tau\rfloor_\Upsilon \in \lfloor\Gamma\rfloor_\Upsilon$, so the result follows by the typing rules.
- (Fix expression) In this case $e = (\texttt{fix}\ f[\alpha](x{:}\tau_1){:}\tau_2.e')^i$. The typing rule requires that both $\tau = (\forall\alpha.\tau_1 \to \tau_2)^i$ and $\Delta; \Gamma, f{:}\tau, x{:}\tau_1 \vdash e' : \tau_2$. The assumption $\mathrm{C}; \varrho \vdash e$ requires $\varrho(f) = i$, $\varrho(x) = \mathrm{lbl}(\tau_1)$, $\mathrm{C} \vdash \tau$, and $\mathrm{C}; \varrho \vdash e'$. From $\mathrm{C} \vdash \tau$ and the rules for acceptability, $\mathrm{C} \vdash \tau_1$. From these facts, $\mathrm{C}; \varrho \vdash \Gamma, f{:}\tau, x{:}\tau_1$. The assumption $\Upsilon \vdash e$ requires that $\Upsilon \vdash e'$. By the induction hypothesis, $\Delta; \lfloor\Gamma, f{:}\tau, x{:}\tau_1\rfloor_\Upsilon \vdash \lfloor e'\rfloor_\Upsilon : \lfloor\tau_2\rfloor_\Upsilon$. Since:

$$\lfloor\Gamma, f{:}\tau, x{:}\tau_1\rfloor_\Upsilon = \lfloor\Gamma\rfloor_\Upsilon, f{:}(\forall\alpha.\lfloor\tau_1\rfloor_\Upsilon \to \lfloor\tau_2\rfloor_\Upsilon)^i, x{:}\lfloor\tau_2\rfloor_\Upsilon$$

by the typing rules:

$$\Delta; \lfloor\Gamma\rfloor_\Upsilon \vdash (\texttt{fix}\ f[\alpha](x{:}\lfloor\tau_1\rfloor_\Upsilon){:}\lfloor\tau_2\rfloor_\Upsilon.\lfloor e'\rfloor_\Upsilon)^i : (\forall\alpha.\lfloor\tau_1\rfloor_\Upsilon \to \lfloor\tau_2\rfloor_\Upsilon)^i$$

The result follows since:

$$\lfloor e\rfloor_\Upsilon = (\texttt{fix}\ f[\alpha](x{:}\lfloor\tau_1\rfloor_\Upsilon){:}\lfloor\tau_2\rfloor_\Upsilon.\lfloor e'\rfloor_\Upsilon)^i$$
$$\lfloor\tau\rfloor_\Upsilon = (\forall\alpha.\lfloor\tau_1\rfloor_\Upsilon \to \lfloor\tau_2\rfloor_\Upsilon)^i$$

- (Application) In this case, $e = (e_1[\tau']\ e_2)^i$. The typing rule, $\mathrm{C}; \varrho \vdash e$, and $\Upsilon \vdash e$ require that:

$$\begin{array}{ll}
\Delta; \Gamma \vdash e_1 : (\forall\alpha.\tau_1 \to \tau_3)^j & (1)\\
\tau = \tau_3[\tau'/\alpha] & \\
\Delta; \Gamma \vdash e_2 : \tau_2 & (2)\\
\Delta \vdash \tau'\ wf & (3)\\
\vdash \tau_1[\tau'/\alpha] = \tau_2 & (4)\\
\mathrm{C}; \varrho \vdash e_1 & (5)\\
\mathrm{C}; \varrho \vdash \tau' & (6)\\
\mathrm{C}; \varrho \vdash e_2 & (7)\\
funC(\mathrm{lbl}(e_1), \mathrm{lbl}(\tau), \mathrm{lbl}(e_2), i) & (8)\\
\Upsilon \vdash e_1 & (10)\\
\Upsilon \vdash e_2 & (11)\\
\Upsilon \vdash \tau' : \mathtt{r} & (12)
\end{array}$$

for some $\tau_1$, $\tau_3$, $j$, and $\tau_2$. By (1), (5), (10), (2), (7), (11), and the induction hypothesis:

$$\begin{array}{ll}
\Delta; \lfloor\Gamma\rfloor_\Upsilon \vdash \lfloor e_1\rfloor_\Upsilon : \lfloor(\forall\alpha.\tau_1 \to \tau)^j\rfloor_\Upsilon & (13)\\
\Delta; \lfloor\Gamma\rfloor_\Upsilon \vdash \lfloor e_2\rfloor_\Upsilon : \lfloor\tau_2\rfloor_\Upsilon & (14)
\end{array}$$

By definition $\lfloor(\forall\alpha.\tau_1 \to \tau_3)^j\rfloor_\Upsilon = (\forall\alpha.\lfloor\tau_1\rfloor_\Upsilon \to \lfloor\tau_3\rfloor_\Upsilon)^j$. By (3) and Lemma 19, $\Delta \vdash \lfloor\tau'\rfloor_\Upsilon\ wf$ (15). By (12) and Lemma 21, $tr(\lfloor\tau'\rfloor_\Upsilon) = \mathtt{r}$ (16). By (1), (2), and Lemma 16:

$$\begin{array}{ll}
\mathrm{C}(j) \subseteq \mathrm{C}(\mathrm{lbl}(e_1)) & (17)\\
\mathrm{C}; \varrho \vdash (\forall\alpha.\tau_1 \to \tau_3)^j & (18)\\
\mathrm{C}(\mathrm{lbl}(\tau_2)) \subseteq \mathrm{C}(\mathrm{lbl}(e_2)) & (19)\\
\mathrm{C} \vdash \tau_2 & (20)
\end{array}$$

By (18) and the rules for acceptability:

$$\begin{array}{ll}
\mathrm{C}; \varrho \vdash \tau_1 & (21)\\
\mathrm{C}; \varrho \vdash \tau_3 & (22)\\
(\forall\varrho(\alpha).\mathrm{lbl}(\tau_1) \to \mathrm{lbl}(\tau_3))^k_{\mathtt{t}} \in \mathrm{C}(j) & (23)
\end{array}$$

By (23), (17), and (8), $\mathrm{C}(\mathrm{lbl}(\tau')) = \mathrm{C}(\varrho(\alpha))$ (24) and $\mathrm{C}(\mathrm{lbl}(e_2)) \subseteq \mathrm{C}(\mathrm{lbl}(\tau_1))$. Hence by (19), $\mathrm{C}(\mathrm{lbl}(\tau_2)) \subseteq \mathrm{C}(\mathrm{lbl}(\tau_1))$. Since $\mathrm{lbl}(\tau_1[\tau'/\alpha]) = \mathrm{lbl}(\tau_1)$, $\mathrm{C}(\mathrm{lbl}(\tau_2)) \subseteq \mathrm{C}(\mathrm{lbl}(\tau_1[\tau'/\alpha]))$ (25). By (21), (6), (24), and Lemma 13, $\mathrm{C}; \varrho \vdash \tau_1[\tau'/\alpha]$ (26). By (4), (26), (20), (25), and Lemma 22, $\vdash \lfloor\tau_1[\tau'/\alpha]\rfloor_\Upsilon = \lfloor\tau_2\rfloor_\Upsilon$. By (21), (6), (24), and Lemma 20, $\vdash \lfloor\tau_1\rfloor_\Upsilon[\lfloor\tau'\rfloor_\Upsilon/\alpha] = \lfloor\tau_2\rfloor_\Upsilon$ (27). Thus by (13), (14), (15), (16), (27), and the typing rules, $\Delta; \lfloor\Gamma\rfloor_\Upsilon \vdash (\lfloor e_1\rfloor_\Upsilon[\lfloor\tau'\rfloor_\Upsilon]\ \lfloor e_2\rfloor_\Upsilon)^i : \lfloor\tau_3\rfloor_\Upsilon[\lfloor\tau'\rfloor_\Upsilon/\alpha]$. By definition, (22), (6), (24), and Lemma 20, $\Delta; \lfloor\Gamma\rfloor_\Upsilon \vdash \lfloor(e_1[\tau']\ e_2)^i\rfloor_\Upsilon : \lfloor\tau\rfloor_\Upsilon$, as required.

- (Box expression) In this case, $e = (\texttt{box}_{\tau''}\ e')^i$ for some $e'$ and $i$. The typing rule requires that $\tau = \texttt{box}(\tau'')^i$, $\Delta \vdash \tau''\ wf$, $\Delta; \Gamma \vdash e' : \tau'$, and $\vdash \tau'' = \tau'$ for some $\tau'$. The assumption $\mathrm{C}; \varrho \vdash e$ requires that $\mathrm{C} \vdash \tau$ and $\mathrm{C}; \varrho \vdash e'$. The assumption $\Upsilon \vdash e$ requires that $\Upsilon \vdash e'$. By the induction hypothesis, $\Delta; \lfloor\Gamma\rfloor_\Upsilon \vdash \lfloor e'\rfloor_\Upsilon : \lfloor\tau'\rfloor_\Upsilon$. There are two subcases:
  - If $i \in \mathrm{T}$ then $\lfloor e\rfloor_\Upsilon = \lfloor e'\rfloor_\Upsilon$ and $\lfloor\tau\rfloor_\Upsilon = \lfloor\tau'\rfloor_\Upsilon$ and the result is immediate.
  - If $i \notin \mathrm{T}$ then $\lfloor e\rfloor_\Upsilon = (\texttt{box}_{\lfloor\tau''\rfloor_\Upsilon}\lfloor e'\rfloor_\Upsilon)^i$ and $\lfloor\tau\rfloor_\Upsilon = \texttt{box}(\lfloor\tau'\rfloor_\Upsilon)^i$. The result follows by the typing rules if $\Delta \vdash \lfloor\tau'\rfloor_\Upsilon\ wf$, which holds by Lemma 19, and $\vdash \lfloor\tau''\rfloor_\Upsilon = \lfloor\tau'\rfloor_\Upsilon$, which holds by Lemma 22 if its other three premises hold. Since $\mathrm{C} \vdash \tau$, by the rules for acceptability, $boxC(i, \mathrm{lbl}(\tau''))$ (1) and $\mathrm{C} \vdash \tau''$, showing the first premise. Since $\Delta; \Gamma \vdash e' : \tau'$, by Lemma 16, $\mathrm{C}(\mathrm{lbl}(\tau')) \subseteq \mathrm{C}(\mathrm{lbl}(e'))$ (2) and $\mathrm{C} \vdash \tau'$, showing the second premise. By $\mathrm{C}; \varrho \vdash e$, $(\texttt{box}_{tr(\tau')}\,\mathrm{lbl}(e'))^j_{\mathtt{v}} \in \mathrm{C}(i)$. Thus by (1), $\mathrm{C}(\mathrm{lbl}(e')) \subseteq \mathrm{C}(\mathrm{lbl}(\tau''))$, so by (2), $\mathrm{C}(\mathrm{lbl}(\tau')) \subseteq \mathrm{C}(\mathrm{lbl}(\tau''))$, showing the third premise, as required.

- (Unbox) In this case, $e = (\texttt{unbox}\, e')^i$ for some $e'$ and $i$. The typing rule requires that $\Delta;\Gamma \vdash e' : \texttt{box}(\tau)^j$ for some $j$. The assumption $\mathrm{C};\varrho \vdash e$ requires $\mathrm{C};\varrho \vdash e'$. The assumption $\Upsilon \vdash e$ requires $\Upsilon \vdash e'$. By the induction hypothesis, $\Delta;\lfloor\Gamma\rfloor_\Upsilon \vdash \lfloor e'\rfloor_\Upsilon : \lfloor\texttt{box}(\tau)^j\rfloor_\Upsilon$. By Lemma 16, $\mathrm{C}(j) \subseteq \mathrm{C}(\mathrm{lbl}(e'))$. By Lemmas 17 and 18, $j \overset{\Upsilon}{\simeq} \mathrm{lbl}(e')$. There are two subcases:
  - If $j \in \Upsilon$ then $\lfloor e\rfloor_\Upsilon = \lfloor e'\rfloor_\Upsilon$ and $\lfloor\texttt{box}(\tau)^j\rfloor_\Upsilon = \lfloor\tau\rfloor_\Upsilon$ and the result is immediate.
  - If $j \notin \Upsilon$ then $\lfloor e\rfloor_\Upsilon = (\texttt{unbox}\,\lfloor e'\rfloor_\Upsilon)^i$ and $\lfloor\texttt{box}(\tau)^j\rfloor_\Upsilon = \texttt{box}(\lfloor\tau\rfloor_\Upsilon)^j$. The result then follows by the typing rules.
- (Frame) In this case, $e = \rho'(e')^i$ for some $\rho'$, $e'$, and $i$. The typing rule requires that $\vdash \rho' : \Gamma'$ and $\Delta;\Gamma' \vdash e' : \tau$ for some $\Gamma'$. The assumption $\mathrm{C};\varrho \vdash e$ requires that $\mathrm{C};\varrho \vdash \rho'$ and $\mathrm{C};\varrho \vdash e'$. The former requires that $\mathrm{C};\varrho \vdash \Gamma'$. The assumption $\Upsilon \vdash e$ requires $\Upsilon \vdash e'$. By the induction hypothesis, $\vdash \lfloor\rho'\rfloor_\Upsilon : \lfloor\Gamma'\rfloor_\Upsilon$ and $\Delta;\lfloor\Gamma'\rfloor_\Upsilon \vdash \lfloor e'\rfloor_\Upsilon : \lfloor\tau\rfloor_\Upsilon$. So by the typing rules, $\Delta;\lfloor\Gamma\rfloor_\Upsilon \vdash \lfloor\rho'\rfloor_\Upsilon(\lfloor e'\rfloor_\Upsilon)^i : \lfloor\tau\rfloor_\Upsilon$. The result follows since $\lfloor e\rfloor_\Upsilon = \lfloor\rho'\rfloor_\Upsilon(\lfloor e'\rfloor_\Upsilon)^i$.
- (Constant) In this case $e = c^i$ for some $c$ and $i$. The typing rule requires that $\tau = \texttt{B}^i$. Clearly, $\lfloor e\rfloor_\Upsilon = c^i$, $\lfloor\tau\rfloor_\Upsilon = \texttt{B}^i$, and the result follows by the typing rules.
- (Fix value) In this case $e = \langle\rho, \texttt{fix}\ f[\alpha](x{:}\tau_1){:}\tau_2.e'\rangle^i$. The typing rule require that $\tau = (\forall\alpha.\tau_1 \to \tau_2)^i$, $\vdash \rho : \Gamma'$, and $\Delta,\alpha;\Gamma',f{:}\tau,x{:}\tau_1 \vdash e' : \tau_2$. The assumption $\mathrm{C};\varrho \vdash e$ requires $\mathrm{C};\varrho \vdash \rho$, from which $\mathrm{C};\varrho \vdash \Gamma'$, $\varrho(f) = i$, $\varrho(x) = \mathrm{lbl}(\tau_1)$, $\mathrm{C} \vdash \tau$, and $\mathrm{C};\varrho \vdash e'$. From $\mathrm{C} \vdash \tau$ and the rules for acceptability, $\mathrm{C} \vdash \tau_1$. From these facts, $\mathrm{C};\varrho \vdash \Gamma',f{:}\tau,x{:}\tau_1$. The assumption $\Upsilon \vdash e$ requires $\Upsilon \vdash e'$. By the induction hypothesis, $\vdash \lfloor\rho\rfloor_\Upsilon : \lfloor\Gamma'\rfloor_\Upsilon$ and $\Delta,\alpha;\lfloor\Gamma',f{:}\tau,x{:}\tau_1\rfloor_\Upsilon \vdash \lfloor e'\rfloor_\Upsilon : \lfloor\tau_2\rfloor_\Upsilon$. Since:
  $$\lfloor\Gamma',f{:}\tau,x{:}\tau_1\rfloor_\Upsilon = \lfloor\Gamma'\rfloor_\Upsilon, f{:}(\forall\alpha.\lfloor\tau_1\rfloor_\Upsilon \to \lfloor\tau_2\rfloor_\Upsilon)^i, x{:}\lfloor\tau_2\rfloor_\Upsilon$$
  by the typing rules:
  $$\Delta;\lfloor\Gamma\rfloor_\Upsilon \vdash \langle\lfloor\rho\rfloor_\Upsilon, \texttt{fix}\ f[\alpha](x{:}\lfloor\tau_1\rfloor_\Upsilon){:}\lfloor\tau_2\rfloor_\Upsilon.\lfloor e'\rfloor_\Upsilon\rangle^i : (\forall\alpha.\lfloor\tau_1\rfloor_\Upsilon \to \lfloor\tau_2\rfloor_\Upsilon)^i$$
  The result follows since:
  $$\lfloor e\rfloor_\Upsilon = \langle\lfloor\rho\rfloor_\Upsilon, \texttt{fix}\ f[\alpha](x{:}\lfloor\tau_1\rfloor_\Upsilon){:}\lfloor\tau_2\rfloor_\Upsilon.\lfloor e'\rfloor_\Upsilon\rangle^i$$
  $$\lfloor\tau\rfloor_\Upsilon = (\forall\alpha.\lfloor\tau_1\rfloor_\Upsilon \to \lfloor\tau_2\rfloor_\Upsilon)^i$$
- (Box value) In this case, $e = \langle v^j{:}\tau''\rangle^i$ for some $v$, $i$, and $j$. The typing rule requires that $\tau = \texttt{box}(\tau'')^i$, $\Delta \vdash \tau''\ \mathit{wf}$, $\Delta;\Gamma \vdash v^j : \tau'$, and $\vdash \tau'' = \tau'$ for some $\tau'$. The assumption $\mathrm{C};\varrho \vdash e$ requires that $\mathrm{C} \vdash \tau$ and $\mathrm{C};\varrho \vdash v^j$. The assumption $\Upsilon \vdash e$ requires $\Upsilon \vdash v^j$. By the induction hypothesis, $\Delta;\lfloor\Gamma\rfloor_\Upsilon \vdash \lfloor v^j\rfloor_\Upsilon : \lfloor\tau'\rfloor_\Upsilon$. There are two subcases:
  - If $i \in \mathrm{T}$ then $\lfloor e\rfloor_\Upsilon = \lfloor v^j\rfloor_\Upsilon$ and $\lfloor\tau\rfloor_\Upsilon = \lfloor\tau'\rfloor_\Upsilon$ and the result is immediate.
  - If $i \notin \mathrm{T}$ then $\lfloor e\rfloor_\Upsilon = \langle\lfloor v^j\rfloor_\Upsilon{:}\lfloor\tau''\rfloor_\Upsilon\rangle^i$ and $\lfloor\tau\rfloor_\Upsilon = \texttt{box}(\lfloor\tau'\rfloor_\Upsilon)^i$. The result follows by the typing rules if $\Delta \vdash \lfloor\tau''\rfloor_\Upsilon\ \mathit{wf}$, which holds by Lemma 19, and $\vdash \lfloor\tau''\rfloor_\Upsilon = \lfloor\tau'\rfloor_\Upsilon$, which holds by Lemma 22 if its other three premises hold. Since $\mathrm{C} \vdash \tau$, by the rules for acceptability, $\mathit{boxC}(i, \mathrm{lbl}(\tau''))$ (1) and $\mathrm{C} \vdash \tau''$, showing the first premise. Since $\Delta;\Gamma \vdash v^j : \tau'$, by Lemma 16, $\mathrm{C}(\mathrm{lbl}(\tau')) \subseteq \mathrm{C}(j)$ (2) and $\mathrm{C} \vdash \tau'$, showing the second premise. By $\mathrm{C};\varrho \vdash e$, $(\texttt{box}_{tr(\tau')}\, j)^k_{\mathsf{v}} \in \mathrm{C}(i)$. Thus by (1), $\mathrm{C}(j) \subseteq \mathrm{C}(\mathrm{lbl}(\tau''))$, so by (2), $\mathrm{C}(\mathrm{lbl}(\tau')) \subseteq \mathrm{C}(\mathrm{lbl}(\tau''))$, showing the third premise, as required.
- (Environment) In this case $\rho = x_1{:}\tau_1 = {v_1}^{i_1}, \ldots, x_n{:}\tau_n = {v_n}^{i_n}$ and $\Gamma = x_1{:}\tau_1, \ldots, x_n{:}\tau_n$. The typing rule requires that $\emptyset;\emptyset \vdash {v_j}^{i_j} : \tau'_j$ and $\vdash \tau_j = \tau'_j$ for $1 \le j \le n$ and some $\tau'_j$s. The assumption $\mathrm{C};\varrho \vdash \rho$ requires $\mathrm{C};\varrho \vdash \tau_j$ and $\mathrm{C};\varrho \vdash {v_j}^{i_j}$ for $1 \le j \le n$. Clearly $\mathrm{C};\varrho \vdash \Gamma'$ where $\Gamma'$ is empty. The assumption $\Upsilon \vdash \rho$ requires $\Upsilon \vdash {v_j}^{i_j}$ for $1 \le j \le n$. By the induction hypotheis, $\emptyset;\emptyset \vdash \lfloor {v_j}^{i_j}\rfloor_\Upsilon : \lfloor\tau'_j\rfloor_\Upsilon$ for $1 \le j \le n$. By the rules for acceptability, $\mathrm{C}(i_j) \subseteq \mathrm{C}(\mathrm{lbl}(\tau_j))$ for $1 \le j \le n$. By Lemma 16, $\mathrm{C}(\tau'_j) \subseteq \mathrm{C}(i_j)$ and $\mathrm{C} \vdash \tau'_j$ for $1 \le j \le n$. Thus $\mathrm{C}(\tau'_j) \subseteq \mathrm{C}(\tau_j)$ for $1 \le j \le n$. By Lemma 22, $\vdash \lfloor\tau_j\rfloor_\Upsilon = \lfloor\tau'_j\rfloor_\Upsilon$ for $1 \le j \le n$. Then by the typing rules $\vdash x_1{:}\lfloor\tau_1\rfloor_\Upsilon = \lfloor {v_1}^{i_1}\rfloor_\Upsilon, \ldots, x_n{:}\lfloor\tau_n\rfloor_\Upsilon = \lfloor {v_n}^{i_n}\rfloor_\Upsilon : x_1{:}\lfloor\tau_1\rfloor_\Upsilon, \ldots, x_n{:}\lfloor\tau_n\rfloor_\Upsilon$. The result follows since $\lfloor\rho\rfloor_\Upsilon = x_1{:}\lfloor\tau_1\rfloor_\Upsilon = \lfloor {v_1}^{i_1}\rfloor_\Upsilon, \ldots, x_n{:}\lfloor\tau_n\rfloor_\Upsilon = \lfloor {v_n}^{i_n}\rfloor_\Upsilon$ and $\lfloor\Gamma\rfloor_\Upsilon = x_1{:}\lfloor\tau_1\rfloor_\Upsilon, \ldots, x_n{:}\lfloor\tau_n\rfloor_\Upsilon$.
- (Machine state) In this case $M = (\rho, e)$. By the typing rule, $\vdash \rho : \Gamma$ and $\emptyset;\Gamma \vdash e : \tau$ for some $\Gamma$. The assumption $\mathrm{C};\varrho \vdash M$ requires both $\mathrm{C};\varrho \vdash \rho$ and $\mathrm{C};\varrho \vdash e$. The former requires $\mathrm{C};\varrho \vdash \Gamma$. The assumption $\Upsilon \vdash M$ requires $\Upsilon \vdash \rho$ and $\Upsilon \vdash e$. By the induction hypothesis, $\vdash \lfloor\rho\rfloor_\Upsilon : \lfloor\Gamma\rfloor_\Upsilon$ and $\emptyset;\lfloor\Gamma\rfloor_\Upsilon \vdash \lfloor e\rfloor_\Upsilon : \lfloor\tau\rfloor_\Upsilon$. So by the typing rules, $\vdash (\lfloor\rho\rfloor_\Upsilon, \lfloor e\rfloor_\Upsilon) : \lfloor\tau\rfloor_\Upsilon$. The result follows since $\lfloor M\rfloor_\Upsilon = (\lfloor\rho\rfloor_\Upsilon, \lfloor e\rfloor_\Upsilon)$.

■

A consequence of type preservation is that unboxed well typed programs are traceable.

**Theorem 2**
*If* $\vdash M : \tau$, $\mathrm{C} \vdash \Upsilon$, *and* $\mathrm{C};\varrho \vdash M$ *then* $\vdash \lfloor M\rfloor_\Upsilon\ \mathbf{tr}$.

**Proof:** The proof follows from Theorem 1 and Lemma 11. ■

### 4.4 Coherence

The other part of proving correctness is to show that unboxing preserves semantics in some appropriate sense. That requires two key lemmas—that a step of the program can be matched by zero or more steps of the unboxed program and that consistency is preserved under reduction.

To show the first lemma, we need three technical lemmas—that a value's cache is nonempty, that reduction preserves the unboxing or not of the outermost label, and a multistep compositionality property.

**Lemma 23 (Inhabitance)**
*If* $\mathrm{C};\varrho \vdash v^k$ *then* $\exists s \in \mathrm{C}(k)$ *such that* $\mathrm{lbl}(s) = k$.

**Proof:** By inspection of the acceptable analysis and acceptable instantiation rules. ■

**Lemma 24 (Unboxing set preservation)**
*If* $\mathrm{C};\varrho \vdash \rho$, $\mathrm{C};\varrho \vdash e$, $\mathrm{C} \vdash \Upsilon$, *and* $(\rho, e_1) \longmapsto (\rho, e_2)$ *then* $\mathrm{lbl}(e) \overset{\Upsilon}{\simeq} \mathrm{lbl}(e')$.

**Proof:** All of the cases for which $\mathrm{lbl}(e_1) = \mathrm{lbl}(e_2)$ follow immediately. For the remaining cases:

- If $(\rho, x^k) \longmapsto (\rho, v^j)$ where $x{:}\tau = v^j \in \rho$ then by the assumptions we have that $\varrho(x) = \mathrm{lbl}(\tau)$ (1), $\mathrm{C}(j) \subseteq \mathrm{C}(\mathrm{lbl}(\tau))$ (2) and $\mathrm{C}(\varrho(x)) \subseteq \mathrm{C}(k)$ (3), so by transitivity we have $\mathrm{C}(j) \subseteq \mathrm{C}(k)$ (4). By Inhabitance (Lemma 23) we have an $s \in \mathrm{C}(j)$ (5) such that $\mathrm{lbl}(s) = j$ (6), and so by Agreement (Lemma 18) we have $k \overset{\Upsilon}{\simeq} j$. Since $\mathrm{lbl}(e) = k$ and $\mathrm{lbl}(e') = j$, the result follows.

- If $(\rho, (\texttt{unbox}\,\langle v^i{:}t\rangle^j)^k) \longmapsto (\rho, v^i)$ then we must show that $k \overset{\Upsilon}{\simeq} i$. By Inhabitance we have $s \in \mathrm{C}(i)$ with $\mathrm{lbl}(s) = i$, so by Agreement, it suffices to show that $\mathrm{C}(i) \subseteq \mathrm{C}(k)$. By the box rule for an acceptable analysis, there is a $s = (\texttt{box}_t\; l)^j_{\mathsf{v}} \in \mathrm{C}(j)$ such that $\mathrm{C}(i) \subseteq \mathrm{C}(l)$. Since $s \in \mathrm{C}(j)$, by the rule for unbox, $\mathrm{C}(l) \subseteq \mathrm{C}(k)$, so $\mathrm{C}(i) \subseteq \mathrm{C}(k)$ and we're done.
- If $(\rho, \rho'(v^i)^j) \longmapsto (\rho, v^i)$ then we must show that $j \overset{\Upsilon}{\simeq} i$. By Inhabitance, there is an $s \in \mathrm{C}(i)$, and by the acceptable analysis rule for frames we have that $\mathrm{C}(i) \subseteq \mathrm{C}(j)$, so by Agreement we have that $j \overset{\Upsilon}{\simeq} i$.

■

**Lemma 25 (Many step compositionality)**
*If* $(\rho, e_1) \longmapsto^* (\rho, e_2)$ *then:*

- $(\rho, (e_1\; e)^i) \longmapsto^* (\rho, (e_2\; e)^i)$
- $(\rho, (v^j\; e_1)^i) \longmapsto^* (\rho, (v^j\; e_2)^i)$
- $(\rho, (\texttt{box}_\tau\; e_1)^i) \longmapsto^* (\rho, (\texttt{box}_\tau\; e_2)^i)$
- $(\rho, (\texttt{unbox}\, e_1)^i) \longmapsto^* (\rho, (\texttt{unbox}\, e_2)^i)$
- $(\rho, \rho'(e_1)^i) \longmapsto^* (\rho, \rho'(e_2)^i)$

**Proof:** The proof is by an easy induction on the length of the reduction sequences. ■

**Theorem 3 (Single step reduction coherence)**
*If* $\vdash M : \tau$, $\mathrm{C}; \varrho \vdash M$, $\mathrm{C} \vdash \Upsilon$, $\Upsilon \vdash M$, *and* $M \longmapsto M'$ *then* $\downharpoonleft M \downharpoonright_\Upsilon \longmapsto^* \downharpoonleft M' \downharpoonright_\Upsilon$.

**Proof:** The proof is by induction on the derivation of $M \longmapsto M'$, consider the cases for the last rule used to derive it:

- If $(\rho, x^k) \longmapsto (\rho, v^j)$ where $x{:}\tau = v^j \in \rho$ then by definition $\downharpoonleft M \downharpoonright_\Upsilon = (\downharpoonleft \rho \downharpoonright_\Upsilon, x^k)$, $\downharpoonleft M' \downharpoonright_\Upsilon = (\downharpoonleft \rho \downharpoonright_\Upsilon, \downharpoonleft v^j \downharpoonright_\Upsilon)$, and $x{:}\downharpoonleft \tau \downharpoonright_\Upsilon = \downharpoonleft v^j \downharpoonright_\Upsilon \in \downharpoonleft \rho \downharpoonright_\Upsilon$. Thus $\downharpoonleft M \downharpoonright_\Upsilon \longmapsto \downharpoonleft M' \downharpoonright_\Upsilon$ by the same rule.
- If:
$$(\rho, (\texttt{fix}\; f[\alpha](x{:}\tau_1){:}\tau_2.e_1)^j) \longmapsto$$
$$(\rho, \langle \rho, \texttt{fix}\; f[\alpha](x{:}\tau_1){:}\tau_2.e_1\rangle^j)$$
then the unboxings of the $e$ and $e'$ are of the same form, and the same reduction step applies.
- If $(\rho, (\texttt{box}_{\tau'}\; {v_t}^i)^j) \longmapsto (\rho, \langle {v_t}^i{:}\tau'\rangle^j)$ where $tr(\tau') = t$ then:
  - If $j \notin \Upsilon$ then:
  
    By the definition of unboxing, $\downharpoonleft e \downharpoonright_\Upsilon = (\texttt{box}_{\downharpoonleft \tau' \downharpoonright_\Upsilon}\; {v'_{t'}}^k)^j$ where ${v'_{t'}}^k = \downharpoonleft {v_t}^i \downharpoonright_\Upsilon$.
    
    By hypothesis, $\vdash \rho : \Gamma$, $;\Gamma \vdash {v_t}^i : \tau''$ and $\vdash \tau' = \tau''$ for some $\tau''$. By hypothesis, $\Upsilon \vdash {v_t}^i$. By Theorem 1, $\emptyset; \downharpoonleft \Gamma \downharpoonright_\Upsilon \vdash {v'_{t'}}^k : \downharpoonleft \tau'' \downharpoonright_\Upsilon$. The proof of that theorem also showed that $\vdash \downharpoonleft \tau' \downharpoonright_\Upsilon = \downharpoonleft \tau'' \downharpoonright_\Upsilon$, so by Lemma 4, $tr(\downharpoonleft \tau' \downharpoonright_\Upsilon) = tr(\downharpoonleft \tau'' \downharpoonright_\Upsilon)$. By Lemma 6, $tr(\downharpoonleft \tau'' \downharpoonright_\Upsilon) = t'$, so $tr(\downharpoonleft \tau' \downharpoonright_\Upsilon) = t'$. By definition of reduction $(\downharpoonleft \rho \downharpoonright_\Upsilon, (\texttt{box}_{\downharpoonleft \tau' \downharpoonright_\Upsilon}\; {v'_{t'}}^k)^j) \longmapsto (\downharpoonleft \rho \downharpoonright_\Upsilon, \langle {v'_{t'}}^k{:}\downharpoonleft \tau' \downharpoonright_\Upsilon\rangle^j)$.
  - If $j \in \Upsilon$ then:
  
    By definition of unboxing
    $\downharpoonleft e \downharpoonright_\Upsilon = {v'_{t'}}^k$ where ${v'_{t'}}^k = \downharpoonleft {v_t}^i \downharpoonright_\Upsilon$
    
    By definition of reduction
    $(\downharpoonleft \rho \downharpoonright_\Upsilon, {v'_{t'}}^k) \longmapsto^* (\downharpoonleft \rho \downharpoonright_\Upsilon, {v'_{t'}}^k)$
- If $(\rho, (e_1[\tau']\; e_2)^j) \longmapsto (\rho, (e_1'[\tau']\; e_2)^j)$ then:

  By definition of $\mathrm{C}; \varrho \vdash M$ we have that $\mathrm{C}; \varrho \vdash \rho$ and $\mathrm{C}; \varrho \vdash e_1$. Hence we have that $\mathrm{C}; \varrho \vdash (\rho, e_1)$. By the typing rules we also have that $\vdash \rho : \Gamma$ and $\emptyset; \Gamma \vdash e_1 : \tau_1$ for some $\Gamma$ and $\tau_1$, so $\vdash (\rho, e_1) : \tau_1$. By the rules for consistency, $\Upsilon \vdash \rho$ and $\Upsilon \vdash e_1$, so $\Upsilon \vdash (\rho, e_1)$. Hence by induction we have that $(\downharpoonleft \rho \downharpoonright_\Upsilon, \downharpoonleft e_1 \downharpoonright_\Upsilon) \longmapsto^* (\downharpoonleft \rho \downharpoonright_\Upsilon, \downharpoonleft e_1' \downharpoonright_\Upsilon)$.
  
  By Lemma 25
  $$(\downharpoonleft \rho \downharpoonright_\Upsilon, (\downharpoonleft e_1 \downharpoonright_\Upsilon[\downharpoonleft \tau' \downharpoonright_\Upsilon]\; \downharpoonleft e_2 \downharpoonright_\Upsilon)^j) \longmapsto^*$$
  $$(\downharpoonleft \rho \downharpoonright_\Upsilon, (\downharpoonleft e_1' \downharpoonright_\Upsilon[\downharpoonleft \tau' \downharpoonright_\Upsilon]\; \downharpoonleft e_2 \downharpoonright_\Upsilon)^j)$$
  By definition of unboxing
  $$(\downharpoonleft \rho \downharpoonright_\Upsilon, \downharpoonleft (e_1[\tau']\; e_2)^j \downharpoonright_\Upsilon) \longmapsto^* (\downharpoonleft \rho \downharpoonright_\Upsilon, \downharpoonleft (e_1'[\tau']\; e_2)^j \downharpoonright_\Upsilon)$$
- If $(\rho, (e_1[\tau']\; e_2)^j) \longmapsto (\rho, (e_1[\tau']\; e_2')^j)$ then the argument follows by the symmetric argument to the previous case.
- If $(\rho, ({v_f}^j[\tau']\; {v_t}^k)^l) \longmapsto (\rho, \rho''(e''[\tau'/\alpha])^l)$ where:
$$\begin{aligned} &v_f = \langle \rho', \texttt{fix}\; f[\alpha](x{:}\tau_1){:}\tau_2.e''\rangle \\ &\rho'' = \rho', f{:}\tau = {v_f}^j, x{:}\tau_1' = {v_t}^k \\ &\tau = (\forall \alpha.\tau_1 \to \tau_2)^j \\ &\tau_1' = \tau_1[\tau'/\alpha] \end{aligned}$$
  then:
  
  By definition of unboxing we have that:
$$\begin{aligned} \downharpoonleft e \downharpoonright_\Upsilon &= ({v_f'}^j[\downharpoonleft \tau' \downharpoonright_\Upsilon]\; {v_{t'}}^{k'})^l \\ v_f' &= \langle \downharpoonleft \rho' \downharpoonright_\Upsilon, \texttt{fix}\; f[\alpha](x{:}\downharpoonleft \tau_1 \downharpoonright_\Upsilon){:}\downharpoonleft \tau_2 \downharpoonright_\Upsilon.\downharpoonleft e'' \downharpoonright_\Upsilon\rangle \\ {v_{t'}}^{k'} &= \downharpoonleft {v_t}^k \downharpoonright_\Upsilon \\ \downharpoonleft e' \downharpoonright_\Upsilon &= (\downharpoonleft \rho' \downharpoonright_\Upsilon, f{:}\downharpoonleft \tau \downharpoonright_\Upsilon = {v_f'}^j, x{:}\downharpoonleft \tau_1' \downharpoonright_\Upsilon = {v_{t'}}^{k'}) \\ &\qquad (\downharpoonleft e''[\tau'/\alpha] \downharpoonright_\Upsilon)^l \end{aligned}$$
  By hypothesis, $\vdash \rho : \Gamma$, $\emptyset; \Gamma \vdash {v_t}^k : \tau_1''$, and $\vdash \tau_1' = \tau_1''$. By Theorem 1, $\emptyset; \downharpoonleft \Gamma \downharpoonright_\Upsilon \vdash {v_{t'}}^{k'} : \downharpoonleft \tau_1'' \downharpoonright_\Upsilon$. The proof of that theorem also showed that $\vdash \downharpoonleft \tau_1' \downharpoonright_\Upsilon = \downharpoonleft \tau_1'' \downharpoonright_\Upsilon$. By Lemma 4, $tr(\downharpoonleft \tau_1' \downharpoonright_\Upsilon) = tr(\downharpoonleft \tau_1'' \downharpoonright_\Upsilon)$. By Lemma 6, $tr(\downharpoonleft \tau_1'' \downharpoonright_\Upsilon) = t'$. Thus $tr(\downharpoonleft \tau_1' \downharpoonright_\Upsilon) = t'$. So by the application beta rule:
$$(\downharpoonleft \rho \downharpoonright_\Upsilon, \downharpoonleft e \downharpoonright_\Upsilon) \longmapsto (\downharpoonleft \rho \downharpoonright_\Upsilon, \rho'''(e''')^l)$$
  where:
$$\begin{aligned} \rho''' &= \downharpoonleft \rho \downharpoonright_\Upsilon, f{:}\tau'' = {v_f'}^j, x{:}\tau_1'' = {v_{t'}}^{k'} \\ \tau'' &= (\forall \alpha.\downharpoonleft \tau_1 \downharpoonright_\Upsilon \to \downharpoonleft \tau_2 \downharpoonright_\Upsilon)^j \\ \tau_1'' &= \downharpoonleft \tau_1 \downharpoonright_\Upsilon[\downharpoonleft \tau' \downharpoonright_\Upsilon/\alpha] \\ e''' &= \downharpoonleft e'' \downharpoonright_\Upsilon[\downharpoonleft \tau' \downharpoonright_\Upsilon/\alpha] \end{aligned}$$
  By definition of unboxing, $\tau'' = \downharpoonleft \tau \downharpoonright_\Upsilon$. The proof of the theorem above also showed that $\mathrm{C}; \varrho \vdash \tau_1$, $\mathrm{C}; \varrho \vdash \tau'$, and $\mathrm{C}(\mathrm{lbl}(\tau')) = \mathrm{C}(\varrho(\alpha))$. By the rules for acceptability, clearly $\mathrm{C}; \varrho \vdash e''$. By Lemma 20, $\tau_1'' = \downharpoonleft \tau_1' \downharpoonright_\Upsilon$ and $e''' = \downharpoonleft e''[\tau'/\alpha] \downharpoonright_\Upsilon$. Putting that altogether, $\rho'''(e''')^l = \downharpoonleft e' \downharpoonright_\Upsilon$, as required.
- If $(\rho, (\texttt{box}_{\tau'}\; e)^j) \longmapsto (\rho, (\texttt{box}_{\tau'}\; e')^j)$ then: By definition of acceptability $\mathrm{C}; \varrho \vdash \rho$ and $\mathrm{C}; \varrho \vdash e$, so $\mathrm{C}; \varrho \vdash (\rho, e)$. By the typing rules, $\vdash \rho : \Gamma$ and $\emptyset; \Gamma \vdash e : \tau''$ for some $\Gamma$ and $\tau''$, so $\vdash (\rho, e) : \tau''$. By the rules for consistency, $\Upsilon \vdash \rho$ and $\Upsilon \vdash e$, so $\Upsilon \vdash (\rho, e)$. So by the induction hypothesis, $(\downharpoonleft \rho \downharpoonright_\Upsilon, \downharpoonleft e \downharpoonright_\Upsilon) \longmapsto^* (\downharpoonleft \rho \downharpoonright_\Upsilon, \downharpoonleft e' \downharpoonright_\Upsilon)$.
  - If $j \in \Upsilon$ then:
  
    By definition of unboxing
    $\downharpoonleft (\texttt{box}_{\tau'}\; e)^j \downharpoonright_\Upsilon = \downharpoonleft e \downharpoonright_\Upsilon$
    
    By definition of unboxing
    $\downharpoonleft (\texttt{box}_{\tau'}\; e')^j \downharpoonright_\Upsilon = \downharpoonleft e' \downharpoonright_\Upsilon$
    
    By induction
    $(\downharpoonleft \rho \downharpoonright_\Upsilon, \downharpoonleft e \downharpoonright_\Upsilon) \longmapsto^* (\downharpoonleft \rho \downharpoonright_\Upsilon, \downharpoonleft e' \downharpoonright_\Upsilon)$

- If $j \notin \Upsilon$ then:

  By definition of unboxing:

$$\begin{array}{rcl} \downharpoonleft(\mathtt{box}_{\tau'}\, e)^j \downharpoonright_\Upsilon & = & (\mathtt{box}_{\downharpoonleft\tau'\downharpoonright_\Upsilon} \downharpoonleft e \downharpoonright_\Upsilon)^j \\ \downharpoonleft(\mathtt{box}_{\tau'}\, e')^j \downharpoonright_\Upsilon & = & (\mathtt{box}_{\downharpoonleft\tau'\downharpoonright_\Upsilon} \downharpoonleft e' \downharpoonright_\Upsilon)^j \end{array}$$

  Hence by the induction hypothesis and Lemma 25 we have that:

$$(\downharpoonleft\rho\downharpoonright_\Upsilon, (\mathtt{box}_{\downharpoonleft\tau'\downharpoonright_\Upsilon} \downharpoonleft e\downharpoonright_\Upsilon)^j) \longmapsto^* \\ (\downharpoonleft\rho\downharpoonright_\Upsilon, (\mathtt{box}_{\downharpoonleft\tau'\downharpoonright_\Upsilon} \downharpoonleft e'\downharpoonright_\Upsilon)^j)$$

- If $(\rho, (\mathtt{unbox}\, e)^j) \longmapsto (\rho, (\mathtt{unbox}\, e')^j)$ then let $i = \mathrm{lbl}(e)$ and $i' = \mathrm{lbl}(e')$. By Lemma 24 we have that $i \overset{\Upsilon}{\simeq} i'$. By the rules for acceptability, $\mathrm{C}; \varrho \vdash \rho$ and $\mathrm{C}; \varrho \vdash e$, so $\mathrm{C}; \varrho \vdash (\rho, e)$. By the typing rules, $\vdash \rho : \Gamma$ and $\emptyset; \Gamma \vdash e : \tau'$ for some $\Gamma$ and $\tau'$, so $\vdash (\rho, e) : \tau'$. By the rules for consistency, $\Upsilon \vdash \rho$ and $\Upsilon \vdash e$, so $\Upsilon \vdash (\rho, e)$. So by the induction hypothesis, $(\downharpoonleft\rho\downharpoonright_\Upsilon, \downharpoonleft e\downharpoonright_\Upsilon) \longmapsto^* (\downharpoonleft\rho\downharpoonright_\Upsilon, \downharpoonleft e'\downharpoonright_\Upsilon)$.
  - If $i, i' \in \Upsilon$ then:

    By definition of unboxing
    $\downharpoonleft(\mathtt{unbox}\, e)^j\downharpoonright_\Upsilon = \downharpoonleft e\downharpoonright_\Upsilon$

    By definition of unboxing
    $\downharpoonleft(\mathtt{unbox}\, e')^j\downharpoonright_\Upsilon = \downharpoonleft e'\downharpoonright_\Upsilon$

    By induction
    $(\downharpoonleft\rho\downharpoonright_\Upsilon, \downharpoonleft e\downharpoonright_\Upsilon) \longmapsto^* (\downharpoonleft\rho\downharpoonright_\Upsilon, \downharpoonleft e'\downharpoonright_\Upsilon)$
  - If $i, i' \notin \Upsilon$ then:

    By definition of unboxing
    $\downharpoonleft(\mathtt{unbox}\, e)^j\downharpoonright_\Upsilon = (\mathtt{unbox} \downharpoonleft e\downharpoonright_\Upsilon)^j$

    By definition of unboxing
    $\downharpoonleft(\mathtt{unbox}\, e')^j\downharpoonright_\Upsilon = (\mathtt{unbox} \downharpoonleft e'\downharpoonright_\Upsilon)^j$

    By induction
    $(\downharpoonleft\rho\downharpoonright_\Upsilon, \downharpoonleft e\downharpoonright_\Upsilon) \longmapsto^* (\downharpoonleft\rho\downharpoonright_\Upsilon, \downharpoonleft e'\downharpoonright_\Upsilon)$

    By Lemma 25
    $(\downharpoonleft\rho\downharpoonright_\Upsilon, \downharpoonleft(\mathtt{unbox}\, e)^j\downharpoonright_\Upsilon) \longmapsto^*$
    $(\downharpoonleft\rho\downharpoonright_\Upsilon, \downharpoonleft(\mathtt{unbox}\, e')^j\downharpoonright_\Upsilon)$

- If $(\rho, (\mathtt{unbox}\, \langle v^i{:}\tau\rangle^j)^k) \longmapsto (\rho, v^i)$ then:
  - If $j \in \Upsilon$ then:

    By definition of unboxing
    $\downharpoonleft(\mathtt{unbox}\, \langle v^i{:}\tau\rangle^j)^k\downharpoonright_\Upsilon = \downharpoonleft\langle v^i{:}\tau\rangle^j\downharpoonright_\Upsilon = \downharpoonleft v^i\downharpoonright_\Upsilon$

    So in zero steps
    $(\downharpoonleft\rho\downharpoonright_\Upsilon, \downharpoonleft(\mathtt{unbox}\, \langle v^i{:}\tau\rangle^j)^k\downharpoonright_\Upsilon) \longmapsto^* (\downharpoonleft\rho\downharpoonright_\Upsilon, \downharpoonleft v^i\downharpoonright_\Upsilon)$
  - If $j \notin \Upsilon$ then:

    By definition of unboxing
    $\downharpoonleft(\mathtt{unbox}\, \langle v^i{:}\tau\rangle^j)^k\downharpoonright_\Upsilon = (\mathtt{unbox} \downharpoonleft\langle v^i{:}\tau\rangle^j\downharpoonright_\Upsilon)^k$

    By definition of unboxing
    $(\mathtt{unbox} \downharpoonleft\langle v^i{:}\tau\rangle^j\downharpoonright_\Upsilon)^k = (\mathtt{unbox}\, \langle \downharpoonleft v^i\downharpoonright_\Upsilon{:}\downharpoonleft\tau\downharpoonright_\Upsilon\rangle^j)^k$

    By definition of reduction
    $(\downharpoonleft\rho\downharpoonright_\Upsilon, (\mathtt{unbox}\, \langle \downharpoonleft v^i\downharpoonright_\Upsilon{:}\downharpoonleft\tau\downharpoonright_\Upsilon\rangle^j)^k) \longmapsto$
    $(\downharpoonleft\rho\downharpoonright_\Upsilon, \downharpoonleft v^i\downharpoonright_\Upsilon)$

- If $(\rho, \rho'(e_1)^i) \longmapsto (\rho, \rho'(e_2)^i)$ then:

  By the rules for acceptability, $\mathrm{C}; \varrho \vdash \rho'$ and $\mathrm{C}; \varrho \vdash e_1$, so $\mathrm{C}; \varrho \vdash (\rho', e_1)$. By the typing rules, $\vdash \rho' : \Gamma'$ and $\emptyset; \Gamma' \vdash e_1 : \tau$ for some $\Gamma'$, so $\vdash (\rho', e_1) : \tau$. By the rules for consistency, $\Upsilon \vdash \rho'$ and $\Upsilon \vdash e_1$, so $\Upsilon \vdash (\rho', e_1)$. So by the induction hypothesis, $(\downharpoonleft\rho'\downharpoonright_\Upsilon, \downharpoonleft e_1\downharpoonright_\Upsilon) \longmapsto^* (\downharpoonleft\rho'\downharpoonright_\Upsilon, \downharpoonleft e_2\downharpoonright_\Upsilon)$.

  By definition of unboxing
  $\downharpoonleft\rho'(e_1)^i\downharpoonright_\Upsilon = \downharpoonleft\rho'\downharpoonright_\Upsilon(\downharpoonleft e_1\downharpoonright_\Upsilon)^i$

  By definition of unboxing
  $\downharpoonleft\rho'(e_2)^i\downharpoonright_\Upsilon = \downharpoonleft\rho'\downharpoonright_\Upsilon(\downharpoonleft e_2\downharpoonright_\Upsilon)^i$

  By induction
  $(\downharpoonleft\rho'\downharpoonright_\Upsilon, \downharpoonleft e_1\downharpoonright_\Upsilon) \longmapsto^* (\downharpoonleft\rho'\downharpoonright_\Upsilon, \downharpoonleft e_2\downharpoonright_\Upsilon)$

  By Lemma 25
  $(\downharpoonleft\rho\downharpoonright_\Upsilon, \downharpoonleft\rho'\downharpoonright_\Upsilon(\downharpoonleft e_1\downharpoonright_\Upsilon)^i) \longmapsto^* (\downharpoonleft\rho\downharpoonright_\Upsilon, \downharpoonleft\rho'\downharpoonright_\Upsilon(\downharpoonleft e_2\downharpoonright_\Upsilon)^i)$

- If $(\rho, \rho'(v^i)^j) \longmapsto (\rho, v^i)$ then:

  By definition of unboxing
  $\downharpoonleft\rho'(v^i)^j\downharpoonright_\Upsilon = \downharpoonleft\rho'\downharpoonright_\Upsilon(\downharpoonleft v^i\downharpoonright_\Upsilon)^j$

  Unboxed value is a value, so by reduction rules
  $(\downharpoonleft\rho\downharpoonright_\Upsilon, \downharpoonleft\rho'\downharpoonright_\Upsilon(\downharpoonleft v^i\downharpoonright_\Upsilon)^j) \longmapsto (\downharpoonleft\rho\downharpoonright_\Upsilon, \downharpoonleft v^i\downharpoonright_\Upsilon)$

■

To show preservation of consistency we need a type substitution lemma.

**Lemma 26**
*If* $\mathrm{C} \vdash \Upsilon$, $\Upsilon \vdash \tau : \mathtt{r}$, $\tau$ *is not a type variable, and* $\mathrm{C}(\mathrm{lbl}(\tau)) = \mathrm{C}(\varrho(\alpha))$ *then:*

- *If* $\Upsilon \vdash \tau' : \mathtt{r}$ *and* $\mathrm{C}; \varrho \vdash \tau'$ *then* $\Upsilon \vdash \tau'[\tau/\alpha] : \mathtt{r}$.
- *If* $\Upsilon \vdash e$ *and* $\mathrm{C}; \varrho \vdash e$ *then* $\Upsilon \vdash e[\tau/\alpha]$.
- *If* $\Upsilon \vdash \rho$ *and* $\mathrm{C}; \varrho \vdash \rho$ *then* $\Upsilon \vdash \rho[\tau/\alpha]$.

**Proof:** The proof is by simultaneous induction on the derivation of $\Upsilon \vdash \tau' : t$, $\Upsilon \vdash e$, and $\Upsilon \vdash \rho$. The cases for expressions and environments are straight forward. Consider the cases for types:

- Case 1, $\tau' = \alpha^i$: If $\tau = \sigma^j$ then $\tau'[\tau/\alpha] = \sigma^i$. Since $\mathrm{C}; \varrho \vdash \tau'$, $\mathrm{C}(i) = \mathrm{C}(\varrho(\alpha))$, so $\mathrm{C}(i) = \mathrm{C}(j)$. By Lemmas 17 and 18, $i \overset{\Upsilon}{\simeq} j$. Then by inspection of the rules, $\Upsilon \vdash \sigma^i : \mathtt{r}$ as $\Upsilon \vdash \sigma^j : \mathtt{r}$.
- Case 2, $\tau' = \beta^i$ and $\alpha \neq \beta$: Then $\tau'[\tau/\alpha] = \tau'$ and the result is immediate.
- Case 3, $\tau' = \mathtt{B}^i$: Then $\Upsilon \vdash \tau' : \mathtt{r}$ is not possible.
- Case 4, $\tau' = (\forall\beta.\tau_1 \to \tau_2)^i$: Then $\Upsilon \vdash \tau'[\tau/\alpha] : \mathtt{r}$, as required.
- Case 5, $\tau' = (\mathtt{box}(\tau''))^i$, $i \in \Upsilon$: By the rule, $\Upsilon \vdash \tau'' : \mathtt{r}$. Assumption $\mathrm{C}; \varrho \vdash \tau'$ requires $\mathrm{C}; \varrho \vdash \tau''$. By the induction hypothesis, $\Upsilon \vdash \tau''[\tau/\alpha] : \mathtt{r}$. By the consistency rules, $\Upsilon \vdash \mathtt{box}(\tau''[\tau/\alpha])^i : \mathtt{r}$. By definition of substitution, $\Upsilon \vdash \mathtt{box}(\tau'')^i[\tau/\alpha] : \mathtt{r}$, as required.
- Case 6, $\tau' = (\mathtt{box}(\tau''))^i$, $i \notin \Upsilon$: Then $\Upsilon \vdash \tau'[\tau/\alpha] : \mathtt{r}$, as required.

■

**Lemma 27**
*If* $\Upsilon \vdash M_1$, $\mathrm{C} \vdash \Upsilon$, $\mathrm{C}; \varrho \vdash M_1$, $\vdash M_1 : \tau$, *and* $M_1 \longmapsto M_2$ *then* $\Upsilon \vdash M_2$.

**Proof:** The proof is by induction on the derivation of $M_1 \longmapsto M_2$. Let $M_1 = (\rho, e_1)$ and $M_2 = (\rho, e_2)$. By the rules for consistency, $\Upsilon \vdash \rho$ and $\Upsilon \vdash e_1$. The result follows if $\Upsilon \vdash e_2$. The typing rules require $\vdash \rho : \Gamma$ and $\emptyset; \Gamma \vdash e_1 : \tau'$ for some $\Gamma$ and $\tau'$. Assumption $\mathrm{C}; \varrho \vdash M$ requires $\mathrm{C}; \varrho \vdash \rho$ and $\mathrm{C}; \varrho \vdash e_1$. Consider the cases for the last rule used (in the same order as the figure):

- (Variable) In this case: $e_1 = x^i$, $e_2 = v^j$, and $x{:}\tau' = v^j \in \rho$. By $\Upsilon \vdash \rho$, $\Upsilon \vdash v^j$, as required.

- (Fix expression) In this case: $e_1 = (\texttt{fix}\ f[\alpha](x{:}\tau_1){:}\tau_2.e)^i$ and $e_2 = \langle \rho, \texttt{fix}\ f[\alpha](x{:}\tau_1){:}\tau_2.e\rangle^i$. Then $\Upsilon \vdash e_1$ requires $\Upsilon \vdash e$, then since $\Upsilon \vdash \rho$, $\Upsilon \vdash e_2$, as required.
- (Box expression) In this case: $e_1 = (\texttt{box}_\tau\, v^i)^j$ and $e_2 = \langle v^i{:}\tau\rangle^j$. Then $\Upsilon \vdash e_1$ requires $\Upsilon \vdash v^i$, so $\Upsilon \vdash e_2$, as required.
- (Application left) In this case, we have $e_1 = (e_3[\tau]\ e_4)^i$, $e_2 = (e_5[\tau]\ e_4)^i$, and $(\rho, e_3) \longmapsto (\rho, e_5)$ is a subderivation. Then $\Upsilon \vdash e_1$ requires $\Upsilon \vdash e_3$, $\Upsilon \vdash e_4$, and $\Upsilon \vdash \tau : \mathtt{r}$; $\emptyset; \Gamma \vdash e_1 : \tau'$ requires $\emptyset; \Gamma \vdash e_3 : \tau''$ for some $\tau''$; $\mathrm{C}; \varrho \vdash e_1$ requires $\mathrm{C}; \varrho \vdash e_3$. By the induction hypothesis, $\Upsilon \vdash e_5$, so by the consistency rules, $\Upsilon \vdash e_2$, as required.
- (Application right) In this case: $e_1 = (e_3[\tau]\ e_4)^i$, $e_2 = (e_3[\tau]\ e_5)^i$, and $(\rho, e_3) \longmapsto (\rho, e_5)$ is a subderivation. Then $\Upsilon \vdash e_1$ requires $\Upsilon \vdash e_3$, $\Upsilon \vdash e_4$, and $\Upsilon \vdash \tau : \mathtt{r}$; $\emptyset; \Gamma \vdash e_1 : \tau'$ requires $\emptyset; \Gamma \vdash e_4 : \tau''$ for some $\tau''$; $\mathrm{C}; \varrho \vdash e_1$ requires $\mathrm{C}; \varrho \vdash e_4$. By the induction hypothesis, $\Upsilon \vdash e_5$, so by the consistency rules, $\Upsilon \vdash e_2$, as required.
- (Application beta) In this case:

$$
\begin{array}{rcl}
e_1 & = & ({v_1}^i[\tau]\ {v_2}^j)^k \\
v_1 & = & \langle \rho', \texttt{fix}\ f[\alpha](x{:}\tau_1){:}\tau_2.e\rangle \\
e_2 & = & \rho''(e[\tau/\alpha])^k \\
\rho'' & = & \rho', f{:}\tau' = {v_1}^i, x{:}\tau_1' = {v_2}^j \\
\tau' & = & (\forall\alpha.\tau_1 \to \tau_2)^i \\
\tau_1' & = & \tau_1[\tau/\alpha]
\end{array}
$$

  By $\Upsilon \vdash e_1$ and the rules for consistency, $\Upsilon \vdash {v_1}^i$, $\Upsilon \vdash \rho'$, $\Upsilon \vdash e$, $\Upsilon \vdash \tau : \mathtt{r}$, and $\Upsilon \vdash {v_2}^j$. Thus by the rules for consistency, $\Upsilon \vdash \rho''$. By the typing rule $\emptyset \vdash \tau'\ \ wf$, so $\tau'$ cannot be a type variable. Assumption $\mathrm{C}; \varrho \vdash M$ requires $\mathrm{C}; \varrho \vdash e$ and, as in previous proofs, $\mathrm{C}(\mathrm{lbl}(\tau)) = \mathrm{C}(\varrho(i))$. By Lemma 26, $\Upsilon \vdash e[\tau/\alpha]$. By the rules for consistency, $\Upsilon \vdash e_2$, as required.
- (Under box) Similar to application left.
- (Under unbox) Similar to appliction left.
- (Unbox beta) In this case: $e_1 = (\texttt{unbox}\,\langle v^i{:}\tau\rangle^j)^k$ and $e_2 = v^i$. Then $\Upsilon \vdash e_1$ requires $\Upsilon \vdash v^i$, as required.
- (Under frame) Similar to application left.
- (Frame return) In this case: $e_1 = \rho'(v^i)^j$ and $e_2 = v^i$. Then $\Upsilon \vdash e_1$ requires $\Upsilon \vdash v^i$, as required.

■

With these lemmas we can prove our semantics preservation result.

**Theorem 4 (Coherence)**
- *If* $\vdash M : \tau$, $\mathrm{C}; \varrho \vdash M$, $\mathrm{C} \vdash \Upsilon$, $\Upsilon \vdash M$, *and* $M \longmapsto^* (\rho, v^i)$ *then* $\lfloor M \rfloor_\Upsilon \longmapsto^* (\lfloor\rho\rfloor_\Upsilon, \lfloor v^i\rfloor_\Upsilon)$.
- *If* $\vdash M : \tau$, $\mathrm{C}; \varrho \vdash M$, $\mathrm{C} \vdash \Upsilon$, $\Upsilon \vdash M$, *and* $M \longmapsto \cdots$ *then* $\lfloor M \rfloor_\Upsilon \longmapsto \cdots$.

**Proof:**

- By induction on reduction derivations, using Theorem 3.
  1. If $M \longmapsto^* (\rho, v^i)$ in zero steps, then the result follows immediately.
  2. If $M \longmapsto^* (\rho, v^i)$ in $n$ steps, then by definition, $M \longmapsto M'$ and $M' \longmapsto^* (\rho, v^i)$ in $n-1$ steps.

     By Theorem 3
     $\lfloor M\rfloor_\Upsilon \longmapsto \lfloor M'\rfloor_\Upsilon$

     By Theorem 1
     $\vdash M' : \tau'$

     By Lemma 15
     $\mathrm{C}; \varrho \vdash M'$

     By Lemma 27
     $\Upsilon \vdash M'$

     By induction
     $\lfloor M'\rfloor_\Upsilon \longmapsto^* (\lfloor\rho\rfloor_\Upsilon, \lfloor v^i\rfloor_\Upsilon)$

     By the defininition of many step reduction
     $\lfloor M\rfloor_\Upsilon \longmapsto^* (\lfloor\rho\rfloor_\Upsilon, \lfloor v^i\rfloor_\Upsilon)$
- In the operational semantics, there are six leaf reductions. Two of them take expression forms to value forms, but otherwise leave the term unchanged. One of the them takes unbox of box of a value to that value. One of them takes a frame of a value to that value. Thus if we measure a term by adding its size, number of lambda expressions, and number of box expressions, then this metric strictly decreases for these three leaf reductions. Therefore, in any infinite reduction sequence, there must be an infinite number of steps whose leaf reduction is a variable reduction or an application beta reduction. Then observe in the proof of Theorem 3 that the unboxing of a variable redex or of an application beta redex will always take a step, and that Lemma 25 preserves this. Thus the unboxing will also take an infinite number of steps.

■

Theorem 4 shows that if two terms are related by reduction, then their images under the unboxing function are also related by the many step reduction relation given that the unboxing pair is acceptable; and that if a term diverges under reduction, then its image under the unboxing function also diverges. In other words, for an acceptable analysis and an acceptable unboxing, the induced unboxing function preserves the semantics of the original program up to elimination of boxes. Since the semantics of the core language only defines reduction steps that preserve GC safety, this theorem implies that the image of a GC safe program under unboxing is also GC safe.

## 5. Construction of an acceptable unboxing

The previous section gives a declarative specification for when an unboxing set $\Upsilon$ is correct but does not specify how such a set might be chosen. In this section we give a simple algorithm for constructing an acceptable unboxing given an arbitrary acceptable flow analysis.

The idea behind the algorithm is that given a program and an acceptable flow analysis for it, we use the results of the flow analysis to construct the connected components of the interprocedural flow graph of the program. All of the elements of a connected component will then either be unboxed together, or not unboxed at all. Any such choice of unboxing (as we will show) satisfies the cache coherence property. The only remaining requirement is that the choice of unboxing set be consistent, which is easily satisfied by ensuring that any connected component which includes a type passed to a polymorphic function is only unboxed if the unboxing of the type argument still has traceability $\mathtt{r}$. In the rest of the section, we make this informal algorithm concrete and show that the choice of unboxing that it produces is in fact acceptable.

For the purposes of this section we ignore environments and the intermediate forms $\rho(e)$, $\langle \rho, \texttt{fix}\ f[\alpha](x{:}\tau_1){:}\tau_2.e\rangle^j$ and $\langle v^i{:}t\rangle^j$. These constructs are present in the language solely as mechanisms to discuss the dynamic semantics—in this sense they can be thought of as intermediate terms, rather than source terms. It is straightforward to incorporate these into the algorithm if desired.

Given a flow analysis $(\mathrm{C}, \varrho)$ and program $e$ such that $\mathrm{C}; \rho \vdash e$, we define the induced undirected flow graph $\mathcal{FG}$ as an undirected graph with a node for every label in C, and edges as follows:

- For every label $i$ and every shape $s \in \mathrm{C}(i)$, we add an edge between $i$ and $\mathrm{lbl}(s)$.

The edges simply connect up each program point with all of its reaching definitions.

Given a flow graph $\mathcal{FG}$, we can find the connected components in the usual way. Let $\mathcal{CC}$ be a mapping which maps labels to the connected component in which they occur. Note that by definition each label occurs in exactly one connected component. It is easy to show that any connected component is cache consistent.

**Lemma 28 (Cache consistency of a connected component)**
*Given any acceptable analysis $(\mathrm{C}, \varrho)$ with induced flow graph $\mathcal{FG}$, and any connected component $S$ of $\mathcal{FG}$, $S$ is cache consistent: that is,* $\mathrm{C} \vdash S$.

**Proof:** To show that $\mathrm{C} \vdash S$ we must show that $\forall i, s : s \in \mathrm{C}(i) \implies i \overset{S}{\simeq} \mathrm{lbl}(s)$. But note that by the construction of the induced flow graph $\mathcal{FG}$, whenever $s \in \mathrm{C}(i)$ there is an edge between $i$ and $\mathrm{lbl}(s)$, and consequently by definition of a connected component, $i$ and $\mathrm{lbl}(s)$ must be in the same connected component. Since every label occurs in exactly one connected component, either both $i$ and $\mathrm{lbl}(s)$ are in $S$ or both are not in $S$. By definition then, $i \overset{S}{\simeq} \mathrm{lbl}(s)$. ■

It is also easy to show that the union of any two disjoint cache consistent sets is also cache consistent.

**Lemma 29 (Cache consistency (unary) closure)**
*Given any acceptable analysis $(\mathrm{C}, \varrho)$ and disjoint label sets $S_1$ and $S_2$, then if $\mathrm{C} \vdash S_1$ and $\mathrm{C} \vdash S_2$ then $\mathrm{C} \vdash S_1 \cup S_2$*

**Proof:** To show that $\mathrm{C} \vdash S_1 \cup S_2$ we must show that $\forall i, s : s \in \mathrm{C}(i) \implies i \overset{S_1 \cup S_2}{\simeq} \mathrm{lbl}(s)$. Consider an abitrary label $i$. If $i$ is not in $S_1 \cup S_2$, then we have that $i$ is not in $S_1$ and not in $S_2$, and hence by assumption, $\mathrm{lbl}(s)$ is not in $S_1$ and not in $S_2$, and hence we have agreement. If $i$ is in $S_1 \cup S_2$, then it must be in either $S_1$ or $S_2$. WLOG, assume that $i \in S_1$. By assumption, $i \overset{S_1}{\simeq} \mathrm{lbl}(s)$, and so $\mathrm{lbl}(s) \in S_1$, and hence $\mathrm{lbl}(s) \in S_1 \cup S_2$ and we have agreement. ■

Consequently, we can show that any set consisting of a union of connected components of the induced flow graph is cache consistent.

**Lemma 30 (Cache consistency closure)**
*Given any acceptable analysis $(\mathrm{C}, \varrho)$ with induced flow graph $\mathcal{FG}$, and any set $\mathcal{SS}$ of connected components of $\mathcal{FG}$, $\bigcup \mathcal{SS}$ is cache consistent.*

**Proof:** By Lemma 28, each connected component is cache consistent. By definition, any two connected components are disjoint, and so by Lemma 29 the union of any two connected components are cache consistent, and are disjoint from any other connected component. The cache consistency of $\bigcup \mathcal{SS}$ follows directly by induction. ■

### 5.1 The algorithm

Given the set of connected components for the induced flow graph, the algorithm begins with an initial unboxing set $\Upsilon$ consisting of the union of all of the connected components. By Lemma 30, we have that $\mathrm{C} \vdash \Upsilon$. The algorithm then proceeds by considering in turn each application sub-term $e_1[\tau]e_2$ as follows:

- For each sub-term of $e$ of the form $e_1[\tau]e_2$:
  - if $\mathrm{lbl}(\tau) \in \Upsilon$, and if $\Upsilon \vdash \tau : \mathtt{b}$, then:
    - $\Upsilon \leftarrow \Upsilon - \mathcal{CC}(\mathrm{lbl}(\tau))$.

That is, for any application for which the current unboxing results in the type argument being unboxed to a non-reference type, we remove the connected component for the type from the unboxing set. Note that after removing a connected component from $\Upsilon$, the new unboxing set $\Upsilon$ is still cache consistent since it is still a union of connected components (just a union of one less connected component).

With the help of some technical lemmas, it is straightforward to show that the final unboxing set $\Upsilon$ computed by the algorithm is an acceptable unboxing for the program.

To begin with, we observe that if a type's label is not in the unboxing set $\Upsilon$, then it is consistent and its traceability is unchanged by the unboxing.

**Lemma 31 (Type consistency)**
*For any unboxing set $\Upsilon$ and type $\tau$, if $\mathrm{lbl}(\tau) \notin \Upsilon$ then $\Upsilon \vdash \tau : tr(\tau)$.*

**Proof:** By inspection.

- (Variable) $tr(\alpha^i) = \mathtt{r}$, and $\Upsilon \vdash \alpha^i : \mathtt{r}$.
- (Base type) $tr(\mathtt{B}^i) = \mathtt{b}$, and $\Upsilon \vdash \mathtt{B}^i : \mathtt{b}$.
- (Fun type) $tr(\forall\alpha.\tau_1 \to {\tau_2}^i) = \mathtt{r}$, and $\Upsilon \vdash \forall\alpha.\tau_1 \to {\tau_2}^i : \mathtt{r}$.
- (Box type) $tr(\mathtt{box}(\tau')^i) = \mathtt{r}$, and by assumption $i \notin \Upsilon$, so we have that $\Upsilon \vdash \mathtt{box}(\tau')^i : \mathtt{r}$.

■

It is also the case that the consistent type judgement defines a total function on types, and hence for any type we either have that it is consistent at traceability $\mathtt{r}$ or that it is consistent at traceability $\mathtt{b}$.

**Lemma 32 (Type consistency is a total function)**
*For any unboxing set $\Upsilon$ and type $\tau$, either $\Upsilon \vdash \tau : \mathtt{b}$, or $\Upsilon \vdash \tau : \mathtt{r}$.*

**Proof:** By induction on types. All of the cases follow immediately except when $\tau = \mathtt{box}(\tau')^i$ and $i \in \Upsilon$. In that case, by induction we have that either $\Upsilon \vdash \tau' : \mathtt{b}$, or $\Upsilon \vdash \tau' : \mathtt{r}$, and so by construction either $\Upsilon \vdash \tau : \mathtt{b}$, or $\Upsilon \vdash \tau : \mathtt{r}$. ■

**Theorem 5**
*If $\Delta; \Gamma \vdash e : \tau$, $\mathrm{C}; \varrho \vdash \Gamma$, and $\mathrm{C}; \varrho \vdash e$ and if $\Upsilon$ is the unboxing set computed by the algorithm in this section, then $\Upsilon$ is an acceptable unboxing for $e$. That is, $\mathrm{C} \vdash \Upsilon$ and $\Upsilon \vdash e$.*

**Proof:** The conclusion that $\mathrm{C} \vdash \Upsilon$ follows almost immediately from Lemma 30. The initial choice of $\Upsilon$ is a union of connected components, and hence is cache consistent. At every step of the algorithm, we may remove a single connected component from $\Upsilon$. The result is still a union of connected components (since connected components are disjoint), and hence the result of removing a connected component is still cache consistent by Lemma 30.

The conclusion that $\Upsilon \vdash e$ follows by induction on the structure of the typing derivation.

- (Variable) In this case, $e = x^i$, consistency is immediate.
- (Fix) In this case $e = (\mathtt{fix}\ f[\alpha](x{:}\tau_1){:}\tau_2.e')^i$. To get consistency, we must show that $\Upsilon \vdash e'$. The last rule applied in the typing judgement must have been the fix rule, and by its premises we have that $\Delta \vdash \forall\alpha.\tau_1 \to {\tau_2}^i\ \ wf$ (1), and that $\Delta, \alpha; \Gamma, f{:}\forall\alpha.\tau_1 \to {\tau_2}^i, x{:}\tau_1 \vdash e' : \tau_2$ (2). The last rule applied in the acceptable analysis judgement must also have been the fix rule, and by its premises we have that $\mathrm{C}; \varrho \vdash e'$ (3). To apply the induction hypothesis, we need (1), (3), and that $\mathrm{C}; \varrho \vdash \Gamma, f{:}\forall\alpha.\tau_1 \to {\tau_2}^i, x{:}\tau_1$ (4). To show (4), it is sufficient to show that:

- $\varrho(f) = i$ which is a premise of the acceptable analysis derivation
- $\mathrm{C}; \varrho \vdash \forall\alpha.\tau_1 \to {\tau_2}^i$ which is a premise of the acceptable analysis derivation
- $\varrho(x) = \mathrm{lbl}(\tau_1)$ which is a premise of the acceptable analysis derivation
- $\mathrm{C}; \varrho \vdash \tau_1$ which is a sub-premise of the derivation of $\mathrm{C}; \varrho \vdash \forall\alpha.\tau_1 \to {\tau_2}^i$.

So by (1), (3), and (4), we have by induction that $\Upsilon \vdash e'$.

- (Application) In this case $e = (e_1[\tau]e_2)^i$. To prove consistency, we need that $\Upsilon \vdash e_1$ (1), $\Upsilon \vdash e_2$ (2), and $\Upsilon \vdash \tau : \mathtt{r}$ (3). Inverting the typing derivation and the acceptable analysis derivation immediately gives us the premises we need to apply the induction hypothesis to get (1) and (2). To prove (3), note that a premise of the typing derivation gives us that $tr(\tau) = \mathtt{r}$ (4). If $\mathrm{lbl}(\tau) \notin \Upsilon$, then by Lemma 31 we have that $\Upsilon \vdash \tau : tr(\tau)$ and so by (4) we're done. If $\mathrm{lbl}(\tau) \in \Upsilon$, then by the definition of the algorithm, we must have that $\Upsilon \vdash \tau : \mathtt{b}$ does not hold (since otherwise the algorithm would have removed the connected component containing $\mathrm{lbl}(\tau)$ from $\Upsilon$), and so by Lemma 32 we must have that $\Upsilon \vdash \tau : \mathtt{r}$ and we're done.
- (Box) All of the premises need to apply the induction hypothesis are available immediately by inverting the typing derivation and the acceptable analysis derivation.
- (Unbox) All of the premises need to apply the induction hypothesis are available immediately by inverting the typing derivation and the acceptable analysis derivation.
- (Constant) Follows immediately.

■

Thus we have shown by construction that the specification defined in Section 4 is a useful one in the sense that it is satisfiable.

## 6. Open Terms

The paper so far has considered whole-program optimization and proved that unboxing in that setting is correct. We would like to be able to optimize program fragments where we have part of the program but know nothing about the rest of the program. Such a setting adds one more correctness criteria—since we are not optimizing the rest of the program, anything that flows across the boundary to or from the rest of the program must remain as boxed as it originally was. We can ensure this requirement by simply requiring that nothing on the boundary is in the unboxing set. This section formalizes these ideas and proves them correct.

For our purposes, a program fragment is a module, which is a triple $(\Gamma \Rightarrow e : \tau)$. $\Gamma$ specifies the imports of the module, $e$ specifies the body of the module, which exports only one thing—the value that $e$ evaluates to, and $\tau$ specifies the type of the export. We wish to optimize modules without making any assumptions about the code that the module is linked to. In particular that means we cannot unbox any of the imports nor unbox anything exported. This requirement can be achieved by not unboxing any subterm of any type in $\Gamma$ nor in $\tau$.

We can extend the definitions of well typedness, acceptability of flow analysis, unboxing, and consistency of unboxing to modules,

$\boxed{\vdash (\Gamma \Rightarrow e : \tau)\ wf}$

$$\frac{\emptyset \vdash \Gamma\ wf \quad \emptyset; \Gamma \vdash e : \tau}{\vdash (\Gamma \Rightarrow e : \tau)\ wf}$$

$\boxed{\mathrm{C}; \varrho \vdash (\Gamma \Rightarrow e : \tau)}$

$$\frac{\mathrm{C}; \varrho \vdash^{\mathsf{s}} \Gamma \quad \mathrm{C}; \varrho \vdash^{\mathsf{s}} e \quad \mathrm{C}; \varrho \vdash^{\mathsf{s}} \tau}{\mathrm{C}; \varrho \vdash (\Gamma \Rightarrow e : \tau)}$$

$\boxed{\lfloor(\Gamma \Rightarrow e : \tau)\rfloor_\Upsilon}$

$$\lfloor(\Gamma \Rightarrow e : \tau)\rfloor_\Upsilon = (\Gamma \Rightarrow \lfloor e\rfloor_\Upsilon : \tau)$$

$\boxed{\Upsilon \vdash \tau\ not\ unboxed}$

$$\frac{i \notin \Upsilon}{\Upsilon \vdash \alpha^i\ not\ unboxed} \qquad \frac{i \notin \Upsilon}{\Upsilon \vdash \mathtt{B}^i\ not\ unboxed}$$

$$\frac{i \notin \Upsilon \quad \Upsilon \vdash \tau_1\ not\ unboxed \quad \Upsilon \vdash \tau_2\ not\ unboxed}{\Upsilon \vdash (\forall\alpha.\tau_1 \to \tau_2)^i\ not\ unboxed}$$

$$\frac{i \notin \Upsilon \quad \Upsilon \vdash \tau\ not\ unboxed}{\Upsilon \vdash \mathtt{box}(\tau)^i\ not\ unboxed}$$

$\boxed{\Upsilon \vdash \Gamma\ not\ unboxed}$

$$\frac{\forall 1 \le j \le n : \Upsilon \vdash \tau_j\ not\ unboxed}{\Upsilon \vdash x_1{:}\tau_1, \ldots, x_n{:}\tau_n\ not\ unboxed}$$

$\boxed{\Upsilon \vdash (\Gamma \Rightarrow e : \tau)}$

$$\frac{\Upsilon \vdash \Gamma\ not\ unboxed \quad \Upsilon \vdash e \quad \Upsilon \vdash \tau\ not\ unboxed}{\Upsilon \vdash (\Gamma \Rightarrow e : \tau)}$$

**Figure 10.** Judgements for modules

$\boxed{\mathrm{C}; \varrho \vdash^{\mathsf{s}} \tau}$

$$\frac{\mathrm{C}(\varrho(\alpha)) = \mathrm{C}(i) \quad s \in \mathrm{C}(i)}{\mathrm{C}; \varrho \vdash^{\mathsf{s}} \alpha^i}$$

$\boxed{\mathrm{C}; \varrho \vdash^{\mathsf{s}} e}$

$$\frac{\mathrm{C}; \varrho \vdash^{\mathsf{s}} \mathtt{box}(\tau)^i \quad \mathrm{C}; \varrho \vdash^{\mathsf{s}} e \quad (\mathtt{box}_{tr(\tau)}\ \mathrm{lbl}(e))^i_{\mathsf{v}} \in \mathrm{C}(i)}{\mathrm{C}; \varrho \vdash^{\mathsf{s}} (\mathtt{box}_\tau\ e)^i}$$

$$\frac{\mathrm{C}; \varrho \vdash^{\mathsf{s}} \mathtt{box}(\tau)^i \quad \mathrm{C}; \varrho \vdash^{\mathsf{s}} v^j \quad (\mathtt{box}_{tr(\tau)}\ j)^i_{\mathsf{v}} \in \mathrm{C}(i)}{\mathrm{C}; \varrho \vdash^{\mathsf{s}} \langle v^j{:}\tau\rangle^i}$$

**Figure 11.** Stronger Analysis

and the formal judgements appear in Figure 10. The acceptability of a flow analysis for modules is stronger than that for programs. The rules for type variables, box expressions, and box values are replaced with those in Figure 11, the other rules remain the same. The rules for boxes require a more precise shape in the cache, any actual flow analysis would use such a shape, so this requirement is not a burden. The rules for programs are weaker as the stronger conditions are not closed under reduction whereas the weaker conditions are. The stronger condition for type variables is required to ensure consistency for type variables, and is also not a burden.

The goal is to show that unboxing is correct for modules. A suitable notion of correctness is that a module and its unboxing are contextually equivalent. Rather than define contextual equivalence directly, we will use a notion that is usually proven equivalent to contextually equivalence as our definition. Namely, two expressions are equivalent if in any environment that closes them and any elimination context for their type they are observable equivalent then they are contextually equivalent. The formal definition is in Figure 12.

The strategy is that we will take the context and alpha vary it and relabel it so that it is sufficiently distinct from the module. Then we will argue that we can modify the flow analysis and unboxing to cover the context without unboxing any of it. Then by coherence the module in context will behave the same as the unboxing of the module in context, which because the context is not unboxed, will act the same as the unboxed module in context.

First we formalize and prove that the operational semantics is insensitive to the alpha variant and labels used. Let $x \sim_{\mathsf{s}} y$ mean that $x$ and $y$ are alpha variants and possibly relabelled.

**Lemma 33**
*If $M_1 \sim_{\mathsf{s}} M_2$ and $M_1 \longmapsto M_3$ then there exists $M_4$ such that $M_3 \sim_{\mathsf{s}} M_4$ and $M_2 \longmapsto M_4$.*

**Proof:** The proof is by a straight forward induction on the derivation of $M_1 \longmapsto M_3$. ■

Next we prove three lemmas about unboxing preservation. In the first two we show that something's unboxing is that something because either the not unboxed judgement (the first lemma) or the labels in the something are not in the unboxing set (the second lemma). In the third we show the unboxing of an expression is the same if the unboxing set is the same on the labels in the expression. To state and prove these and subsequent lemmas we need a function to return all the labels in an expressions, type, or environment. It is defined in Figure 13.

**Lemma 34**
- *If $\Upsilon \vdash \tau$ $\mathit{not\ unboxed}$ then $\lfloor\tau\rfloor_\Upsilon = \tau$.*
- *If $\Upsilon \vdash \Gamma$ $\mathit{not\ unboxed}$ then $\lfloor\Gamma\rfloor_\Upsilon = \Gamma$.*

**Proof:**

- The proof is by induction on the structure of $\tau$. Consider the cases:
  - Case 1, $\tau = \alpha^i$: Then by definition $\lfloor\tau\rfloor_\Upsilon = \tau$, as required.
  - Case 2, $\tau = \mathtt{B}^i$: Then by definition $\lfloor\tau\rfloor_\Upsilon = \tau$, as required.
  - Case 3, $\tau = (\forall\alpha.\tau_1 \to \tau_2)^i$: Then $\Upsilon \vdash \tau$ $\mathit{not\ unboxed}$ requires $\Upsilon \vdash \tau_1$ $\mathit{not\ unboxed}$ and $\Upsilon \vdash \tau_2$ $\mathit{not\ unboxed}$. By the induction hypothesis, $\lfloor\tau_1\rfloor_\Upsilon = \tau_1$ and $\lfloor\tau_2\rfloor_\Upsilon = \tau_2$. By definition, $\lfloor\tau\rfloor_\Upsilon = \tau$, as required.
  - Case 4, $\tau = \mathtt{box}(\tau')^i$: Then $\Upsilon \vdash \tau$ $\mathit{not\ unboxed}$ requires $i \notin \Upsilon$ and $\Upsilon \vdash \tau'$ $\mathit{not\ unboxed}$. By the induction hypothesis, $\lfloor\tau'\rfloor_\Upsilon = \tau'$. By definition, $\lfloor\tau\rfloor_\Upsilon = \tau$, as required.
- If $\Gamma = x_1{:}\tau_1, \ldots, x_n{:}\tau_n$ then: $\Upsilon \vdash \Gamma$ $\mathit{not\ unboxed}$ requires $\Upsilon \vdash \tau_j$ $\mathit{not\ unboxed}$ for $1 \leq j \leq n$. So by the first item, $\lfloor\tau_j\rfloor_\Upsilon = \tau_j$ for $1 \leq j \leq n$. Then by definition, $\lfloor\Gamma\rfloor_\Upsilon = \Gamma$, as required.

■

**Lemma 35**
- *If $\mathrm{lbls}(\rho) \cap \Upsilon = \emptyset$ then $\lfloor\rho\rfloor_\Upsilon = \rho$.*
- *If $\mathrm{lbls}(E) \cap \Upsilon = \emptyset$ then $\lfloor E\rfloor_\Upsilon = E$.*

**Proof:** The proof is a straight forward induction on the structure of $\rho$ and $E$. ■

**Lemma 36**
*If $\Upsilon_1 \cap \mathrm{lbls}(e) = \Upsilon_2 \cap \mathrm{lbls}(e)$ then $\lfloor e\rfloor_{\Upsilon_1} = \lfloor e\rfloor_{\Upsilon_2}$.*

**Proof:** The proof is a staight forward induction on the structure of $e$. ■

Next we state and prove our main technical lemma. This lemma states that we can rewrite the context and flow analysis to have certain desirable properties, namely that the flow analysis covers the context and the module, that the context is not unboxed, that the module is unboxed as before, and the unboxing set and flow analysis remain consistent and consistent with the module and context.

**Lemma 37**
*If:*

$$\begin{array}{l}
\emptyset \vdash \Gamma\ \mathit{wf} \\
\emptyset; \Gamma \vdash e : \tau \\
\mathrm{C}; \varrho \vdash \Gamma \\
\mathrm{C}; \varrho \vdash e \\
\mathrm{C}; \varrho \vdash \tau \\
\mathrm{C} \vdash \Upsilon \\
\Upsilon \vdash \Gamma\ \mathit{not\ unboxed} \\
\Upsilon \vdash e \\
\Upsilon \vdash \tau\ \mathit{not\ unboxed} \\
\vdash \rho : \Gamma \\
\Gamma \vdash E : \mathtt{B}^i\langle\tau\rangle
\end{array}$$

*then there exists $\rho'$, $E'$, $\mathrm{C}'$, $\varrho'$, and $\Upsilon'$ such that:*

$$\begin{array}{l}
\rho \sim_{\mathsf{s}} \rho' \\
E \sim_{\mathsf{s}} E' \\
\vdash \rho' : \Gamma \\
\Gamma \vdash E' : \mathtt{B}^j\langle\tau\rangle \\
\mathrm{C}'; \varrho' \vdash (\rho', E'\langle e\rangle) \\
\mathrm{C}' \vdash \Upsilon' \\
\Upsilon' \vdash (\rho', E'\langle e\rangle) \\
\mathrm{lbls}(\rho') \cap \Upsilon' = \emptyset \\
\mathrm{lbls}(E') \cap \Upsilon' = \emptyset \\
\Upsilon \cap \mathrm{lbls}(e) = \Upsilon' \cap \mathrm{lbls}(e)
\end{array}$$

**Proof:** Let $V$ be the set of variables that occur in $e$. Let $A$ be the set of type variables that occur in $\Gamma$, $e$, or $\tau$. Both these sets are finite.

The derivation of $\mathrm{C}; \varrho \vdash \Gamma$, $\mathrm{C}; \varrho \vdash e$, and $\mathrm{C}; \varrho \vdash \tau$ will for each type that is not a type variable require a particular type shape with some label on it in the cache of the label of that type, similarly for each box expression and box value require a box shape with some label of its contents in the cache. Let $L$ be one such label for each such type and such box as well as $\varrho(V) \cup \varrho(A) \cup \mathrm{lbls}(\Gamma) \cup \mathrm{lbls}(e) \cup \mathrm{lbls}(\tau)$. Note that $L$ is a finite set.

Let $\varrho'$ be $\varrho$ on $V$ and $A$ and on every other variable or type variable let it map to a fresh label (distinct from each other and

$$E ::= \langle\rangle \mid (E[\tau]\ e)^i \mid (\texttt{unbox}\, E)^i$$

$$\frac{}{\Gamma \vdash \langle\rangle : \tau\langle\tau\rangle}$$

$$\frac{\Gamma \vdash E : (\forall\alpha.\tau_1 \to \tau_2)^j\langle\tau\rangle \quad \emptyset \vdash \tau\ wf \quad tr(\tau) = \texttt{r} \quad \emptyset;\Gamma \vdash e : \tau_1' \quad \vdash \tau_1[\tau/\alpha] = \tau_1'}{\Gamma \vdash (E[\tau]\ e)^i : \tau_2[\tau/\alpha]\langle\tau\rangle}$$

$$\frac{\Gamma \vdash E : \texttt{box}(\tau')^j\langle\tau\rangle}{\Gamma \vdash (\texttt{unbox}\, E)^i : \tau'\langle\tau\rangle}$$

$$\begin{array}{lcl}
M_1 \overset{obs}{\equiv} M_2 & \overset{\text{def}}{=} & (\forall c, i : (M_1 \longmapsto^* c^i \Leftrightarrow M_2 \longmapsto^* c^i)) \wedge (M_1 \longmapsto \cdots \Leftrightarrow M_2 \longmapsto \cdots) \\
\Gamma \vdash e_1 \equiv e_2 : \tau & \overset{\text{def}}{=} & \emptyset;\Gamma \vdash e_1 : \tau \wedge \emptyset;\Gamma \vdash e_2 : \tau \wedge \\
& & \forall\rho, E : \ \vdash \rho : \Gamma \wedge \Gamma \vdash E : \texttt{B}^i\langle\tau\rangle \implies (\rho, C\langle e_1\rangle) \overset{obs}{\equiv} (\rho, C\langle e_2\rangle) \\
\vdash (\Gamma_1 \Rightarrow e_1 : \tau_1) \equiv (\Gamma_2 \Rightarrow e_2 : \tau_2) & \overset{\text{def}}{=} & \Gamma_1 = \Gamma_2 \wedge \tau_1 = \tau_2 \wedge (\Gamma_1 \vdash e_1 \equiv e_2 : \tau_1)
\end{array}$$

**Figure 12.** Contextual Equivalence

$$\begin{array}{lcl}
\text{lbls}(\sigma^i) & = & \{i\} \cup \text{lbls}(\sigma) \\
\text{lbls}(\alpha) & = & \emptyset \\
\text{lbls}(\texttt{B}) & = & \emptyset \\
\text{lbls}(\forall\alpha.\tau_1 \to \tau_2) & = & \text{lbls}(\tau_1) \cup \text{lbls}(\tau_2) \\
\text{lbls}(\texttt{box}(\tau)) & = & \text{lbls}(\tau) \\
\text{lbls}(m^i) & = & \{i\} \cup \text{lbls}(m) \\
\text{lbls}(v^i) & = & \{i\} \cup \text{lbls}(v) \\
\text{lbls}(x) & = & \emptyset \\
\text{lbls}(\texttt{fix}\ f[\alpha](x{:}\tau_1){:}\tau_2.e) & = & \text{lbls}(\tau_1) \cup \text{lbls}(\tau_2) \cup \text{lbls}(e) \\
\text{lbls}(e_1[\tau]\ e_2) & = & \text{lbls}(e_1) \cup \text{lbls}(\tau) \cup \text{lbls}(e_2) \\
\text{lbls}(\texttt{box}_\tau\, e) & = & \text{lbls}(\tau) \cup \text{lbls}(e) \\
\text{lbls}(\texttt{unbox}\, e) & = & \text{lbls}(e) \\
\text{lbls}(\rho(e)) & = & \text{lbls}(\rho) \cup \text{lbls}(e) \\
\text{lbls}(c) & = & \emptyset \\
\text{lbls}(\langle\rho, \texttt{fix}\ f[\alpha](x{:}\tau_1){:}\tau_2.e\rangle) & = & \text{lbls}(\rho) \cup \text{lbls}(\tau_1) \cup \text{lbls}(\tau_2) \cup \text{lbls}(e) \\
\text{lbls}(\langle v^i{:}\tau\rangle) & = & \text{lbls}(\tau) \cup \text{lbls}(v^i) \\
\text{lbls}(x_1{:}\tau_1 = {v_1}^{i_1}, \ldots, x_n{:}\tau_n = {v_n}^{i_n}) & = & \cup_{1\le i\le n}\text{lbls}(\tau_j) \cup \text{lbls}({v_j}^{i_j})
\end{array}$$

**Figure 13.** The labels in an type, expression, or environment

from $L$). Define:

$$\mathrm{C}''(i) = \begin{cases} \{s \mid s \in \mathrm{C}(i) \wedge \text{lbls}(s) \subseteq L\} & i \in L \\ \emptyset & i \notin L \end{cases}$$

Claim: $\mathrm{C}''; \varrho' \vdash \Gamma$, $\mathrm{C}''; \varrho' \vdash e$, and $\mathrm{C}''; \varrho' \vdash \tau$. The proof is by induction on the derivation, consider the last rule used:

- (Variable) In this case $e = x^i$ and $\mathrm{C}(\varrho(x)) \subseteq \mathrm{C}(\varrho(x))$. Since $x \in V$, $\varrho'(x) = \varrho(x)$ and $\varrho(x) \in L$. Also $i \in L$. Therefore, $\mathrm{C}''(\varrho'(x)) \subseteq \mathrm{C}''(i)$, as required. Thus by the same rule, $\mathrm{C}''; \varrho' \vdash e$, as required.
- (Fix expression) In this case $e = (\texttt{fix}\ f[\alpha](x{:}\tau_1){:}\tau_2.e')^i$, $\varrho(f) = i$, $\varrho(x) = \text{lbl}(\tau_1)$, $\mathrm{C}; \varrho \vdash (\forall\alpha.\tau_1 \to \tau_2)^i$, $\mathrm{C}; \varrho \vdash e'$, and $(\forall\varrho(\alpha).\text{lbl}(\tau_1) \to \text{lbl}(e'))_{\mathsf{v}}^{\in}\mathrm{C}(i)$. Since $f, x \in V$ and $\alpha \in A$, $\varrho'(f) = \varrho(f)$, $\varrho'(x) = \varrho(x)$, $\varrho'(\alpha) = \varrho(\alpha)$, and $\varrho(\alpha) \in L$. By the induction hypothesis, $\mathrm{C}''; \varrho' \vdash (\forall\alpha.\tau_1 \to \tau_2)^i$ and $\mathrm{C}''; \varrho' \vdash e'$. Since $\varrho'(\alpha) \in L$, $\text{lbl}(\tau_1) \in L$, $\text{lbl}(e) \in L$, and $i \in L$, $(\forall\varrho'(\alpha).\text{lbl}(\tau_1) \to \text{lbl}(e'))_{\mathsf{v}}^{\in}\mathrm{C}''(i)$. Thus by the same rule, $\mathrm{C}''; \varrho' \vdash e$, as required.
- (Application) In this case $e = (e_1[\tau]\ e_2)^i$, $\mathrm{C}; \varrho \vdash e_1$, $\mathrm{C}; \varrho \vdash \tau$, $\mathrm{C}; \varrho \vdash e_2$, and $funC(\text{lbl}(e_1), \text{lbl}(\tau), \text{lbl}(e_2), i)$. By the induction hypothesis, $\mathrm{C}'; \varrho' \vdash e_1$, $\mathrm{C}'; \varrho' \vdash \tau$, and $\mathrm{C}'; \varrho' \vdash e_2$. Since $\text{lbl}(e_1) \in L$, $\text{lbl}(\tau) \in L$, $\text{lbl}(e_2) \in L$, and $i \in L$, it is easy to see that $funC(\text{lbl}(e_1), \text{lbl}(\tau), \text{lbl}(e_2), i)$ (for $\mathrm{C}''$). Thus by the same rule, $\mathrm{C}''; \varrho' \vdash e$, as required.
- Other cases are similar …

Let $A'$ be the set of type variables that appear in $\Gamma$ and $\tau$. We construct $\rho'$ and $E'$ as alpha variants and relabellings of $\rho$ and $E$ as follows. Since $\vdash \rho : \Gamma$, $\rho$ contains $\Gamma$, so we keep that part the same. Type variables that are in $A'$ we keep the same. All other type variables and variables we pick an alpha variant that is fresh (distinct from each other and from $A$ respectively $V$). The outermost label on types on variables we relabel to the binding label for that variable. All other labels we relabel to be fresh. Clearly $\rho \sim_{\mathsf{s}} \rho'$ and $E \sim_{\mathsf{s}} E'$.

Claim: $\vdash \rho' : \Gamma$ and $\Gamma \vdash E' : \texttt{B}^j\langle\tau\rangle$ for some $j$. The proof is a straight forward induction on the structure of $\rho'$ and $E'$.

Now we need to build a $\mathrm{C}'$ such that $\mathrm{C}'; \varrho' \vdash (\rho', E'\langle e \rangle)$. We start from $\mathrm{C}''$. First we add into the caches, shapes required directly for the rules for $\mathrm{C}'; \varrho' \vdash \rho'$ and $\mathrm{C}'; \varrho' \vdash E'$ (such things are already there for $e$). In the case of types we add shapes using the label of the type as the label of the shape. In the case of box expressions and values we use the label of the contents of the box as the label of the contents of the shape. What remains is a bunch of subset and equalty constraints between cache entries, so we pick $\mathrm{C}'$ to be the smallest larger cache that satisfies these constraints. Clearly such a $\mathrm{C}'$ exists and by construction, $\mathrm{C}'; \varrho' \vdash (\rho', E'\langle e \rangle)$.

Set $\Upsilon' = \Upsilon \cap L$. Clearly, $\Upsilon \cap \mathrm{lbls}(e) = \Upsilon' \cap \mathrm{lbls}(e)$ as $\mathrm{lbls}(e) \subseteq L$. By construction, the labels of $\rho'$ and $E'$ are in the labels of $\Gamma$ or $\tau$ or are not in $L$. Since $\Upsilon \vdash \Gamma$ *not unboxed* and $\Upsilon \vdash \tau$ *not unboxed*, the labels of $\Gamma$ and $\tau$ are not in $\Upsilon$. Therefore, $\mathrm{lbls}(\Gamma) \cap \Upsilon' = \emptyset$ and $\mathrm{lbls}(\tau) \cap \Upsilon' = \emptyset$. In fact, if $A''$ is a set of type variables in $\rho'$ and $E'$ then $\varrho'(A'') \cap \Upsilon' = \emptyset$ too.

Claim: any flow from the interface to a $boxC(i, j)$ condition has a box type at the interface (**), and similarly for $funC(i, j, k, l)$. The proof is by induction on the flow conditions noting that in all cases the two end points have the same type.

Claim: $\mathrm{C}' \vdash \Upsilon'$. Let $s$ and $i$ be such that $s \in \mathrm{C}'(i)$. If $s \in \mathrm{C}''(i)$ then $s \in \mathrm{C}(i)$, $i \in L$, and $\mathrm{lbls}(s) \subseteq L$, and in particular, $\mathrm{lbl}(s) \in L$. Since $\mathrm{C} \vdash \Upsilon$, $i \overset{\Upsilon}{\simeq} \mathrm{lbl}(s)$. Since $\Upsilon' = \Upsilon \cap L$, $i \in L$, and $\mathrm{lbl}(s) \in L$, $i \overset{\Upsilon'}{\simeq} \mathrm{lbl}(s)$, as required. Otherwise, we claim that $i, \mathrm{lbl}(s) \notin \Upsilon'$. Let $L_C = \mathrm{lbls}(\rho') \cup \mathrm{lbls}(E') \cup \varrho'(A'')$, $L_I = \mathrm{lbls}(\Gamma) \cup \mathrm{lbls}(\tau)$, $L_M = L - L_C$, and $L_A = L_C \cup L$. First notice that $\mathrm{C}''$ has entries only for labels in $L$ and with shapes whose labels are in $L$. The first part of computing $\mathrm{C}'$ added shapes to cache entries for labels in $L_C$ with shapes whose labels are in $L_C$. The second part of computing $\mathrm{C}'$ only propagates existing shapes from one cache entry to another, and only from/to cache entries in $L_A$ or in labels in shapes in the cache entries. Thus, the cache entries of $\mathrm{C}'$ are only for $L_A$ with shapes with labels in $L_A$. If $\mathrm{lbl}(s) \in L_C$ then by previous argument $\mathrm{lbl}(s) \notin \Upsilon'$, as required. If $\mathrm{lbl}(s) \in L_M$ then we will show that $s \in \mathrm{C}''(j)$ for some $j \in L_I$. Then $s \in \mathrm{C}(j)$, and since $\mathrm{C} \vdash \Upsilon$ and $j \notin \Upsilon$, $\mathrm{lbl}(s) \notin \Upsilon$ so $\mathrm{lbl}(s) \notin \Upsilon'$, as required. If $i \in L_C$ then by previous argument $i \notin \Upsilon'$, as required. If $i \in L_M$ then we will show that $\mathrm{C}''(j) \subseteq \mathrm{C}''(i)$ for some $j \in L_I$. Then since $\mathrm{C}''(j)$ is inhabited because it labels a type checked in $\mathrm{C}''; \varrho' \vdash \Gamma$ or $\mathrm{C}''; \varrho' \vdash \tau$, and since that required type shape has labels in $L$, $\emptyset \neq \mathrm{C}(j) \subseteq \mathrm{C}(i)$. Then since $\mathrm{C} \vdash \Upsilon$ and $j \notin \Upsilon$, $i \notin \Upsilon$, so $i \notin \Upsilon'$, as required. It remains to show the two conditions we claimed. Since $\mathrm{C}'$ was computed using a least fixed point, we prove these claims by induction on when $s$ was added to $\mathrm{C}'(i)$. Consider the cases:

- Case 1, $s$ was added to $i$ because $s \in \mathrm{C}'(j)$, $s \notin \mathrm{C}'(i)$, and $\mathrm{C}'(j) \subseteq \mathrm{C}(i)$ is required by the rules for variables, box expressions, frames, box values, or environments. In this case, $i$ and $j$ have to come from the same term, that is, either $i, j \in L_C$ or $i, j \in L$. If $\mathrm{lbl}(s) \in L_M$ then $s$ must have been added to $\mathrm{C}(j)$ previously in the second phase of constructing $\mathrm{C}'$, so by the induction hypothesis, $\mathrm{lbl}(s) \in \mathrm{C}''(k)$ for some $k \in L_I$. If $i \in L_M$ then $j \in L$. First note that the condition on $j$ and $i$ is also required to show that $\mathrm{C}''; \varrho' \vdash \Gamma$, $\mathrm{C}''; \varrho' \vdash e$, or $\mathrm{C}''; \varrho' \vdash \tau$, so $\mathrm{C}''(j) \subseteq \mathrm{C}''(i)$. If $j \in L_I$ then we have what we need. Otherwise $j \in L_M$, so $s$ was added to $\mathrm{C}'(j)$ previously in the second phase of constructing $\mathrm{C}'(j)$, so by the induction hypothesis, $\mathrm{C}''(k) \subseteq \mathrm{C}''(j)$ for some $k \in L_I$. Then $\mathrm{C}''(k) \subseteq \mathrm{C}''(i)$, as required.
- Case 2, $s$ was added to $i$ because $\mathrm{C}(\varrho'(\alpha)) = \mathrm{C}(i)$, required by the rule for type variables, did not hold and $s \in \mathrm{C}(\varrho'(\alpha))$. Similar to Case 1.
- Case 3, $s$ was added to $i$ because $boxC(j, i)$ is required, either $(\mathtt{box}_t\, i')^{j'}_{\mathtt{v}} \in \mathrm{C}'(j)$ or $(\mathtt{box}\, i')^{j'}_{\mathtt{t}} \in \mathrm{C}'(j)$, and $s$ was already in $\mathrm{C}(i')$. In this case, $i$ and $j$ have to come from the same term, that is, either $i, j \in L_C$ or $i, j \in L$. Note that the labels in any shape under consideration come from the same term, that is, either they are all in $L$ or they are all in $L_C$.
  - If $\mathrm{lbl}(s) \in L_M$ then:
    - If $s$ was added to $\mathrm{C}'(i')$ previously in the second phase of constructing $\mathrm{C}'$ then by the induction hypothesis, $s \in \mathrm{C}''(k)$ for some $k \in L_I$.
    - Otherwise $s \in \mathrm{C}''(i')$ and $i', j' \in L$. Since $s$ was not already in $\mathrm{C}'(i)$ then the box shape was added to $\mathrm{C}'(j)$ previously in the second phase of constructing $\mathrm{C}'$, so by the induction hypothesis, the box shape is in $\mathrm{C}''(k)$ for some $k \in L_I$. By (**), $k$ labels a box type. By the rules for acceptability, $boxC(k, k')$ for $\mathrm{C}''$ for some $k' \in L_I$. Thus, $\mathrm{C}''(i') \subseteq \mathrm{C}''(k')$. Thus $s \in \mathrm{C}''(k')$, as required.
  - If $i \in L_M$ then $j \in L$ and $boxC(j, i)$ holds for $\mathrm{C}''$.
    - If the box shape was added to $\mathrm{C}'(j)$ previously in the second phase of the constructing $\mathrm{C}'$ then by the induction hypothesis, $\mathrm{C}''(k) \subseteq \mathrm{C}''(j)$ for some $k \in L_I$. By (**), $k$ labels a box type. Then by the rules for acceptability, $(\mathtt{box}\, i'')^{j''}_{\mathtt{t}} \in \mathrm{C}''(k)$ for some $i'' \in L_I$, so $(\mathtt{box}\, i'')^{j''}_{\mathtt{t}} \in \mathrm{C}''(j)$. By $boxC(j, i)$, $\mathrm{C}''(i'') \subseteq \mathrm{C}''(i)$, as required.
    - Otherwise, the box shape was in $\mathrm{C}''(j)$ and $i', j' \in L$. By $boxC(j, i)$, $\mathrm{C}''(i') \subseteq \mathrm{C}''(i)$. Since $s$ was not already in $\mathrm{C}'(i)$ then it was previously added to $\mathrm{C}'(i')$ in the second phase of constructing $\mathrm{C}'$, so by the induction hypothesis, $\mathrm{C}''(k) \subseteq \mathrm{C}''(i')$ for some $k \in L_I$. Then by transitivity $\mathrm{C}''(k) \subseteq \mathrm{C}''(i)$, as required.
- Case 4, $s$ was added because a $funC(j_1, j_2, j_3, j_4)$ is required. Similar to Case 3.

Claim: $\Upsilon' \vdash (\rho', E'\langle e \rangle)$. The proof is by a straight forward induction on the structure of $(\rho', E'\langle e \rangle)$. The only interesting case is application. In that case, we have $(e_1[\tau]\ e_2)^i$. By the induction hypothesis we get $\Upsilon' \vdash e_1$ and $\Upsilon' \vdash e_2$. We just need to show that $\Upsilon' \vdash \tau : \mathtt{r}$. If the application came from $e$ then since the labels of $\tau$ are in $L$, the result follows from $\Upsilon \vdash \tau : \mathtt{r}$, which holds by assumption ($\Upsilon \vdash e$). Otherwise, the labels of $\tau$ are not in $\Upsilon'$ so clearly $\Upsilon' \vdash \tau tr(\tau)$. Then $tr(\tau) = \mathtt{r}$ holds by the typing rules.

■

We these definitions we can prove that unboxing for modules is correct.

**Theorem 6**
*If* $\vdash (\Gamma \Rightarrow e : \tau)$ *wf*, $\mathrm{C}; \varrho \vdash (\Gamma \Rightarrow e : \tau)$, $\mathrm{C} \vdash \Upsilon$, *and* $\Upsilon \vdash (\Gamma \Rightarrow e : \tau)$ *then* $\vdash (\Gamma \Rightarrow e : \tau) \equiv \lfloor(\Gamma \Rightarrow e : \tau)\rfloor_\Upsilon$.

**Proof:** By definition, $\lfloor(\Gamma \Rightarrow e : \tau)\rfloor_\Upsilon = (\Gamma \Rightarrow \lfloor e\rfloor_\Upsilon : \tau)$. Clearly $\Gamma = \Gamma$ and $\tau = \tau$, so it remains to show that $\Gamma \vdash e \equiv \lfloor e\rfloor_\Upsilon : \tau$. By $\vdash (\Gamma \Rightarrow e : \tau)$ *wf*, $\emptyset \vdash \Gamma$ *wf* and $\Gamma \vdash e : \tau$. By $\mathrm{C}; \varrho \vdash (\Gamma \Rightarrow e : \tau)$, $\mathrm{C}; \varrho \vdash \Gamma$, $\mathrm{C}; \varrho \vdash e$, and $\mathrm{C}; \varrho \vdash \tau$. By $\Upsilon \vdash (\Gamma \Rightarrow e : \tau)$, $\Upsilon \vdash \Gamma$ *not unboxed*, $\Upsilon \vdash e$, and $\Upsilon \vdash \tau$ *not unboxed*. By Theorem 1, $\lfloor\Gamma\rfloor_\Upsilon \vdash \lfloor e\rfloor_\Upsilon : \lfloor\tau\rfloor_\Upsilon$. By Lemma 34, $\Gamma \vdash \lfloor e\rfloor_\Upsilon : \tau$. Let $\rho$ and $E$ be such that $\vdash \rho : \Gamma$ and $\Gamma \vdash E : \mathtt{B}^i\langle\tau\rangle$. Then by Lemma 37, there exists $\rho'$, $E'$, $\mathrm{C}'$, $\varrho'$, and

$\Upsilon'$ such that:

$$\begin{array}{l} \rho \sim_{\mathsf{s}} \rho' \\ E \sim_{\mathsf{s}} E' \\ \vdash \rho' : \Gamma \\ \Gamma \vdash E' : \mathtt{B}^{i'}\langle\tau\rangle \\ \mathrm{C}'; \varrho' \vdash (\rho', E'\langle e\rangle) \\ \mathrm{C}' \vdash \Upsilon' \\ \Upsilon' \vdash (\rho', E'\langle e\rangle) \\ \mathrm{lbls}(\rho') \cap \Upsilon' = \emptyset \\ \mathrm{lbls}(E') \cap \Upsilon' = \emptyset \\ \Upsilon \cap \mathrm{lbls}(e) = \Upsilon' \cap \mathrm{lbls}(e) \end{array}$$

Since the operational semantics is deterministic, we just need to show that $(\rho, E\langle\downharpoonleft e\downharpoonright_{\Upsilon}\rangle)$ matches $(\rho, E\langle e\rangle)$ in behaviour. There are two cases:

- If $(\rho, E\langle e\rangle) \longmapsto^* (\rho, c^j)$ then by Lemma 33, $(\rho', E'\langle e\rangle) \longmapsto^* (\rho', c^{j'})$ for some $j'$. By Theorem 4, $\downharpoonleft(\rho', E'\langle e\rangle)\downharpoonright_{\Upsilon'} \longmapsto^* \downharpoonleft(\rho', c^{j'})\downharpoonright_{\Upsilon'}$. By both Lemma 35 and definition of unboxing, $(\rho', E'\langle\downharpoonleft e\downharpoonright_{\Upsilon'}\rangle) \longmapsto^* (\rho', c^{j'})$. Hence by Lemma 36, $(\rho', E'\langle\downharpoonleft e\downharpoonright_{\Upsilon}\rangle) \longmapsto^* (\rho', c^{j'})$. Therefore by Lemma 33 again, $(\rho, E\langle\downharpoonleft e\downharpoonright_{\Upsilon}\rangle) \longmapsto^* (\rho', c^{j''})$, for some $j''$. It is not too hard to see that $j = j''$, as required.
- If $(\rho, E\langle e\rangle) \longmapsto \cdots$ then by Lemma 33, $(\rho', E'\langle e\rangle) \longmapsto \cdots$. By Theorem 4, $\downharpoonleft(\rho', E'\langle e\rangle)\downharpoonright_{\Upsilon'} \longmapsto \cdots$. By Lemma 35, $(\rho', E'\langle\downharpoonleft e\downharpoonright_{\Upsilon'}\rangle) \longmapsto \cdots$. By Lemma 36, $(\rho', E'\langle\downharpoonleft e\downharpoonright_{\Upsilon}\rangle) \longmapsto \cdots$. By Lemma 33, $(\rho, E\langle\downharpoonleft e\downharpoonright_{\Upsilon}\rangle) \longmapsto \cdots$, as required.

■

## 7. Related work

This paper provides a modular approach to showing correctness of a realistic compiler optimization that rewrites the structure of program data structures in significant ways. Our approach uses an arbitrary inter-procedural reaching definitions analysis to eliminate unnecessary heap allocation in an intermediate representation in which object representation has been made explicit. Our optimization can be staged freely with other optimizations. Unlike any previous work that we are aware of, we account for correctness with respect to the meta-data requirements of the garbage collector. For presentational purposes, we have restricted our attention to the core concern of GC safety, but additional issues such as value size, dynamic type tests, etc. are straightforward to incorporate.

There has been substantial previous work addressing the problem of unboxing. Peyton Jones [3] introduced an explicit distinction between boxed and unboxed objects to provide a linguistic account of unboxing, and hence to allow a high-level compiler to locally eliminate unboxes of syntactically apparent box introduction operations. Leroy [4] defined a type-driven approach to adding coercions into and out of specialized representations. The type driven translation represented monomorphic objects natively (unboxed, in our terminology), and then introduced wrappers to coerce polymorphic uses into an appropriate form. To a first-order approximation, instead of boxing at definition sites this approach boxes objects at polymorphic use sites. This style of approach has the problem that it is not necessarily beneficial, since allocation is introduced in places where it would not otherwise be present. This is reflected in the slowdowns observed on some benchmarks described in the original paper. This approach also has the potential to introduce space leaks. In a later paper [5] Leroy argued that a simple untyped approach gives better and more predictable results.

Henglein and Jørgensen [2] defined a formal notion of optimality for local unboxings and gave two different choices of coercion placements that satisfy their notion of optimality. Their definition of optimality explicitly does not correspond in any way to reduced allocation or reduced instruction count and does not seem to provide uniform improvement over Leroy's approach.

The MLton compiler [11] largely avoids the issue of a uniform object representation by completely monomorphizing programs before compilation. This approach requires whole-program compilation. More limited monomorphization schemes could be considered in an incremental compilation setting. Monomorphization does not eliminate the need for boxing in the presence of dynamic type tests or reflection. Just in time compilers (e.g. for .NET) may monomorphize dynamically at runtime.

The TIL compiler [1, 10] uses intensional type analysis in a whole-program compiler to allow native data representations without committing to whole-program compilation. As with the Leroy coercion approach, polymorphic uses of objects require conditionals and boxing coercions to be inserted at use sites, and consequently there is the potential to slow down, rather than speed up, the program.

Serrano and Feeley [9] described a flow analysis for performing unboxing substantially similar in spirit to our approach. Their algorithm attempts to find a monomorphic typing for a program in which object representations have not been made explicit, which they then use selectively to choose whether to use a uniform or non-uniform representation for each particular object. Their approach differs in that they define a dedicated analysis rather than using a generic reaching definitions analysis. They assume a conservative garbage collector and hence do not need to account for the requirements of GC safety, and they do not prove a correctness result.